\documentclass[aps,preprint,floats,epsf,epsfig,nofootinbib,letter]{revtex4}

\usepackage{epsfig}
\usepackage{graphicx}
\usepackage{dcolumn}
\usepackage{bm}
\usepackage{subfigure}

\def\DESepsf(#1 width #2){\epsfxsize=#2 \epsfbox{#1}}

\begin{document}
\def\be{\begin{eqnarray}}
\def\en{\end{eqnarray}}
\def\non{\nonumber}
\def\la{\langle}
\def\ra{\rangle}
\def\Br{{\mathcal B}}
\def\BB{{{\cal B} \overline {\cal B}}}
\def\BD{{{\cal B} \overline {\cal D}}}
\def\DB{{{\cal D} \overline {\cal B}}}
\def\DD{{{\cal D} \overline {\cal D}}}
\def\sq{\sqrt}


\title{Charmless Two-body Baryonic $B_{u,d,s}$ Decays Revisited}

\author{Chun-Khiang Chua}
\affiliation{Department of Physics and Center for High Energy Physics,
Chung Yuan Christian University,
Chung-Li, Taiwan 320, Republic of China}

\date{\today}

\begin{abstract}
We study charmless two-body baryonic $B$ decays using the topological amplitude approach.
We extend a previous work to include all ground state octet and decuplet final states with full topological amplitudes. Relations on rates and CP asymmetries are obtained.
The 
number of independent topological amplitudes is significantly reduced in the large $m_B$ asymptotic limit.
With the long awaited $\overline B{}^0\to p\bar p$ data, we can finally extract information on the topological amplitudes and predict rates of all other modes. 
The predicted rates are in general with uncertainties of a factor of two by including corrections to the asymptotic relations and from sub-leading contributions.
We point out some modes that will cascadely decay to all charged final states and have large decay rates.
For example, 
$\overline B {}^0_s\to\Omega^-\overline{\Omega^-}$, $B^-\to p\overline{\Delta^{++}}$,
$B^-\to \Lambda\bar p$ and $\overline B{}^0_s\to \Lambda\overline{\Lambda}$ decays are interesting modes to search for.
We find that the $\overline B{}^0\to p\bar p$ mode is the most accessible one among octet-anti-octet final states in the $\Delta S=0$ transition. It is not surprise that it is the first $\overline B_q\to \BB$ mode being observed. 
With the detection of $\pi^0$ and/or $\gamma$ many other unsuppressed modes can be searched for.
The predicted $\overline B{}^0_s\to p\bar p$ rate is several order smaller than the present experimental result.
The central value of the experimental result can be reproduced only with unnaturally scaled up ``subleading contributions", which will affect other modes including the $\overline B{}^0\to p\bar p$ decay. We need more data to clarify the situation.
The analysis presented in this work can be systematically improved when more measurements on decay rates become available.
\end{abstract}


\maketitle

\section{Introduction}

Recently, following the observation of $B^-\to\Lambda(1520) \bar p$ decay~\cite{Aaij:2013fla},
LHCb collaboration found the evidence
for the charmless two-body baryonic mode, 
$\overline B{}^0\to p\bar p$, with~\cite{Aaij:2013fta}
\be
{\cal B}(\overline B {}^0\to p \overline{p})=
          1.47{}^{+0.62+0.35}_{-0.51-0.14}\times 10^{-8}.
\en 
and also obtained
\be
{\cal B}(\overline B{}^0_s\to p\bar p)
=(2.84{}^{+2.03+0.85}_{-1.68-0.18})\times 10^{-8}.
\en
The present experimental situation for charmless two-body baryonic decay rates is shown in Table~\ref{tab:expt} \cite{Aaij:2013fla,Aaij:2013fta,Tsai:2007pp,Wang:2007as,Wei:2007fg}.
Many three-body baryonic modes have been observed~\cite{PDG}, and show threshold enhancement behavior, with the baryon pair moving colinearly, in their spectra. It has been conjectured that the threshold enhancement is the underlying reason of the large three body rates from the two-body ones~\cite{Hou:2000bz}. 
The rates and threshold enhancement can be understood and reproduced theoretically with factorization approach right after the observations of some of the three-body modes~\cite{Chua:2001vh,Chua:2003it,Chua:2001xn,Cheng:2001tr,Cheng:2002fp,Chua:2002wn,Chua:2002yd,Geng}.
For reviews, see~\cite{review, review1}.

On the other hand, progress on the study of two-body modes is slow and on a smaller scale~\cite{review, review1}.
The two-body baryonic decays are in general non-factorizable, which makes the theoretical study difficult. In general, one has to resort to model calculations. There are pole
model~\cite{Deshpande:1987nc,Jarfi:1990ej,Cheng:2001tr,Cheng:2001ub},
sum rule~\cite{Chernyak:ag}, diquark
model~\cite{Ball:1990fw,Chang:2001jt} and flavor symmetry
related~\cite{Gronau:1987xq,He:re,Sheikholeslami:fa,Luo:2003pv}
studies.
Predictions from various models usually differ a lot, and explicit calculations
usually give too large rates on the charmless modes.
For example, all existing predictions on $\overline B{}^0\to p\bar p$ rate are off by several order of magnitude comparing to the LHCb result~\cite{Aaij:2013fta,review, review1}.  
%

\begin{table}[t!]
\caption{\label{tab:expt} Current experimental status of rates of two-body baryonic modes. Upper limits are at 90\% C.L..}
\begin{ruledtabular}
\begin{tabular}{lcc}
Mode
          & ${\mathcal B}(10^{-8})$
          & Reference
          \\
\hline 
$B{}^-\to \Lambda \overline{\Lambda}$
          & $<32$
          & \cite{Tsai:2007pp}
          \\                    
$B^-\to \Lambda\overline{p}$
          & $<32$
          & \cite{Tsai:2007pp}
          \\
$B^-\to \Lambda \overline{\Delta^+}$
          & $<82$
          & \cite{Wang:2007as}
          \\
$B^-\to \Delta^0 \overline{p}$
          & $<138$
          & \cite{Wei:2007fg}
          \\
$B^-\to p \overline{\Delta^{++}}$
          & $<14$
          & \cite{Wei:2007fg}
          \\
          \hline  
$\overline B {}^0\to p \overline{p}$
          & $1.47{}^{+0.62+0.35}_{-0.51-0.14}$
          &\cite{Aaij:2013fta}
          \\         
$\overline B{}^0\to \Lambda \overline{\Delta^0}$
          & $<93$
          & \cite{Wang:2007as}
          \\ 
          \hline
$\overline B {}^0_s\to p \overline{p}$
          & $2.84{}^{+2.03+0.85}_{-1.68-0.18}$
          &\cite{Aaij:2013fta}
          \\                             
\end{tabular}
\end{ruledtabular}
\end{table}

Given that direct computation is not reliable at this moment, it is thus useful to use symmetry related approach to relate modes and make use of the newly measured $B\to p\bar p$ rate to give information on other modes.
In \cite{Chua:2003it}, we use the quark diagram or the so-called topological
approach, which was proposed in and has been used extensively in mesonic modes~\cite{Zeppenfeld:1980ex,Chau:tk,Chau:1990ay,Gronau:1994rj,Gronau:1995hn} 
(for a recent review, see \cite{review}), 
to the charmless two-body baryonic decays and obtained predictions on relative rates.
In fact, the same approach was also applied to charmful baryonic $\overline B {}^0\to \Lambda_c^+\overline p$ decay~\cite{Luo:2003pv}. 
Note that the quark diagram approach is closely related to the SU(3) flavor
symmetry~\cite{Zeppenfeld:1980ex,Gronau:1994rj,Savage:ub}. 
It is important to stress that the topological approach
does not rely on any factorization assumption and, hence, 
is applicable to the study of non-factorizable decay modes, 
such as charmless two-body baryonic modes 
that we are interested to in this study. 
With the evidence on the $\overline B{}^0\to p\bar p$ mode, 
it is timely to revisit the subject.
In this work we will extend the previous work to include all topological amplitudes, where only dominant ones were considered previously~\cite{Chua:2003it}.
We can now make use of the newly observed $\overline B{}^0\to p\bar p$ rate to extract information on decay amplitudes and proceed to provide predictions on rates of all other charmless two-body baryonic modes of ground state octet and decuplet baryons. 

As a first step towards numerical study, we
use asymptotic relations in the large $m_B$ limit~\cite{Brodsky:1980sx} to relate various
topological amplitudes~\cite{Chua:2003it}. 
The number of independent amplitudes are significantly reduced.
It should be note that
the same technics has been used in the study of the
three-body case~\cite{Cheng:2001tr,Chua:2002wn,Chua:2002yd}. It
leads to encouraging results. For example, the experiment finding
of ${\mathcal B}(\Lambda\overline p\pi^-)>{\mathcal
B}(\Sigma^0\overline p\pi^-)$~\cite{Wang:2003yi} can be
understood~\cite{Chua:2002yd} and three-body decay spectra are
consistent with the QCD counting rule~\cite{Lepage:1979za}
expectations. Due to the large energy release, we expect the
asymptotic relations to work even better in the two-body case than in
the three-body case. The smallness of two-body decay rates may due
to some $1/m_B^2$ suppression as expected from QCD counting rules.
We will extract the asymptotic amplitude from the $\overline B{}^0\to p\bar p$ data.

We then try to relax the asymptotic relations and estimate uncertainties on rates. 
As we shall see, with the present situation, rates can only be predicted or estimated at best within a factor of two following the above procedure. 
However, even order of magnitude estimation on rates is useful, as it can single out several prominent modes that our experimental colleagues may be interested to search for. 
Furthermore, the results can be systematically improved when the measurements of other modes become available in the future.

The layout of this paper is as following.
In Sec. II, we give our formulation for baryonic decays modes,
including all ground state decuplet-decuplet, octet-decuplet and octet-octet final
states. 
Full topological amplitudes are given for these charmless two-body baryonic modes. 
Asymptotic relations are provided at the end of the section.
In Sec. III, we discuss the phenomenology of
the charmless two-body baryonic decays. Relations on rates and $A_{CP}$ using the full topological amplitudes are obtained. 
We give predictions on all charmless two-body baryonic modes with the input from the $\overline B{}^0\to p\bar p$ data. Some suggestion on
the experimental searching are put forward. In Sec IV we give the
conclusion followed by three appendices on a brief derivation of the
asymptotic relations, the decomposition of amplitudes into independent amplitudes and a collection of baryon decay rates.

\section{Formalism}\label{sec:formalism}

In this section, we first develop the formalism of topological amplitudes of charmless two-body baryonic $B_{u,d,s}$ decays. The full amplitudes of all ground state octet (${\cal B}$) and decuplet (${\cal D}$) baryon final states are given using the formulas. Simplification can be obtained in the large $m_B$ limit and the asymptotic forms of the amplitudes will be shown before we end this section.

\subsection{Effective Hamiltonian for topological decay amplitudes of charmless two-body baryonic $B$ decays}

\begin{figure}[t]
\centering
 \subfigure[]{
  \includegraphics[width=0.5\textwidth]{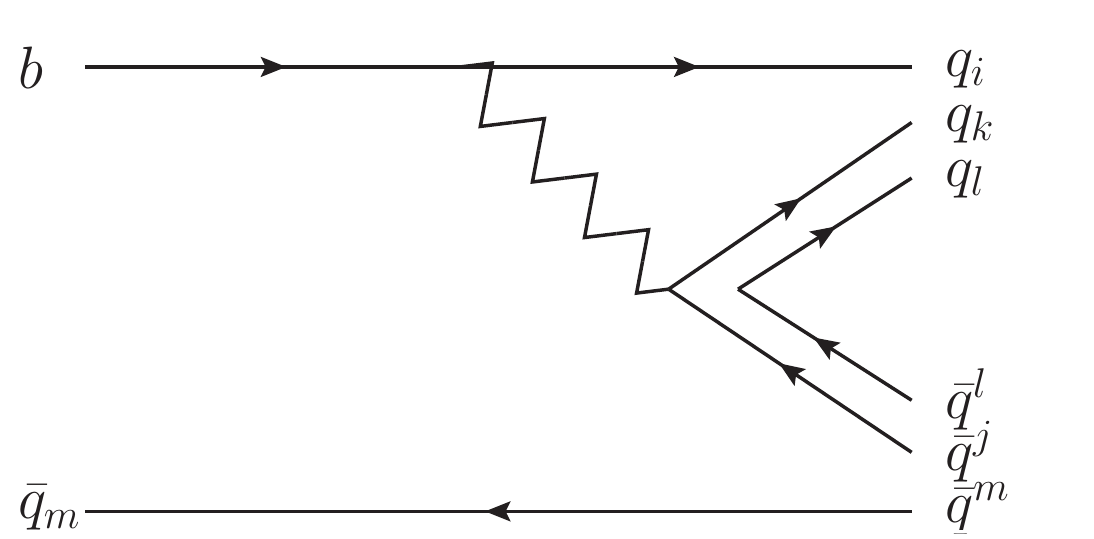}
}\subfigure[]{
  \includegraphics[width=0.5\textwidth]{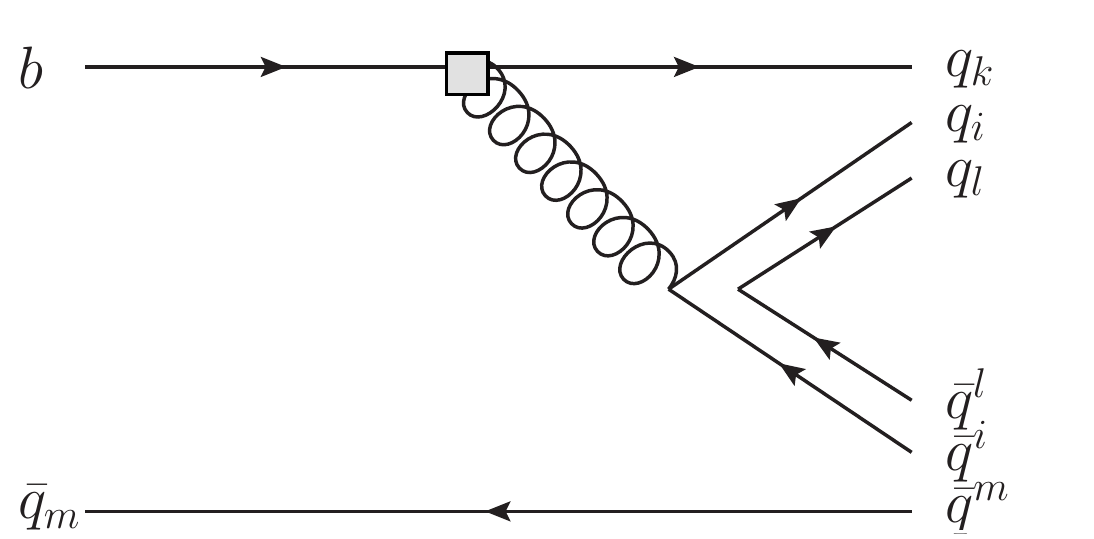}
}\\\subfigure[]{
  \includegraphics[width=0.5\textwidth]{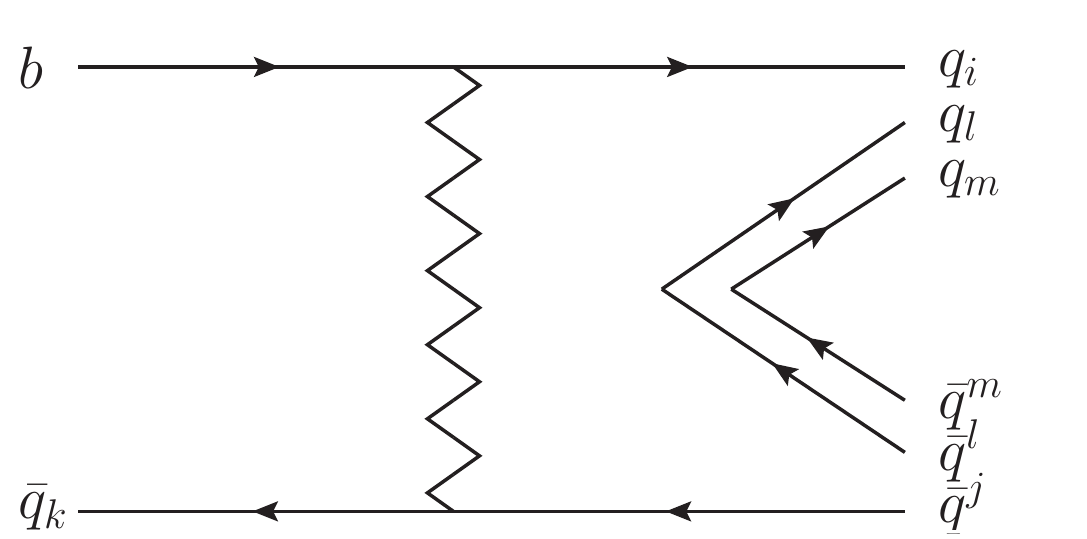}
}\subfigure[]{
  \includegraphics[width=0.5\textwidth]{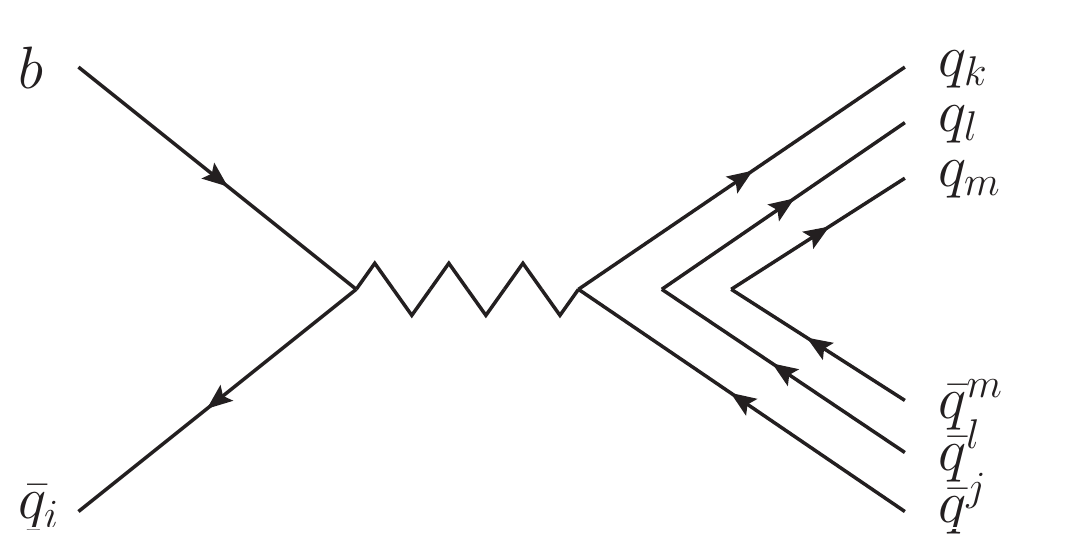}
}
\\\subfigure[]{
  \includegraphics[width=0.5\textwidth]{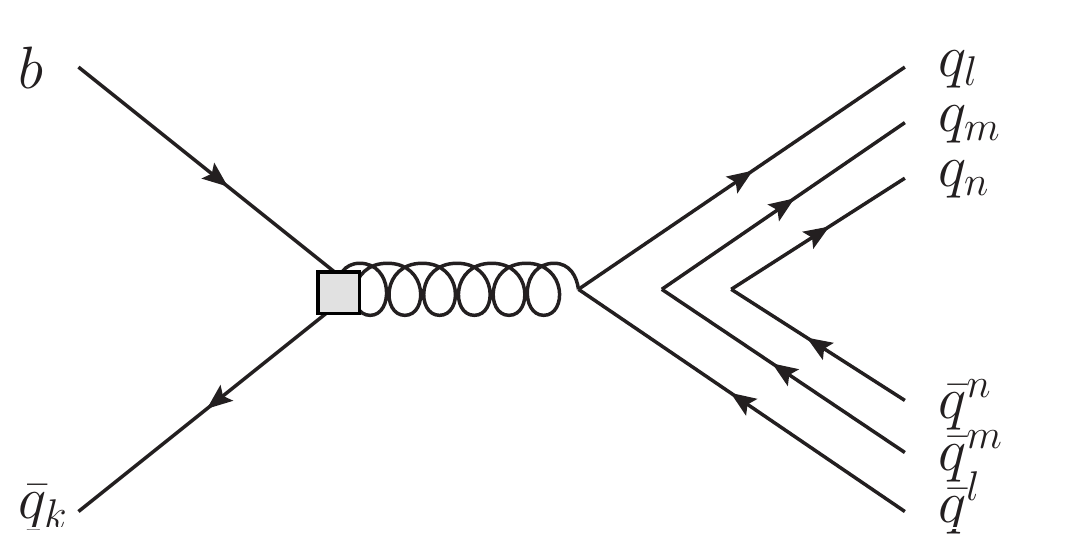}
}\subfigure[]{
  \includegraphics[width=0.5\textwidth]{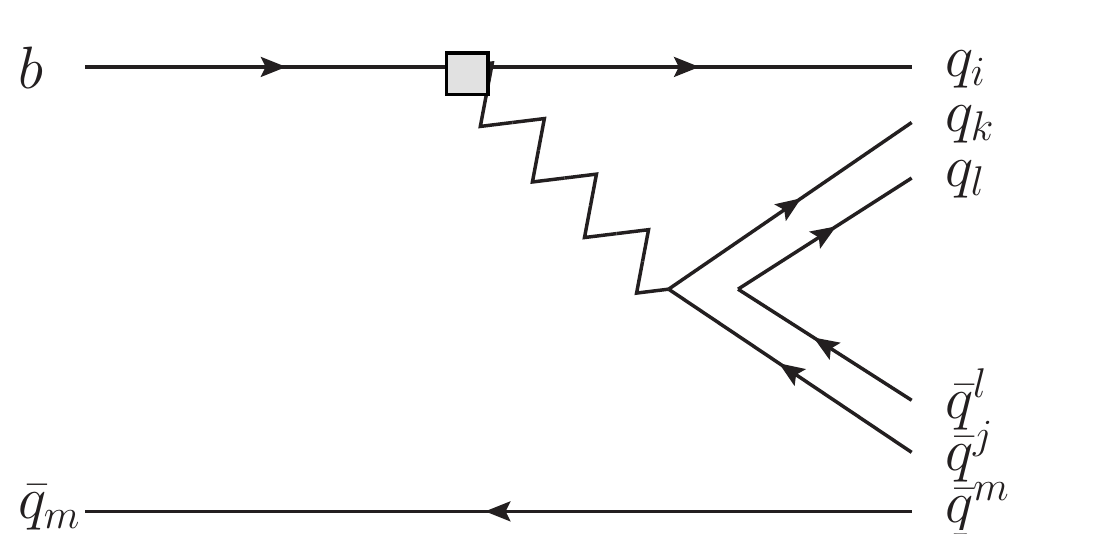}
}
\caption{Pictorial representation of
  (a) $T$ (tree), (b) $P$ (penguin), (c) $E$ ($W$-exchange),
  (d) $A$ (annihilation), (e) $PA$ (penguin annihilation) and (f) $P_{EW}$ (electroweak penguin)
  amplitudes in $\overline B$ to baryon pair decays. These are flavor flow diagrams. We use subscript and superscript according
to the field convention. For example, we assign a subscript
(superscript) to the initial (final) state anti-quark $\bar
q_m$~($\bar q^m$).
} \label{fig:TA}
\end{figure}

The effective weak Hamiltonian for charmless $B$ decays is~\cite{Buras}
\be
H_{\rm eff}=\frac{G_f}{\sq2}
                   \Big\{\sum_{r=u,c} V_{qb}V^*_{uq}[c_1 O^r_1+c_2 O^r_2]
                         -V_{tb} V^*_{tq}\sum_{i=3}^{10} c_i O_i\Big\}+{\rm H.c.},
\label{eq:H_eff1}                         
\en
where $q=d,s$, and
\be
O^r_1=(\bar r b)_{V-A}(\bar q r)_{V-A},
\quad
O^r_2=(\bar r_\alpha b_\beta)_{V-A}(\bar q_\beta r_\alpha)_{V-A}, 
\non\\
O_{3(5)}=(\bar q b)_{V-A}\sum_{q'}(\bar q' q')_{V\mp A},
\quad
O_{4(6)}=(\bar q_\alpha b_\beta)_{V-A}\sum_{q'}(\bar q_\beta' q_\alpha')_{V\mp A},
\non\\
O_{7(9)}=\frac{3}{2}(\bar q b)_{V-A}\sum_{q'}e_{q'}(\bar q' q')_{V\pm A},
\quad
O_{8(10)}=\frac{3}{2}(\bar q_\alpha b_\beta)_{V-A}\sum_{q'}e_{q'}(\bar q_\beta' q_\alpha')_{V\pm A},
\label{eq:H_eff2}   
\en
with $O_{3-6}$ the QCD penguin operators, $O_{7-10}$ the electroweak penguin operators, and $(\bar q' q)_{V\pm A}\equiv \bar q'\gamma_\mu(1+\pm\gamma_5)q$. The next-to-leading order Wilson coefficients,  
\be
c_1 = 1.081,\, 
c_2 = -0.190,\,
c_3 = 0.014,\,
c_4 = -0.036,\,
c_5 = 0.009,\, 
c_6 = -0.042,
\non\\ 
c_7 = -0.011\alpha_{EM},\, 
c_8 = 0.060\alpha_{EM},\, 
c_9 = -1.254\alpha_{EM},\,
c_{10} = 0.223\alpha_{EM},
\en
are evaluated in the naive dimensional regularization scheme at scale $\mu=4.2$GeV~\cite{Beneke:2001ev}.

We will concentrate on the flavor structure of the effective Hamiltonian first. We follow the approach of \cite{Chua:2003it}. As shown in Fig.~\ref{fig:TA}, we
have tree ($T$), penguin ($P$), electroweak penguin($P_{EW}$), $W$-exchange ($E$) annihilation and penguin annihilation ($PA$) amplitudes.
%
%
It is straightforward to obtain the coefficients of these
topological amplitudes. 
We recall that for the $b\to u\bar u d$ and $b\to q\bar q d$ processes, the
tree (${\cal O}_T=O_{1,2}$), penguin (${\cal O}_P=O_{3-6}$) and electroweak penguin (${\cal O}_{EWP}=O_{7-10}$) operators 
have the following flavor quantum numbers~[see Eqs.~(\ref{eq:H_eff1}) and (\ref{eq:H_eff2})]
\begin{eqnarray}
&&{\cal O}_T\sim (\bar u b )(\bar d u )
 =H^{ik}_j (\bar q_i b) (\bar q_k q^j),
\quad
{\cal O}_P\sim(\bar d b) (\bar q_i  q^i )
 =H^k (\bar q_k b) (\bar q_i q^i) ,
\non\\
&&{\cal O}_{EWP}\sim Q_j(\bar d b) ( \bar q_j q^j)
 =H_{EW}{}^{ik}_j (\bar q_i b) (\bar q_k q^j), 
  \nonumber\\
&&H^{12}_1=1=H^2,H_{EW}{}^{2k}_j=Q_j\delta^k_j\,\,\,
{\rm otherwise}\,\,\,H^{ik}_j=H_{EW}{}^{ik}_j=H^k=0,
\end{eqnarray}
respectively.~\footnote{Note that
$H^{ik}_i(=H^k)$ does not lead to any additional term.}
Note that the above equations also apply to the $|\Delta S|=1$
case, with $d$, $H^{12}_1=1=H^2$ and $H_{EW}{}^{2k}_j=Q_j\delta^k_j$ 
replaced by $s$,
$H^{13}_1=1=H^3$ and $H_{EW}{}^{3k}_j=Q_j\delta^k_j$, respectively.

We are now ready to proceed to $\overline B$ to decuplet-anti-decuplet decays.
A decuplet with $q_k q_i q_l$ flavor as
shown in Fig.~\ref{fig:TA} is produced by a $\overline {\cal D}_{kil}$
field, while a decuplet with $\bar q^l \bar q^j \bar q^m$ flavor
is created by a ${\cal D}^{jlm}$ field, where ${\cal D}^{jlm}$ is
the familiar decuplet field with 
${\cal D}^{111}=\Delta^{++}$,
${\cal D}^{112}=\Delta^{+}/\sqrt3$, 
${\cal D}^{122}=\Delta^0/\sq3$,
${\cal D}^{222}=\Delta^-$, 
${\cal D}^{113}=\Sigma^{*-}/\sq3$,
${\cal D}^{123}=\Sigma^{*0}/\sq6$, 
${\cal D}^{223}=\Sigma^{*-}/\sq3$,
${\cal D}^{133}=\Xi^{*0}/\sq3$, 
${\cal D}^{233}=\Xi^{*-}/\sq3$
and ${\cal D}^{333}=\Omega^-$
(see, for example~\cite{text}).
Hence by using the correspondent rule,
we have
\begin{eqnarray}
H_{\rm eff}&=& 6\,T_{{\cal D}\overline{\cal D}}\,\overline B_m
H^{ik}_j \overline {\cal D}_{ikl} {\cal D}^{ljm}
              +2P_{{\cal D}\overline{\cal D}}\,\overline B_m H^k
             \overline {\cal D}_{kil} {\cal D}^{lim}
             +6E_\DD\,\overline B_k H^{ik}_j
              \overline {\cal D}_{ilm} {\cal D}^{mlj}
\nonumber\\
            && +6A_\DD\,\overline B_i H^{ik}_j
               \overline {\cal D}_{klm} {\cal D}^{mlj}
              +6PA_\DD\,\overline B_k H^k
               \overline {\cal D}_{lmn} {\cal D}^{nml}
               +6\,P_{EW{\cal D}\overline{\cal D}}\,\overline B_m
H^{ik}_{EW\,j} \overline {\cal D}_{ikl} {\cal D}^{ljm},
\non\\
              \label{eq:DD}
\end{eqnarray}
with ${\overline B}_m=\left(
B^- ,\overline B {}^0, \overline B {}^0_s\right)$.
Without lost of generality,
the pre-factors before the above terms are assigned for
latter purpose. 



For $\overline B$ to octet-anti-decuplet baryonic decays, the anti-decuplet part is as before, 
while for the octet part, we have~\cite{text}
\begin{eqnarray}
{\mathcal B}= \left(
\begin{array}{ccc}
{{\Sigma^0}\over\sqrt2}+{{\Lambda}\over\sqrt6}
       &{\Sigma^+}
       &{p}
       \\
{\Sigma^-}
       &-{{\Sigma^0}\over\sqrt2}+{{\Lambda}\over\sqrt6}
       &{n}
       \\
{\Xi^-}
       &{\Xi^0}
       &-\sqrt{2\over3}{\Lambda}
\end{array}
\right), \label{eq:octet}
\end{eqnarray}
and note that the ${\cal B}^j_k$ has the flavor structure $q^j q^a
q^b \epsilon_{a,b,k}-\frac{1}{3}\,\delta^j_k q^c q^a
q^b$~\cite{text}. To match the flavor of $q_i q_k q_l$, $\bar
q^l\bar q^j\bar q^m$ final states as shown in Fig.~\ref{fig:TA},
we use
\begin{eqnarray}
q_i q_k q_l&\to& \epsilon_{ika} \overline {\cal B}^a_l,\,\
                 \epsilon_{ial} \overline {\cal B}^a_k,\,\
                 (\epsilon_{akl} \overline {\cal B}^a_i),
\nonumber\\
\bar q^l\bar q^j\bar q^m&\to& \epsilon^{ljb} {\cal B}^m_b,\,\
                              \epsilon^{lbm} {\cal B}^j_b,\,\
                              (\epsilon^{bjm} {\cal B}^l_b),
\label{eq:qqq}
\end{eqnarray}
as corresponding rules in obtaining $H_{\rm eff}$.
Since not all terms shown in the above equation are
independent,
$
\epsilon_{ika} \overline {\cal B}^a_l+\epsilon_{ial} \overline
{\cal B}^a_k+\epsilon_{akl} \overline {\cal
B}^a_i=0=\epsilon^{ljb} {\cal B}^m_b+
               \epsilon^{lbm} \overline {\cal B}^j_b+
               \epsilon^{bjm} \overline {\cal B}^l_b,
$
for each of the $q_k q_j q_l$ and $\bar q^l\bar q^j\bar q^m$
configurations we only need two independent terms. To be specific those in the parentheses in Eq.~(\ref{eq:qqq}) will not be used. 

To obtain the effective Hamiltonian for the
$B\to{\cal B}\overline{\cal D}$ decays, we replace
$\overline{\cal D}_{ikl}$ in Eq.~(\ref{eq:DD}) by
$(\overline {\cal B}_1)_{ikl}\equiv \epsilon_{ika}\overline {\cal B}^a_l$ and
$(\overline {\cal B}_2)_{ikl}\equiv\epsilon_{akl}\overline {\cal B}^a_i$ and get
\begin{eqnarray}
H_{\rm eff}&=& -\sqrt{6}\,T_{1\BD}\,
                        \overline B_m H^{ik}_j
                        \epsilon_{ika}\overline {\cal B}^a_l
                        {\cal D}^{ljm}
               -2\sqrt{6}\,T_{2\BD}\,
                        \overline B_m H^{ik}_j
                        \epsilon_{akl}\overline {\cal B}^a_i
                       {\cal D}^{ljm}
\nonumber\\
           && -\sqrt{6} P_\BD\,
                        \overline B_m H^k
                        \epsilon_{kia} \overline {\cal B}^a_{l}
                       {\cal D}^{lim}
                  -\sq6 E_\BD\,
                        \overline B_k H^{ik}_j
                        \epsilon_{ila}\overline {\cal B}^a_m
                        {\cal D}^{mlj}
                  -\sq6 A_\BD\,
                        \overline B_i H^{ik}_j 
                        \epsilon_{kla}\overline{\cal B}^a_m
                        {\cal D}^{mlj}
                        \non\\
&&-\sqrt{6}\,P_{1EW\BD}\,
                        \overline B_m H_{EW}{}^{ik}_j
                        \epsilon_{ika}\overline {\cal B}^a_l
                        {\cal D}^{ljm}
               -2\sqrt{6}\,P_{2EW\BD}\,
                        \overline B_m H_{EW}{}^{ik}_j
                        \epsilon_{akl}\overline {\cal B}^a_i
                       {\cal D}^{ljm},
\label{eq:BD}
\end{eqnarray}
where some pre-factors are introduced for later purpose.  
Note that terms obtained with the replacement $\overline {\cal D}\to \overline {\cal B}_2$ from penguin, exchange and annihilation topologies of Eq.~(\ref{eq:DD}) are vanishing. 
We have two tree, one penguin, one exchange, one annihilation, two electroweak penguin and no penguin annihilation amplitudes. 
For example, penguin annihilation amplitude cannot exist in this case as the decuplet is symmetric in flavor index, while the octet part comes in through antisymmetric combination.
%


For the $\overline B\to{\cal D}\overline{\cal B}$ case, by
replacing ${\cal D}^{ljm}$ in Eq.~(\ref{eq:DD}) by 
$({\cal B}_1)^{lim}\equiv\epsilon^{ljb}{\cal B}^m_b$ and 
$({\cal B}_2)^{lim}\equiv\epsilon^{bjm} {\cal B}^l_b$, we
have
\begin{eqnarray}
H_{\rm eff}&=& -\sqrt{6}\,T_{1\DB}\,
                        \overline B_m H^{ik}_j \overline{\cal D}_{ikl}
                        \epsilon^{ljb} {\cal B}^m_b
               +\sqrt{6}\,T_{2\DB}\,
                        \overline B_m H^{ik}_j \overline{\cal D}_{ikl}
                        \epsilon^{bjm} {\cal B}^l_b
\nonumber\\
           && +\sqrt{6} P_\DB\,
                        \overline B_m H^k \overline{\cal D}_{kil} 
                        \epsilon^{bim}{\cal B}^l_{b}
                   +\sq6 E_\DB\,
                        \overline B_k H^{ik}_j \overline{\cal D}_{ilm}
                        \epsilon^{bli} {\cal B}^m_b
                     +\sq6 A_\DB\,
                        \overline B_i H^{ik}_j \overline{\cal D}_{klm}
                        \epsilon^{bli} {\cal B}^m_b   
\non\\
           &&-\sqrt{6}\,P_{1EW\DB}\,
                        \overline B_m H_{EW}{}^{ik}_j \overline{\cal D}_{ikl}
                        \epsilon^{ljb} {\cal B}^m_b
               +\sqrt{6}\,P_{2EW\DB}\,
                        \overline B_m H_{EW}{}^{ik}_j \overline{\cal D}_{ikl}
                        \epsilon^{bjm} {\cal B}^l_b.
\label{eq:DB}
\end{eqnarray}
Without lost of generality, we introduce some pre-factors for later purpose. 
Note that terms obtained with the replacement ${\cal D}\to {\cal B}_1$ from penguin, exchange and annihilation topologies of Eq.~(\ref{eq:DD}) are vanishing.  
We have two tree, one penguin, one exchange, one annihilation, two electroweak penguin and no penguin annihilation amplitudes.

To obtain $\overline B\to {\cal B}\overline {\cal B}$ decays
effective Hamiltonian, we first replace 
$\overline{\cal D}_{ikl}$ and ${\cal D}^{ljm}$ in Eq.~(\ref{eq:DD}) by 
$(\overline{\cal B}_1)_{ikl}\equiv\epsilon_{ika}\overline
{\cal B}^a_l$ 
and $(\overline{\cal B}_2)_{ikl}\equiv\epsilon_{akl}\overline {\cal B}^a_i$, and
$({\cal B}_1)^{lim}\equiv\epsilon^{ljb} {\cal B}^m_b$ and $({\cal B}_2)^{lim}\equiv\epsilon^{bjm} {\cal B}^l_b$,
respectively, and obtain
\be
H_{\rm eff}&=&(H_{\rm eff})_{11}-(H_{\rm eff})_{12}+2(H_{\rm eff})_{21}-2(H_{\rm eff})_{22},
\non\\
(H_{\rm eff})_{pq}&\equiv& T_{pq\BB}\,
                        \overline B_m H^{ik}_j
                        (\overline {\cal B}_p)_{ikl}({\cal B}_q)^{ljm}
                  +P_{pq\BB}\,
                        \overline B_m H^k
                        (\overline {\cal B}_p)_{kil} ({\cal B}_q)^{lim}
\non\\                        
            && +E_{pq\BB}\,
                        \overline B_k H^{ik}_j
                          (\overline {\cal B}_p)_{ilm}
                        ({\cal B}_q)^{mlj}
                    + A_{pq\BB}\,
                        \overline B_i H^{ik}_j
                         (\overline {\cal B}_p)_{klm}
                        ({\cal B}_q)^{mlj}
\non\\                        
            &&+P_{pqEW\BB}\,
                        \overline B_m H_{EW}{}^{ik}_j
                         (\overline {\cal B}_p)_{ikl}
                        ({\cal B}_q)^{ljm}
               +PA_{pq\BB}\,
                        \overline B_k H^k
                        (\overline {\cal B}_p)_{lmn}
                         ({\cal B}_q)^{nml},                                    
\label{eq:BB0}
\end{eqnarray}
where without lost of generality the coefficients in front of $(H_{\rm eff})_{pq}$ are assigned for later purpose.
Using identities,  
$-2(\overline {\cal B}_1)_{kil} ({\cal B}_1)^{lim}
=(\overline {\cal B}_2)_{kil} ({\cal B}_1)^{lim}
=-2(\overline {\cal B}_2)_{kil} ({\cal B}_2)^{lim}$,
$-2(\overline {\cal B}_1)_{lmn}({\cal B}_1)^{nml}
=(\overline {\cal B}_1)_{lmn}({\cal B}_2)^{nml}
=(\overline {\cal B}_2)_{lmn}({\cal B}_1)^{nml}
=-2(\overline {\cal B}_2)_{lmn}({\cal B}_2)^{nml}$, and redefinding
topological fields,~\footnote{Explicitly, we redefine
$T_{1\BB}\equiv T_{11\BB}$, $T_{2\BB}\equiv T_{12\BB}$, $T_{3\BB}\equiv T_{21\BB}$, $T_{4\BB}\equiv T_{22\BB}$ (and similarly for $P_{i EW\BB}$),
$-5P_{1\BB}\equiv P_{11\BB}-4P_{21\BB}-2P_{22\BB}$, $P_{2\BB}\equiv P_{12\BB}$ (and similarly for $A_{i\BB}$ and $E_{i\BB}$) and $-3PA_\BB=PA_{11\BB}+2PA_{12\BB}-4PA_{21\BB}-2PA_{22\BB}$.
}
we finally get
\begin{eqnarray}
H_{\rm eff}&=& T_{1\BB}\,
                        \overline B_m H^{ik}_j
                        \epsilon_{ika}\overline {\cal B}^a_l
                        \epsilon^{ljb} {\cal B}^m_b
                   -T_{2\BB}\,
                        \overline B_m H^{ik}_j
                        \epsilon_{ika}\overline {\cal B}^a_l
                        \epsilon^{bjm} {\cal B}^l_b
\non\\
            && +2T_{3\BB}\,
                        \overline B_m H^{ik}_j
                        \epsilon_{akl}\overline {\cal B}^a_i
                        \epsilon^{ljb} {\cal B}^m_b
                 -2T_{4\BB}\,
                        \overline B_m H^{ik}_j
                        \epsilon_{akl}\overline {\cal B}^a_i
                        \epsilon^{bjm} {\cal B}^l_b
\non\\
           && -5 P_{1\BB}\,
                        \overline B_m H^k
                        \epsilon_{kia} \overline {\cal B}^a_{l}
                        \epsilon^{lib} {\cal B}^m_{b}
                 -P_{2\BB}\,
                        \overline B_m H^k
                        \epsilon_{kia} \overline {\cal B}^a_{l}
                        \epsilon^{bim} {\cal B}^l_{b}
\non\\
         &&-5 E_{1\BB}\,
                        \overline B_k H^{ik}_j
                        \epsilon_{ila}\overline {\cal B}^a_m
                        \epsilon^{mlb} {\cal B}^j_b
              -E_{2\BB}\,
                        \overline B_k H^{ik}_j
                        \epsilon_{ila}\overline {\cal B}^a_m
                        \epsilon^{blj} {\cal B}^l_m        
\non\\
         &&-5 A_{1\BB}\,
                        \overline B_i H^{ik}_j
                        \epsilon_{kla}\overline {\cal B}^a_m
                        \epsilon^{mlb} {\cal B}^j_b
                -A_{2\BB}\,
                        \overline B_i H^{ik}_j
                        \epsilon_{kla}\overline {\cal B}^a_m
                        \epsilon^{blj} {\cal B}^l_m 
\non\\
            && +P_{1EW\BB}\,
                        \overline B_m H_{EW}{}^{ik}_j
                        \epsilon_{ika}\overline {\cal B}^a_l
                        \epsilon^{ljb} {\cal B}^m_b
                 -P_{2EW\BB}\,
                        \overline B_m H_{EW}{}^{ik}_j
                        \epsilon_{ika}\overline {\cal B}^a_l
                        \epsilon^{bjm} {\cal B}^l_b
\non\\
            &&+2P_{3EW\BB}\,
                        \overline B_m H_{EW}{}^{ik}_j
                        \epsilon_{akl}\overline {\cal B}^a_i
                        \epsilon^{ljb} {\cal B}^m_b
                 -2P_{4EW\BB}\,
                        \overline B_m H_{EW}{}^{ik}_j
                        \epsilon_{akl}\overline {\cal B}^a_i
                        \epsilon^{bjm} {\cal B}^l_b 
\non\\
              &&-3 PA_\BB\,
                        \overline B_k H^k
                        \epsilon_{lma}\overline {\cal B}^a_n
                        \epsilon^{nmb} {\cal B}^l_b.                                    
\label{eq:BB}
\end{eqnarray}
We have four tree, two penguin, two exchange, one annihilation and four electroweak penguin amplitudes.

All of the above results are for $\Delta S=0$ transitions. For $\Delta S=-1$ transitions, we use $T'$, $P'$ and so on for the corresponding topological amplitudes. 

\subsection{Topological amplitudes of two-body charmless baryonic $B$ decays}

Here we collect all the $\overline B\to\DD$, $\DB$, $\BD$, $\BB$ decay amplitudes expressed in term of topological amplitudes as obtained using formulas in the previous subsection. These are some of the main results of this work.

\subsubsection{$\overline B$ to decuplet-anti-decuplet baryonic decays}

The full $\bar B\to\DD$ decay amplitudes for $\Delta S=0$ processes are given by
\be
A(B^-\to \Delta^+\overline{\Delta^{++}})
   &=&2\sq3 T_\DD+2\sq3 P_\DD+\frac{4}{\sq3}P_{EW\DD}+2\sq3A_\DD,
   \non\\
A(B^-\to\Delta^0\overline{\Delta^+})
   &=&2T_\DD+4 P_\DD+\frac{2}{3}P_{EW\DD}+4A_\DD,
   \non\\
A(B^-\to\Delta^-\overline{\Delta^0})
   &=&2\sq3 P_\DD-\frac{2}{\sq3}P_{EW\DD}+2\sq3A_\DD,
   \non\\
A(B^-\to\Sigma^{*0}\overline{\Sigma^{*+}})
   &=&\sq2T_\DD+2\sq2 P_\DD+\frac{\sq2}{3}P_{EW\DD}+2\sq2A_\DD,
   \non\\   
A(B^-\to\Sigma^{*-}\overline{\Sigma^{*0}})
   &=&2\sq2 P_\DD-\frac{2\sq2}{3}P_{EW\DD}+2\sq2A_\DD,
   \non\\
A(B^-\to\Xi^{*-}\overline{\Xi^{*0}})
   &=&2 P_\DD-\frac{2}{3}P_{EW\DD}+2A_\DD,
\en
\be
A(\bar B^0\to \Delta^{++}\overline{\Delta^{++}})
   &=&6E_\DD+18PA_\DD,
   \non\\
A(\bar B^0\to\Delta^+\overline{\Delta^+})
   &=&2T_\DD+2 P_\DD+\frac{4}{3}P_{EW\DD}+4E_\DD+18PA_\DD,
   \non\\
A(\bar B^0\to\Delta^0\overline{\Delta^0})
   &=&2T_\DD+4 P_\DD+\frac{2}{3}P_{EW\DD}+2E_\DD+18PA_\DD,
   \non\\
A(\bar B^0\to\Delta^-\overline{\Delta^-})
   &=&6P_\DD-2P_{EW\DD}+18PA_\DD,
   \non\\
A(\bar B^0\to\Sigma^{*+}\overline{\Sigma^{*+}})
   &=&\frac{2}{3}6E_\DD+18PA_\DD,
   \non\\
A(\bar B^0\to\Sigma^{*0}\overline{\Sigma^{*0}})
   &=&T_\DD+2P_\DD+\frac{1}{3}P_{EW\DD}+2E_\DD+18PA_\DD,
   \non\\
A(\bar B^0\to\Sigma^{*-}\overline{\Sigma^{*-}})
   &=&4P_\DD-\frac{4}{3}P_{EW\DD}+18PA_\DD,
   \non\\   
A(\bar B^0\to\Xi^{*0}\overline{\Xi^{*0}})
   &=&\frac{1}{3}E_{\DD}+18PA_\DD,
   \non\\
A(\bar B^0\to\Xi^{*-}\overline{\Xi^{*-}})
   &=&2 P_\DD-\frac{2}{3}P_{EW\DD}+18PA_\DD,
   \non\\
A(\bar B^0\to\Omega^{-}\overline{\Omega^{-}})
   &=&18PA_\DD,      
\en
and
\be
A(\bar B^0_s\to \Delta^+\overline{\Sigma^{*+}})
   &=&2 T_\DD+2 P_\DD+\frac{4}{3}P_{EW\DD},
   \non\\
A(\bar B^0_s\to\Delta^0\overline{\Sigma^{*0}})
   &=&\sq2T_\DD+2\sq2 P_\DD+\frac{\sq2}{3}P_{EW\DD},
   \non\\
A(\bar B^0_s\to\Delta^-\overline{\Sigma^{*-}})
   &=&2\sq3 P_\DD-\frac{2}{\sq3}P_{EW\DD},
   \non\\
A(\bar B^0_s\to\Sigma^{*0}\overline{\Xi^{*0}})
   &=&\sq2T_\DD+2\sq2 P_\DD+\frac{\sq2}{3}P_{EW\DD},
   \non\\
A(\bar B^0_s\to\Sigma^{*-}\overline{\Xi^{*-}})
   &=&4P_\DD-\frac{4}{3}P_{EW\DD},
   \non\\   
A(\bar B^0_s\to\Xi^{*-}\overline{\Omega^-})
   &=&2\sq3 P_\DD-\frac{2}{\sq3}P_{EW\DD},
\en
while those for $\Delta S=1$ transitions are given by
\be
A(B^-\to \Sigma^{*+}\overline{\Delta^{++}})
   &=&2\sq3 T'_\DD+2\sq3 P'_\DD+\frac{4}{\sq3}P'_{EW\DD}+2{\sq3}A'_\DD,
   \non\\
A(B^-\to\Sigma^{*0}\overline{\Delta^+})
   &=&\sq2T'_\DD+2\sq2 P'_\DD+\frac{\sq2}{3}P'_{EW\DD}+2{\sq2}A'_\DD,
   \non\\
A(B^-\to\Sigma^{*-}\overline{\Delta^0})
   &=&2 P'_\DD-\frac{2}{3}P'_{EW\DD}+2A'_\DD,
   \non\\
A(B^-\to\Xi^{*0}\overline{\Sigma^{*+}})
   &=&2T'_\DD+4 P'_\DD+\frac{2}{3}P'_{EW\DD}+4A'_\DD,
   \non\\
A(B^-\to\Xi^{*-}\overline{\Sigma^{*0}})
   &=&2\sq2 P'_\DD-\frac{2\sq2}{3}P'_{EW\DD}+2{\sq2}A'_\DD,
   \non\\   
A(B^-\to\Omega^-\overline{\Xi^{*0}})
   &=&2\sq3 P'_\DD-\frac{2}{\sq3}P'_{EW\DD}+2{\sq3}A'_\DD,
\en
\be
A(\bar B^0\to \Sigma^{*+}\overline{\Delta^+})
   &=&2 T_\DD+2 P_\DD+\frac{4}{3}P_{EW\DD},
   \non\\
A(\bar B^0\to\Sigma^{*0}\overline{\Delta^0})
   &=&\sq2T_\DD+2\sq2 P_\DD+\frac{\sq2}{3}P_{EW\DD},
   \non\\
A(\bar B^0\to\Sigma^{*-}\overline{\Delta^-})
   &=&2\sq3 P_\DD-\frac{2}{\sq3}P_{EW\DD},
   \non\\
A(\bar B^0\to\Xi^{*0}\overline{\Sigma^{*0}})
   &=&\sq2T_\DD+2\sq2 P_\DD+\frac{\sq2}{3}P_{EW\DD},
   \non\\
A(\bar B^0\to\Xi^{*-}\overline{\Sigma^{*-}})
   &=&4P_\DD-\frac{4}{3}P_{EW\DD},
   \non\\   
A(\bar B^0\to\Omega^-\overline{\Xi^{*-}})
   &=&2\sq3 P_\DD-\frac{2}{\sq3}P_{EW\DD},
\en
and
\be
A(\bar B^0_s\to \Delta^{++}\overline{\Delta^{++}})
   &=&6E'_\DD+18PA'_\DD,
   \non\\
A(\bar B^0_s\to\Delta^+\overline{\Delta^+})
   &=&4E'_\DD+18PA'_\DD,
   \non\\
A(\bar B^0_s\to\Delta^0\overline{\Delta^0})
   &=&2E'_\DD+18PA'_\DD,
   \non\\
A(\bar B^0_s\to\Delta^-\overline{\Delta^-})
   &=&18PA'_\DD,
   \non\\
A(\bar B^0_s\to\Sigma^{*+}\overline{\Sigma^{*+}})
   &=&2T'_\DD+2P'_\DD+\frac{4}{3}P'_{EW\DD}+4E'_\DD+18PA'_\DD,
   \non\\
A(\bar B^0_s\to\Sigma^{*0}\overline{\Sigma^{*0}})
   &=&T'_\DD+2P'_\DD+\frac{1}{3}P'_{EW\DD}+2E'_\DD+18PA'_\DD,
   \non\\
A(\bar B^0_s\to\Sigma^{*-}\overline{\Sigma^{*-}})
   &=&2P'_\DD-\frac{2}{3}P'_{EW\DD}+18PA'_\DD,
   \non\\   
A(\bar B^0_s\to\Xi^{*0}\overline{\Xi^{*0}})
   &=&2T'_\DD+4P'_\DD+\frac{2}{3}P'_{EW\DD}+2E'_{\DD}+18PA'_\DD,
   \non\\
A(\bar B^0_s\to\Xi^{*-}\overline{\Xi^{*-}})
   &=&4 P'_\DD-\frac{4}{3}P'_{EW\DD}+18PA'_\DD,
   \non\\
A(\bar B^0_s\to\Omega^{-}\overline{\Omega^{-}})
   &=&6 P_\DD-2P_{EW\DD}+18PA'_\DD.
\en

\subsubsection{$\overline B$ to octet-anti-decuplet baryonic decays}

The full $\bar B\to\BD$ decay amplitudes for $\Delta S=0$ processes are given by
\be
A(B^-\to p\overline{\Delta^{++}})
   &=&-\sq6 (T_{1\BD}-2T_{2\BD})+\sq6 P_\BD+2\sq{\frac{2}{3}}P_{1EW\BD}+\sq6 A_\BD,
   \non\\
A(B^-\to n\overline{\Delta^+})
   &=&-\sq2T_{1\BD}+\sq2 P_\BD+\frac{2\sq2}{3}(P_{1EW\BD}-3P_{2EW\BD})+\sq2A_\BD,
   \non\\
A(B^-\to\Sigma^0\overline{\Sigma^{*+}})
   &=&-2T_{2\BD}-P_\BD+\frac{1}{3}(P_{1EW\BD}-6P_{2EW\BD})- A_\BD,
   \non\\ 
A(B^-\to\Sigma^-\overline{\Sigma^{*0}})
   &=&-P_\BD+\frac{1}{3}P_{1EW\BD}- A_\BD,
   \non\\      
A(B^-\to\Xi^{-}\overline{\Xi^{*0}})
   &=&-\sq2 P_\BD+\frac{\sq2}{3}P_{1EW\BD}-\sq2A_\BD,
   \non\\
A(B^-\to\Lambda\overline{\Sigma^{*+}})
   &=&\frac{2}{\sq3}(T_{1\BD}-T_{2\BD})-\sq3 P_\BD-\frac{1}{\sq3}(P_{1EW\BD}-2P_{2EW\BD})-\sq3 A_\BD,
   \non\\
\en
\be
A(\bar B^0\to p\overline{\Delta^+})
   &=&-\sq2(T_{1\BD}-2T_{2\BD})+\sq2 P_\BD+\frac{2\sq2}{3}P_{1EW\BD}-\sq2 E_\BD,
   \non\\
A(\bar B^0\to n\overline{\Delta^0})
   &=&-\sq2T_{1\BD}+\sq2 P_\BD+\frac{2\sq2}{3}(P_{1EW\BD}-3P_{2EW\BD})-\sq2 E_\BD,
   \non\\
A(\bar B^0\to\Sigma^{+}\overline{\Sigma^{*+}})
   &=&\sq2 E_\BD,
   \non\\
A(\bar B^0\to\Sigma^{0}\overline{\Sigma^{*0}})
   &=&-\sq2T_{2\BD}-\frac{1}{\sq2}P_\BD+\frac{1}{3\sq2}(P_{1EW\BD}-6P_ {2EW\BD})-\frac{1}{\sq2} E_\BD,
   \non\\
A(\bar B^0\to\Sigma^{-}\overline{\Sigma^{*-}})
   &=&-\sq2P_\BD+\frac{\sq2}{3}P_{1EW\BD},
   \non\\
A(\bar B^0\to\Xi^{0}\overline{\Xi^{*0}})
   &=&\sq2E_{\BD},
   \non\\
A(\bar B^0\to\Xi^{-}\overline{\Xi^{*-}})
   &=&-\sq2 P_\BD+\frac{\sq2}{3}P_{1EW\BD},
   \non\\
A(\bar B^0\to\Lambda\overline{\Sigma^{*0}})
   &=&\sq{\frac{2}{3}}(T_{1\BD}-T_{2\BD})-\sq{\frac{3}{2}} P_\BD
   -\frac{1}{\sq6}(P_{1EW\BD}-2P_{2EW\BD})+\sq{\frac{3}{2}}E_\BD,
   \non\\   
\en
and
\be
A(\bar B^0_s\to p\overline{\Sigma^{*+}})
   &=&-\sq2 (T_{1\BD}-2T_{2\BD})+\sq2 P_\BD+\frac{2\sq2}{3}P_{1EW\BD},
   \non\\
A(\bar B^0_s\to n\overline{\Sigma^{*0}})
   &=&-T_{1\BD}+P_\BD+\frac{2}{3}(P_{1EW\BD}-3P_{2EW\BD}),
   \non\\
A(\bar B^0_s\to\Sigma^{0}\overline{\Xi^{*0}})
   &=&-2T_{2\BD}-P_\BD+\frac{1}{3}(P_{1EW\BD}-6P_{2EW\BD}),
   \non\\
A(\bar B^0_s\to\Sigma^{-}\overline{\Xi^{*-}})
   &=&-\sq2P_\BD+\frac{\sq2}{3}P_{1EW\BD},
   \non\\   
A(\bar B^0_s\to\Xi^{-}\overline{\Omega^-})
   &=&-\sq6 P_\BD+\sq{\frac{2}{3}}P_{1EW\BD},
   \non\\
A(\bar B^0_s\to\Lambda\overline{\Xi^{*0}})
   &=&\frac{2}{\sq3}(T_{1\BD}-T_{2\BD})
         -\sq3 P_\BD
         -\frac{1}{\sq3}(P_{1EW\BD}-2P_{2EW\BD}),
\en
while those for $\Delta S=1$ transitions are given by
\be
A(B^-\to \Sigma^+\overline{\Delta^{++}})
   &=&\sq6 (T'_{1\BD}-2T'_{2\BD})
          -\sq6P'_\BD
          -2\sq{\frac{2}{3}}P'_{1EW\BD}
          -\sq6 A'_\BD,
   \non\\
A(B^-\to\Sigma^0\overline{\Delta^+})
   &=&-T'_{1\BD}+2T'_{2\BD}
   +2P'_\BD
   +\frac{1}{3}P'_{1EW\BD}
   +2 A'_\BD,
   \non\\
A(B^-\to\Sigma^-\overline{\Delta^0})
   &=&\sq2 P'_\BD
         -\frac{\sq2}{3}P'_{1EW\BD}
         +\sq2 A'_\BD,
   \non\\
A(B^-\to\Xi^{0}\overline{\Sigma^{*+}})
   &=&\sq2T'_{1\BD}
         -\sq2 P'_\BD
         -\frac{2\sq2}{3}(P'_{1EW\BD}-3P'_{2EW\BD})
         -\sq2 A'_\BD,
   \non\\   
A(B^-\to\Xi^{-}\overline{\Sigma^{*0}})
   &=&P'_\BD-\frac{1}{3}P'_{1EW\BD}+ A'_\BD,
   \non\\
A(B^-\to\Lambda\overline{\Delta^{+}})
   &=&\frac{1}{\sq3}(T'_{1\BD}+2T'_{2\BD})
         -\frac{1}{\sq3}(P'_{1EW\BD}-4P'_{2EW\BD}),
\en
\be
A(\bar B^0\to \Sigma^{+}\overline{\Delta^+})
   &=&\sq2 (T'_{1\BD}-2 T'_{2\BD})
          -\sq2 P'_\BD
          -\frac{2\sq2}{3}P'_{1EW\BD},
   \non\\
A(\bar B^0\to\Sigma^{0}\overline{\Delta^0})
   &=&-T'_{1\BD}+2T'_{2\BD}
         +2P'_\BD
         +\frac{1}{3}P'_{1EW\BD},
   \non\\
A(\bar B^0\to\Sigma^{-}\overline{\Delta^-})
   &=&\sq6 P'_\BD
         -\sq{\frac{2}{3}}P'_{1EW\BD},
   \non\\
A(\bar B^0\to\Xi^{0}\overline{\Sigma^{*0}})
   &=&T'_{1\BD}
         -P'_\BD
         -\frac{2}{3}(P'_{1EW\BD}-3P'_{2EW\BD}),
   \non\\
A(\bar B^0\to\Xi^{-}\overline{\Sigma^{*-}})
   &=&\sq2P'_\BD
         -\frac{\sq2}{3}P'_{1EW\BD},
   \non\\   
A(\bar B^0\to\Lambda\overline{\Delta^0})
   &=&\frac{1}{\sq3}(T'_{1\BD}+2T'_{2\BD})
         -\frac{1}{\sq3}(P'_{1EW\BD}-4P'_{2EW\BD}),
\en
and
\be
A(\bar B^0_s\to p\overline{\Delta^+})
   &=&-\sq2E'_\BD,
   \non\\
A(\bar B^0_s\to n\overline{\Delta^0})
   &=&-\sq2E'_\BD,
   \non\\
A(\bar B^0_s\to\Sigma^{+}\overline{\Sigma^{*+}})
   &=&\sq2(T'_{1\BD}-2T'_{2\BD})
         -\sq2P'_\BD
         -\frac{2\sq2}{3}P'_{1EW\BD}
        +\sq2E'_\BD,
   \non\\
A(\bar B^0_s\to\Sigma^{0}\overline{\Sigma^{*0}})
   &=&-\frac{1}{\sq2}(T'_{1\BD}-2T'_{2\BD})
         +\sq2 P'_\BD
         +\frac{1}{3\sq2}P'_{1EW\BD}
         -\frac{1}{\sq2} E'_\BD,
   \non\\
A(\bar B^0_s\to\Sigma^{-}\overline{\Sigma^{*-}})
   &=&\sq2P'_\BD
         -\frac{\sq2}{3}P'_{1EW\BD},
   \non\\   
A(\bar B^0_s\to\Xi^{0}\overline{\Xi^{*0}})
   &=&\sq2T'_{1\BD}
         -\sq2P'_\BD
         -\frac{2\sq2}{3}(P'_{1EW\BD}-3P'_{2EW\BD})
         +\sq2 E'_\BD,
   \non\\
A(\bar B^0_s\to\Xi^{-}\overline{\Xi^{*-}})
   &=&\sq2P'_\BD
         -\frac{\sq2}{3}P'_{1EW\BD},
   \non\\
A(\bar B^0_s\to\Lambda\overline{\Sigma^{*0}})
   &=&\frac{1}{\sq6}(T'_{1\BD}+2T'_{2\BD})
          -\frac{1}{\sq6}(P'_{1EW\BD}-4P'_{2EW\BD})
          +\sq{\frac{3}{2}} E'_\BD.   
\en

\subsubsection{$\overline B$ to decuplet-anti-octet baryonic decays}

The full $\bar B\to\DB$ decay amplitudes for $\Delta S=0$ processes are given by
\be
A(B^-\to\Delta^0\overline{p})
   &=&\sq2T_{1\DB}
          -\sq2 P_\DB
          +\frac{\sq2}{3}(3P_{1EW\DB}+P_{2EW\DB})
          -\sq2 A_\DB,
   \non\\
A(B^-\to\Delta^-\overline{n})
   &=&-\sq6 P_\DB
         +\sq{\frac{2}{3}}P_{2EW\DB}
         -\sq6 A_\DB,
   \non\\
A(B^-\to\Sigma^{*0}\overline{\Sigma^{+}})
   &=&-T_{1\DB}
          +P_\DB
          -\frac{1}{3}(3P_{1EW\DB}+P_{2EW\DB})
          +A_\DB,
   \non\\   
A(B^-\to\Sigma^{*-}\overline{\Sigma^{0}})
   &=&-P_\DB
          +\frac{1}{3}P_{2EW\DB}
          - A_\DB,
   \non\\
A(B^-\to\Xi^{*-}\overline{\Xi^{0}})
   &=&\sq2 P_\DB
         -\frac{\sq2}{3}P_{2EW\DB}
         +\sq2A_\DB,
   \non\\
A(B^-\to\Sigma^{*-}\overline{\Lambda})
   &=&\sq3 P_\DB
          -\frac{1}{\sq3}P_{2EW\DB}
          +\sq3 A_\DB,
\en
\be
A(\bar B^0\to\Delta^+\overline{p})
   &=&\sq2T_{2\DB}
          +\sq2 P_\DB
          +\frac{2\sq2}{3}P_{2EW\DB}
          -\sq2E_\DB,
   \non\\
A(\bar B^0\to\Delta^0\overline{n})
   &=&\sq2(T_{1\DB}+T_{2\DB})
         +\sq2 P_\DB
         +\frac{\sq2}{3}(3P_{1EW\DB}+2P_{2EW\DB})
         -\sq2E_\DB,
   \non\\
A(\bar B^0\to\Sigma^{*+}\overline{\Sigma^{+}})
   &=&\sq2 E_\DB,
   \non\\
A(\bar B^0\to\Sigma^{*0}\overline{\Sigma^{0}})
   &=&\frac{1}{\sq2}T_{1\DB}
         -\frac{1}{\sq2}P_\DB
         +\frac{1}{3\sq2}(3P_{1EW\DB}+P_{2EW\DB})
         -\frac{1}{\sq2} E_\DB,
   \non\\
A(\bar B^0\to\Sigma^{*-}\overline{\Sigma^{-}})
   &=&-\sq2P_\DB
          +\frac{\sq2}{3}P_{2EW\DB},
   \non\\   
A(\bar B^0\to\Xi^{*0}\overline{\Xi^{0}})
   &=&\sq2E_{\DB},
   \non\\
A(\bar B^0\to\Xi^{*-}\overline{\Xi^{-}})
   &=&-\sq2 P_\DB
          +\frac{\sq2}{3}P_{2EW\DB},
   \non\\
A(\bar B^0\to\Sigma^{*0}\overline{\Lambda})
   &=&-\frac{1}{\sq6}(T_{1\DB}+2T_{2\DB})
          -\sq{\frac{3}{2}}P_\DB
          -\frac{1}{\sq6}(P_{1EW\DB}+P_{2EW\DB})
          +\sq{\frac{3}{2}} E_\DB,   
   \non\\
\en
and
\be
A(\bar B^0_s\to \Delta^+\overline{\Sigma^{+}})
   &=&-\sq2 T_{2\DB}
          -\sq2 P_\DB
          -\frac{2\sq2}{3}P_{2EW\DB},
   \non\\
A(\bar B^0_s\to\Delta^0\overline{\Sigma^{0}})
   &=&T_{2\DB}
         +2 P_\DB
         +\frac{1}{3}P_{2EW\DB},
   \non\\
A(\bar B^0_s\to\Delta^-\overline{\Sigma^{-}})
   &=&\sq6 P_\DB
          -\sq{\frac{2}{3}}P_{2EW\DB},
   \non\\
A(\bar B^0_s\to\Sigma^{*0}\overline{\Xi^{0}})
   &=&-(T_{1\DB}+T_{2\DB})
          -P_\DB
          -\frac{1}{3}(3P_{1EW\DB}+2P_{2EW\DB}),
   \non\\
A(\bar B^0_s\to\Sigma^{*-}\overline{\Xi^{-}})
   &=&\sq2P_\DB
          -\frac{\sq2}{3}P_{2EW\DB},
   \non\\   
A(\bar B^0_s\to\Delta^0\overline{\Lambda})
   &=&-\frac{1}{\sq3}(2T_{1\DB}+T_{2\DB})
          -\frac{1}{\sq3}(2P_{1EW\DB}+P_{2EW\DB}),
\en
while those for $\Delta S=1$ transitions are given by
\be
A(B^-\to\Sigma^{*0}\overline{p})
   &=&T'_{1\DB}- P'_\DB
          +\frac{1}{3}(3P'_{1EW\DB}+P'_{2EW\DB})
          -A'_\DB,
   \non\\
A(B^-\to\Sigma^{*-}\overline{n})
   &=&-\sq2 P'_\DB
          +\frac{\sq2}{3}P'_{2EW\DB}
          -\sq2A'_\DB,
   \non\\
A(B^-\to\Xi^{*0}\overline{\Sigma^{+}})
   &=&-\sq2T'_{1\DB}
         +\sq2 P'_\DB
         -\frac{\sq2}{3}(3P'_{1EW\DB}+P'_{2EW\DB})
         +\sq2 A'_\DB,
   \non\\
A(B^-\to\Xi^{*-}\overline{\Sigma^{0}})
   &=&-P'_\DB
          +\frac{1}{3}P'_{2EW\DB}
          -A'_\DB,
   \non\\   
A(B^-\to\Omega^-\overline{\Xi^{0}})
   &=&\sq6 P'_\DB
          -\sq{\frac{2}{3}}P'_{2EW\DB}
          +\sq6 A'_\DB,
   \non\\
A(B^-\to\Xi^{*-}\overline{\Lambda})
   &=&\sq3 P'_\DB
          -\frac{1}{\sq3}P'_{2EW\DB}
          +\sq3A'_\DB,    
\en
\be
A(\bar B^0\to \Sigma^{*+}\overline{p})
   &=&\sq2 T'_{2\DB}
         +\sq2 P'_\DB
         +\frac{2\sq2}{3}P'_{2EW\DB},
   \non\\
A(\bar B^0\to\Sigma^{*0}\overline{n})
   &=&T'_{1\DB}+T'_{2\DB}
          +P'_\DB
          +\frac{1}{3}(3P'_{1EW\DB}+2P'_{2EW\DB}),
   \non\\
A(\bar B^0\to\Xi^{*0}\overline{\Sigma^{0}})
   &=&T'_{1\DB}
          -P'_\DB
          +\frac{1}{3}(3P'_{1EW\DB}+P'_{2EW\DB}),
   \non\\
A(\bar B^0\to\Xi^{*-}\overline{\Sigma^{-}})
   &=&-\sq2P'_\DB
          +\frac{\sq2}{3}P'_{2EW\DB},
   \non\\   
A(\bar B^0\to\Omega^-\overline{\Xi^{-}})
   &=&-\sq6 P'_\DB
          +\sq{\frac{2}{3}}P_{2EW\DB},
      \non\\
A(\bar B^0\to\Xi^{*0}\overline{\Lambda})
   &=&-\frac{1}{\sq3}(T'_{1\DB}+2T'_{2\DB})
          -\sq3 P'_\DB
          -\frac{1}{\sq3}(P'_{1EW\DB}+P'_{2EW\DB}),
\en
and
\be
A(\bar B^0_s\to\Delta^+\overline{p})
   &=&-\sq2 E'_\DB,
   \non\\
A(\bar B^0_s\to\Delta^0\overline{n})
   &=&-\sq2E'_\DB,
   \non\\
A(\bar B^0_s\to\Sigma^{*+}\overline{\Sigma^{+}})
   &=&-\sq2T'_{2\DB}
          -\sq2P'_\DB
          -\frac{2\sq2}{3}P'_{2EW\DB}
          +\sq2 E'_\DB,
   \non\\
A(\bar B^0_s\to\Sigma^{*0}\overline{\Sigma^{0}})
   &=& \frac{1}{\sq2}T'_{2\DB}
         +\sq2P'_\DB
         +\frac{1}{3\sq2}P'_{2EW\DB}
         -\frac{1}{\sq2} E'_\DB,
   \non\\
A(\bar B^0_s\to\Sigma^{*-}\overline{\Sigma^{-}})
   &=&\sq2P'_\DB
          -\frac{\sq2}{3}P'_{2EW\DB},
   \non\\   
A(\bar B^0_s\to\Xi^{*0}\overline{\Xi^{0}})
   &=&-\sq2(T'_{1\DB}+T'_{2\DB})
          -\sq2P'_\DB
          -\frac{\sq2}{3}(3P'_{1EW\DB}+2P'_{2EW\DB})
          +\sq2 E'_\DB,
   \non\\
A(\bar B^0_s\to\Xi^{*-}\overline{\Xi^{-}})
   &=&\sq2 P'_\DB
         -\frac{\sq2}{3}P'_{2EW\DB},
   \non\\
A(\bar B^0_s\to\Sigma^{*0}\overline{\Lambda})
   &=&-\frac{1}{\sq6}(2T'_{1\DB}+T'_{2\DB})
          -\frac{1}{\sq6}(2P'_{1EW\DB}+P'_{2EW\DB})
          +\sq{\frac{3}{2}} E'_\DB.
\en


\subsubsection{$\overline B$ to octet-anti-octet baryonic decays}

The full $\bar B\to\BB$ decay amplitudes for $\Delta S=0$ processes are given by
\be
A(B^-\to n\overline{p})
   &=&-T_{1\BB}-5 P_{1\BB}
           +\frac{2}{3}(P_{1EW\BB}
           -P_{3EW\BB}+P_{4EW\BB})
           -5 A_{1\BB},
   \non\\
A(B^-\to\Sigma^{0}\overline{\Sigma^{+}})
   &=&\sq2T_{3\BB}
         +\frac{1}{\sq2}(5P_{1\BB}-P_{2\BB})
         +\frac{1}{3\sq2}(P_{1EW\BB}+P_{2EW\BB}+2P_{3EW\BB}
   \non\\
   &&-2P_{4EW\BB})
         +\frac{1}{\sq2}(5 A_{1\BB}-A_{2\BB}),
   \non\\   
A(B^-\to\Sigma^{-}\overline{\Sigma^{0}})
   &=&-\frac{1}{\sq2}(5P_{1\BB}-P_{2\BB})
           -\frac{1}{3\sq2}(P_{1EW\BB}+P_{2EW\BB}-4P_{3EW\BB}-2P_{4EW\BB})
   \non\\
   &&-\frac{1}{\sq2}(5 A_{1\BB}- A_{2\BB}),
   \non\\
A(B^-\to\Sigma^{-}\overline{\Lambda})
   &=&-\frac{1}{\sq6}(5P_{1\BB}+P_{2\BB})
           -\frac{1}{3\sq6}(P_{1EW\BB}-P_{2EW\BB}-4P_{3EW\BB}-2P_{4EW\BB})
   \non\\
   &&-\frac{1}{\sq6}(5 A_{1\BB}+ A_{2\BB}),
   \non\\  
A(B^-\to\Xi^{-}\overline{\Xi^{0}})
   &=&-P_{2\BB}+\frac{1}{3}P_{2EW\BB}- A_{2\BB},
   \non\\
A(B^-\to\Lambda\overline{\Sigma^+})
   &=&-\sq{\frac{2}{3}}(T_{1\BB}-T_{3\BB})
          -\frac{1}{\sq6}(5P_{1\BB}+P_{2\BB})
           +\frac{1}{3\sq6}(5P_{1EW\BB}+P_{2EW\BB}
   \non\\
   &&   -4P_{3EW\BB}+2P_{4EW\BB})
           -\frac{1}{\sq6}(5 A_{1\BB}+A_{2\BB}),
\label{eq:BBBm}           
\en
\be
A(\bar B^0\to p\overline{p})
   &=&-T_{2\BB}+2T_{4\BB}
          +P_{2\BB}
           +\frac{2}{3}P_{2EW\BB}
          -5 E_{1\BB}+ E_{2\BB}
          -9PA_\BB,
   \non\\
A(\bar B^0\to n\overline{n})
   &=&-(T_{1\BB}+T_{2\BB})
          -(5P_{1\BB}-P_{2\BB})
           +\frac{2}{3}(P_{1EW\BB}+P_{2EW\BB}
   \non\\
   &&-P_{3EW\BB}-2P_{4EW\BB})
         + E_{2\BB}
         -9 PA_\BB,
   \non\\
A(\bar B^0\to\Sigma^{+}\overline{\Sigma^{+}})
   &=&-5 E_{1\BB}+ E_{2\BB}
          -9 PA_\BB,
   \non\\
A(\bar B^0\to\Sigma^{0}\overline{\Sigma^{0}})
   &=&-T_{3\BB}
          -\frac{1}{2}(5P_{1\BB}-P_{2\BB})
           -\frac{1}{6}(P_{1EW\BB}+P_{2EW\BB}+2P_{3EW\BB}-2P_{4EW\BB})
   \non\\
   &&-\frac{1}{2}(5 E_{1\BB}- E_{2\BB})
          -9 PA_\BB,
   \non\\
A(\bar B^0\to\Sigma^{0}\overline{\Lambda})
   &=&\frac{1}{\sq3}(T_{3\BB}+2T_{4\BB})
          +\frac{1}{2\sq3}(5P_{1\BB}+P_{2\BB})
           +\frac{1}{6\sq3}(P_{1EW\BB}-P_{2EW\BB}
   \non\\
   &&+2P_{3EW\BB}+10P_{4EW\BB})
        -\frac{1}{2\sqrt3}(5 E_{1\BB}+ E_{2\BB}),
   \non\\
A(\bar B^0\to\Sigma^{-}\overline{\Sigma^{-}})
   &=&-(5P_{1\BB}-P_{2\BB})
           -\frac{1}{3}(P_{1EW\BB}+P_{2EW\BB}-4P_{3EW\BB}-2P_{4EW\BB})
   \non\\
   &&-9 PA_\BB,
   \non\\   
A(\bar B^0\to\Xi^{0}\overline{\Xi^{0}})
   &=& E_{2\BB}
          -9 PA_\BB,
   \non\\
A(\bar B^0\to\Xi^{-}\overline{\Xi^{-}})
   &=&P_{2\BB}
           -\frac{1}{3}P_{2EW\BB}
           -9PA_\BB,
   \non\\
A(\bar B^0\to\Lambda\overline{\Sigma^{0}})
   &=&\frac{1}{\sq3}(T_{1\BB}-T_{3\BB})
          +\frac{1}{2\sq3}(5P_{1\BB}+P_{2\BB})
           -\frac{1}{6\sq3}(5P_{1EW\BB}+P_{2EW\BB}
   \non\\
   &&-2P_{3EW\BB}+2P_{4EW\BB})
       -\frac{1}{2\sq3}(5 E_{1\BB}+E_{2\BB}), 
   \non\\
A(\bar B^0\to\Lambda\overline{\Lambda})
   &=&-\frac{1}{3}(T_{1\BB}+2T_{2\BB}-T_{3\BB}-2T_{4\BB})
          -\frac{5}{6}(P_{1\BB}-P_{2\BB})          
   \non\\
   &&+\frac{1}{18}(5P_{1EW\BB}+7P_{2EW\BB}-2P_{3EW\BB}-10P_{4EW\BB})
        -\frac{5}{6}(E_{1\BB}-E_{2\BB})
   \non\\ 
   &&-9PA_\BB,  
\label{eq:BBB0}   
\en
and
\be
A(\bar B^0_s\to p\overline{\Sigma^{+}})
   &=&T_{2\BB}-2T_{4\BB}
          -P_{2\BB}
          -\frac{2}{3}P_{2EW\BB},
   \non\\
A(\bar B^0_s\to n\overline{\Sigma^{0}})
   &=&-\frac{1}{\sq2}T_{2\BB}
          +\frac{1}{\sq2}P_{2\BB}          
          +\frac{\sq2}{3}(P_{2EW\BB}-3P_{4EW\BB}),
   \non\\
A(\bar B^0_s\to n\overline{\Lambda})
   &=&\frac{1}{\sq6}(2T_{1\BB}+T_{2\BB})
          +\frac{1}{\sq6}(10P_{1\BB}-P_{2\BB})          
   \non\\
   &&-\frac{1}{3}\sqrt{\frac{2}{3}}(2P_{1EW\BB}+P_{2EW\BB}-2P_{3EW\BB}-P_{4EW\BB}),
   \non\\   
A(\bar B^0_s\to\Sigma^{0}\overline{\Xi^{0}})
   &=&\sq2(T_{3\BB}+T_{4\BB})
          +\frac{5}{\sq2}P_{1\BB}
          +\frac{1}{3\sq2}(P_{1EW\BB}+2P_{3EW\BB}+4P_{4EW\BB}),
   \non\\
A(\bar B^0_s\to\Sigma^{-}\overline{\Xi^{-}})
   &=&-5P_{1\BB}          
           +\frac{1}{3}(-P_{1EW\BB}+4P_{3EW\BB}+2P_{4EW\BB}),
   \non\\   
A(\bar B^0_s\to\Lambda\overline{\Xi^0})
   &=&-\sq{\frac{2}{3}}(T_{1\BB}+T_{2\BB}-T_{3\BB}-T_{4\BB})
          -\frac{1}{\sq6}(5P_{1\BB}-2P_{2\BB})          
   \non\\
   &&+\frac{1}{3\sq6}(5P_{1EW\BB}+4P_{2EW\BB}-2P_{3EW\BB}-4P_{4EW\BB}),
\label{eq:BBBs}
\en
while those for $\Delta S=1$ transitions are given by
\be
A(B^-\to\Sigma^{0}\overline{p})
   &=&-\frac{1}{\sq2}(T'_{1\BB}-2T'_{3\BB})
           -\frac{1}{\sq2} P'_{2\BB}
           +\frac{1}{3\sq2}(3P'_{1EW\BB}+P'_{2EW\BB})
           -\frac{1}{\sq2}A'_{2\BB},
   \non\\
A(B^-\to\Sigma^{-}\overline{n})
   &=&-P'_{2\BB}
          +\frac{1}{3}P'_{2EW\BB}
          - A'_{2\BB},
   \non\\
A(B^-\to\Xi^{0}\overline{\Sigma^{+}})
   &=&-T'_{1\BB}
           -5 P'_{1\BB}
           +\frac{2}{3}(P'_{1EW\BB}-P'_{3EW\BB}+P'_{4EW\BB})
           -5 A'_{1\BB},
   \non\\
A(B^-\to\Xi^{-}\overline{\Sigma^{0}})
   &=&-\frac{5}{\sq2}P'_{1\BB}
           -\frac{1}{3\sq2}(P'_{1EW\BB}-4P'_{3EW\BB}-2P'_{4EW\BB})
           -\frac{5}{\sq2}A'_{1\BB},
   \non\\  
A(B^-\to\Xi^{-}\overline{\Lambda})
   &=&-\frac{1}{\sqrt6}(5P'_{1\BB}-2P'_{2\BB})
           -\frac{1}{3\sq6}(P'_{1EW\BB}+2P'_{2EW\BB}-4P'_{3EW\BB}-2P'_{4EW\BB})
   \non\\
   &&-\frac{1}{\sq6}(5 A'_{1\BB}-2 A'_{2\BB}),
   \non\\      
A(B^-\to\Lambda\overline{p})
   &=&\frac{1}{\sq6}(T'_{1\BB}+2T'_{3\BB})
           +\frac{1}{\sqrt6}(10P'_{1\BB}-P'_{2\BB})
           -\frac{1}{3\sq6}(P'_{1EW\BB}-P'_{2EW\BB}
   \non\\
   &&-4P'_{3EW\BB}+4P'_{4EW\BB})
         +\frac{1}{\sq6}(10 A'_{1\BB}- A'_{2\BB}),       
\en
\be
A(\bar B^0\to \Sigma^{+}\overline{p})
   &=& T'_{2\BB}-2T'_{4\BB}
           -P'_{2\BB}
           -\frac{2}{3}P'_{2EW\BB},
   \non\\
A(\bar B^0\to\Sigma^{0}\overline{n})
   &=&-\frac{1}{\sq2}(T'_{1\BB}+T'_{2\BB}-2T'_{3\BB}-2T'_{4\BB})
          +\frac{1}{\sq2}P'_{2\BB}
    \non\\
    &&+\frac{1}{3\sq2}(3P'_{1EW\BB}+2P'_{2EW\BB}),
   \non\\
A(\bar B^0\to\Xi^{0}\overline{\Sigma^{0}})
   &=&\frac{1}{\sq2}T'_{1\BB}
           +\frac{5}{\sq2}P'_{1\BB}
           -\frac{\sq2}{3}(P'_{1EW\BB}-P'_{3EW\BB}+P'_{4EW\BB}),
   \non\\
A(\bar B^0\to\Xi^{0}\overline{\Lambda})
   &=&-\frac{1}{\sq6}(T'_{1\BB}+2T'_{2\BB})
           -\frac{1}{\sq6}(5P'_{1\BB}-2P'_{2\BB})
    \non\\
    &&+\frac{1}{3}\sq{\frac{2}{3}}(P'_{1EW\BB}+2P'_{2EW\BB}-P'_{3EW\BB}-5P'_{4EW\BB}),
   \non\\ 
A(\bar B^0\to\Xi^{-}\overline{\Sigma^{-}})
   &=&-5P'_{1\BB}
           -\frac{1}{3}(P'_{1EW\BB}-4P'_{3EW\BB}-2P'_{4EW\BB}),
   \non\\   
A(\bar B^0\to\Lambda\overline{n})
   &=&\frac{1}{\sq6}(T'_{1\BB}+T'_{2\BB}+2T'_{3\BB}+2T'_{4\BB})
           +\frac{1}{\sq6}(10P'_{1\BB}-P'_{2\BB})
    \non\\
    &&-\frac{1}{3\sq6}(P'_{1EW\BB}+2P'_{2EW\BB}-4P'_{3EW\BB}-8P'_{4EW\BB}),
\en
and
\be
A(\bar B^0_s\to p\overline{p})
   &=&-5E'_{1\BB}+ E'_{2\BB}-9 PA'_\BB,
   \non\\
A(\bar B^0_s\to n\overline{n})
   &=& E'_{2\BB}-9PA'_\BB,
   \non\\
A(\bar B^0_s\to\Sigma^{+}\overline{\Sigma^{+}})
   &=&-T'_{2\BB}+2T'_{4\BB}
           +P'_{2\BB}
           +\frac{2}{3}P'_{2EW\BB}
           -5 E'_{1\BB}+ E'_{2\BB}
           -9PA'_\BB,
   \non\\
A(\bar B^0_s\to\Sigma^{0}\overline{\Sigma^{0}})
   &=&-\frac{1}{2}(T'_{2\BB}-2T'_{4\BB})
           +P'_{2\BB}
           +\frac{1}{6}P'_{2EW\BB}
           -\frac{1}{2}(5 E'_{1\BB}- E'_{2\BB})
           -9 PA'_\BB,
   \non\\
A(\bar B^0_s\to\Sigma^{0}\overline{\Lambda})
   &=&\frac{1}{2\sq3}(2T'_{1\BB}+T'_{2\BB}-4T'_{3\BB}-2T'_{4\BB})
           -\frac{1}{2\sq3}(2P'_{1EW\BB}+P'_{2EW\BB})
    \non\\
    &&-\frac{1}{2\sq3}(5 E'_{1\BB}+ E'_{2\BB}),
     \non\\     
A(\bar B^0_s\to\Sigma^{-}\overline{\Sigma^{-}})
   &=&P'_{2\BB}
           -\frac{1}{3}P'_{2EW\BB}
           -9 PA'_\BB,
   \non\\   
A(\bar B^0_s\to\Xi^{0}\overline{\Xi^{0}})
   &=&-T'_{1\BB}-T'_{2\BB}
           -(5P'_{1\BB}-P'_{2\BB})
           +\frac{2}{3}(P'_{1EW\BB}+P'_{2EW\BB}-P'_{3EW\BB}
    \non\\
    &&-2P'_{4EW\BB})
           + E'_{2\BB}
           -9 PA'_\BB,
   \non\\
A(\bar B^0_s\to\Xi^{-}\overline{\Xi^{-}})
   &=&-(5P'_{1\BB}-P'_{2\BB})
          -\frac{1}{3}(P'_{1EW\BB}+P'_{2EW\BB}-4P'_{3EW\BB}-2P'_{4EW\BB})
   \non\\
    &&-9 PA'_\BB,
    \non\\     
A(\bar B^0_s\to\Lambda\overline{\Sigma^{0}})
   &=&\frac{1}{2\sq3}(T'_{2\BB}+2T'_{4\BB})
           +\frac{1}{2\sq3}(-P'_{2EW\BB}+4P'_{4EW\BB})
    \non\\       
    &&-\frac{1}{2\sq3}(5 E'_{1\BB}+E'_{2\BB}),
   \non\\
A(\bar B^0_s\to\Lambda\overline{\Lambda})
   &=&-\frac{1}{6}(2T'_{1\BB}+T'_{2\BB}+4T'_{3\BB}+2T'_{4\BB})
           -\frac{1}{3}(10P'_{1\BB}-P'_{2\BB})
   \non\\       
   &&+\frac{1}{18}(2P'_{1EW\BB}+P'_{2EW\BB}-8P'_{3EW\BB}-4P'_{4EW\BB})
   \non\\      
   &&-\frac{5}{6}(E'_{1\BB}-E'_{2\BB})
         -9PA'_\BB.
\label{eq:BBsBs}
\en

\subsection{Large $m_B$ limit}

Using the chirality structure of $H_{\rm eff}$ in Eq. (3) and large $m_B$ limit, topological amplitudes are related~\cite{Chua:2003it, Brodsky:1980sx}.
As
shown in the appendix we have
 \begin{eqnarray}
 T^{(\prime)}
 &\equiv&
 T^{(\prime)}_{\DD}
 =T^{(\prime)}_{1\BD,2\BD}
 =T^{(\prime)}_{1\DB,2\DB}
 =T^{(\prime)}_{1\BB,2\BB,3\BB,4\BB},
 \non\\
 P^{(\prime)}
 &\equiv&P^{(\prime)}_{{\cal D} \overline{\cal D}}
 =P^{(\prime)}_{{\cal B} \overline{\cal D}}
 =P^{(\prime)}_{{\cal D} \overline{\cal B}}
 =P^{(\prime)}_{1\BB,2\BB},
 \non\\
 P^{(\prime)}_{EW}
 &\equiv&P^{(\prime)}_{EW\DD}
 =P^{(\prime)}_{1EW\BD,2EW\BD}
 =P^{(\prime)}_{1EW\DB,2EW\DB}
 =P^{(\prime)}_{1EW\BB,2EW\BB,3EW\BB,4EW\BB},
 \non\\
&&
\!\!\!\!\!\!\!\!\!\!\!\!\!\!\!\!\!\!\!\!\!\!
 E_{\DD,\BD,\DB,1\BB,2\BB},\,A_{\DD,\BD,\DB,1\BB,2\BB},
 \,PA_{\DD,\BB}\to0,
\label{eq:asymptoticrelations}
\end{eqnarray}
in the large $m_B$ asymptotic limit. In that limit, we need only one tree,
one penguin and one electroweak penguin amplitudes for
all four classes of charmless two-body baryonic modes. 
The asymptotic decay amplitudes 
can be easily read out using the results shown in the previous subsection and Eq.~(\ref{eq:asymptoticrelations}).

Using Eq.~(\ref{eq:H_eff1}) these amplitudes are estimated to be
\be
T^{(\prime)}&=&V_{ub} V^*_{ud(s)}\frac{G_f}{\sq2}(c_1-c_2) \chi \bar u'(1-\gamma_5)v,
\non\\
P^{(\prime)}&=&-V_{tb} V^*_{td(s)}\frac{G_f}{\sq2}[c_3-c_4+c_5-c_6]\chi \bar u'(1-\gamma_5)v,
\non\\
P^{(\prime)}_{EW}&=&-\frac{3}{2}V_{tb} V^*_{td(s)}\frac{G_f}{\sq2}[c_9-c_{10}+c_7-c_8]\chi \bar u'(1-\gamma_5)v.
\label{eq:asymptotic1}
\en
The minus signs between Wilson coefficients are from the color structure. 
Note that $O_{1,3,9}$, and similarly $O_{2,4,10}$, are only different on the flavor structure, their contributions are related in the large $m_B$ limit (see $e_T$, $e_{P_L}$, $e_{P_{EWL}}$ in Appendix A). 
Similarly contributions from $O_{5,6}$ and $O_{7,8}$ are related.

The unknown amplitude $\chi$ will be fitted
from the recent $\overline B{}^0\to p\bar p$ data.

\section{Phenomenology}

\subsection{Relations on rates and $A_{CP}$}

From the full topological amplitude expressions of decay amplitudes, we can obtain relation on averaged rates and rate deferences, as the number of modes are greater than the number of the independent amplitudes (see Appendix~\ref{append:B}). 

For decuplet-antidecuplet modes, we have
\be
2 \Br(B^-\to\Delta^-\overline{\Delta^{0}})
    &=&3 \Br(B^-\to\Sigma^{*-}\overline{\Sigma^{*0}})
      =6 \Br(B^-\to\Xi^{*-}\overline{\Xi^{*0}}),
\non\\
\Br(B^-\to\Delta^0\overline{\Delta^{+}})
      &=&2 \Br(B^-\to\Sigma^{*0}\overline{\Sigma^{*+}}),
\non\\
6 \Br(B^-\to\Sigma^{*-}\overline{\Delta^0})
     &=&   
2 \Br(B^-\to\Omega^-\overline{\Xi^{*0}})
      =3 \Br(B^-\to\Xi^{*-}\overline{\Sigma^{*0}}),
\non\\
2 \Br(B^-\to\Sigma^{*0}\overline{\Delta^+})
     &=&\Br(B^-\to\Xi^{*0}\overline{\Sigma^{*+}}),
\non\\
\Br(\overline{B^0_s}\to\Delta^{0}\overline{\Sigma^{*0}})
     &=&\Br(\overline{B^0_s}\to \Sigma^{*0}\overline{\Xi^{*0}}),
\non\\
4 \Br(\overline{B^0_s}\to\Delta^{-}\overline{\Sigma^{*-}})
     &=&4 \Br(\overline{B^0_s}\to\Xi^{*-}\overline{\Omega^-})
      =3 \Br(\overline{B^0_s}\to\Sigma^{*-}\overline{\Xi^{*-}}),
\non\\
\Br(\bar B^0\to\Sigma^{*0}\overline{\Delta^0})
   &=&\Br(\bar B^0\to\Xi^{*0}\overline{\Sigma^{*0}}),
\non\\
4 \Br(\bar B^0\to\Sigma^{*-}\overline{\Delta^-})
   &=& 4 \Br(\bar B^0\to\Xi^{*-}\overline{\Sigma^{*-}})
   = 3 \Br(\bar B^0\to\Omega^-\overline{\Xi^{*-}}).
\en

Under $U$-spin symmetry,~\cite{Uspin},
using $Im(V_{ub}V^*_{ud} V^*_{tb}V_{td})
=-Im(V_{ub}V^*_{us} V^*_{tb}V_{ts})$ and the expressions of amplitudes, we obtain
\be
2\Delta_{CP}(B^-\to\Delta^-\overline{\Delta^{0}})
    &=&3\Delta_{CP}(B^-\to\Sigma^{*-}\overline{\Sigma^{*0}})
       =\Delta_{CP}(B^-\to\Xi^{*-}\overline{\Xi^{*0}})
\non\\
    &=&-6 \Delta_{CP}(B^-\to\Sigma^{*-}\overline{\Delta^0})
       = -2 \Delta_{CP}(B^-\to\Omega^-\overline{\Xi^{*0}})
\non\\
    &=&-3 \Delta_{CP}(B^-\to\Xi^{*-}\overline{\Sigma^{*0}}),
\non\\
\Delta_{CP}(B^-\to\Delta^0\overline{\Delta^{+}})
     &=&2\Delta_{CP}(B^-\to\Sigma^{*0}\overline{\Sigma^{*+}})
        =-2 \Delta_{CP}(B^-\to\Sigma^{*0}\overline{\Delta^+})
\non\\
      &=&-\Delta_{CP}(B^-\to\Xi^{*0}\overline{\Sigma^{*+}}),
\non\\
\Delta_{CP}(B^-\to\Delta^+\overline{\Delta^{++}})
      &=&-\Delta_{CP}(B^-\to \Sigma^{*+}\overline{\Delta^{++}}),
\non\\
\Delta_{CP}(\overline{B^0_s}\to\Delta^{0}\overline{\Sigma^{*0}})
     &=&\Delta_{CP}(\overline{B^0_s}\to \Sigma^{*0}\overline{\Xi^{*0}})
       =- \Delta_{CP}(\bar B^0\to\Sigma^{*0}\overline{\Delta^0})
\non\\
     &=&-\Delta_{CP}(\bar B^0\to\Xi^{*0}\overline{\Sigma^{*0}}),
\non\\
4 \Delta_{CP}(\overline{B^0_s}\to\Delta^{-}\overline{\Sigma^{*-}})
    &=&4 \Delta_{CP}(\overline{B^0_s}\to\Xi^{*-}\overline{\Omega^-})
       =3 \Delta_{CP}(\overline{B^0_s}\to\Sigma^{*-}\overline{\Xi^{*-}})
\non\\
    &=& -4 \Delta_{CP}(\bar B^0\to\Sigma^{*-}\overline{\Delta^-})
      =-4\Delta_{CP} (\bar B^0\to\Xi^{*-}\overline{\Sigma^{*-}})
\non\\
     &=& -3 \Delta_{CP}(\bar B^0\to\Omega^-\overline{\Xi^{*-}}),
\non\\
\Delta_{CP}(\overline{B^0_s}\to\Delta^+\overline{\Sigma^{*+}})
      &=&-\Delta_{CP}(\bar B^0\to \Sigma^{*+}\overline{\Delta^+}),
\non\\
\Delta_{CP}(\overline{B^0}\to \Xi^{*0}\overline{\Xi^{*0}})
     &=&-\Delta_{CP}(\bar B^0_s\to\Delta^0\overline{\Delta^0}),
\non\\
\Delta_{CP}(\overline{B^0}\to\Xi^{*-}\overline{\Xi^{*-}})
    &=& -\Delta_{CP}(\bar B^0_s\to\Sigma^{*-}\overline{\Sigma^{*-}}),
\non\\
\Delta_{CP}(\overline{B^0}\to\Sigma^{*0}\overline{\Sigma^{*0}})
    &=&-\Delta_{CP}(\bar B^0_s\to\Sigma^{*0}\overline{\Sigma^{*0}}),
\non\\
\Delta_{CP}(\overline{B^0}\to\Omega^-\overline{\Omega^-})
   &=&-\Delta_{CP}(\bar B^0_s\to\Delta^-\overline{\Delta^-}),
\non\\
\Delta_{CP}(\overline{B^0}\to\Delta^{++}\overline{\Delta^{++}})
   &=& -\Delta_{CP}(\bar B^0_s\to \Delta^{++}\overline{\Delta^{++}}),
\non\\
\Delta_{CP}(\overline{B^0}\to\Sigma^{*+}\overline{\Sigma^{*+}})
  &=&-\Delta_{CP}(\bar B^0_s\to\Delta^+\overline{\Delta^+}),
\non\\
\Delta_{CP}(\overline{B^0}\to\Delta^{-}\overline{\Delta^{-}})
  &=&-\Delta_{CP}(\bar B^0_s\to\Omega^{-}\overline{\Omega^{-}}),
\non\\   
\Delta_{CP}(\overline{B^0}\to\Sigma^{*-}\overline{\Sigma^{*-}})
  &=&-\Delta_{CP}(\bar B^0_s\to\Xi^{*-}\overline{\Xi^{*-}}),
\non\\
\Delta_{CP}(\overline{B^0}\to\Delta^+\overline{\Delta^{+}})
  &=&-\Delta_{CP}(\bar B^0_s\to\Sigma^{*+}\overline{\Sigma^{*+}}),
\non\\
\Delta_{CP}(\overline{B^0}\to\Delta^0\overline{\Delta^{0}})
  &=&-\Delta_{CP}(\bar B^0_s\to\Xi^{*0}\overline{\Xi^{*0}}),
\en
where $\Delta_{CP}$ is defined as the $\overline B_q$ decay rate subtracted by the rate of the CP conjugated mode of $B_q$ decay.

For  octet-antidecuplet modes, we have
\be
\Br(B^-\to\Xi^{-}\overline{\Xi^{*0}})
   &=&2 \Br(B^-\to\Sigma^-\overline{\Sigma^{*0}}),
   \non\\
2\Br(B^-\to\Xi^{-}\overline{\Sigma^{*0}})
   &=&\Br(B^-\to\Sigma^-\overline{\Delta^0}),
\non\\
3\tau_{B_d} \Br(\bar B^0\to\Sigma^{-}\overline{\Sigma^{*-}})
   &=&3\tau_{B_d}\Br(\bar B^0\to\Xi^{-}\overline{\Xi^{*-}})=
3\tau_{B_s} \Br(\bar B^0_s\to\Sigma^{-}\overline{\Xi^{*-}})
\non\\
   &=&\tau_{B_s}\Br(\bar B^0_s\to\Xi^{-}\overline{\Omega^-}),
\non\\        
3\tau_{B_s}\Br(\bar B^0_s\to\Sigma^{-}\overline{\Sigma^{*-}})
   &=&3\tau_{B_s}\Br(\bar B^0_s\to\Xi^{-}\overline{\Xi^{*-}})
     =3 \tau_{B_d}\Br(\bar B^0\to\Xi^{-}\overline{\Sigma^{*-}})
\non\\
   &=&\tau_{B_d}\Br(\bar B^0\to\Sigma^{-}\overline{\Delta^-}), 
\non\\      
\Br(\bar B^0\to\Sigma^{+}\overline{\Sigma^{*+}})
    &=&\Br(\bar B^0\to\Xi^{0}\overline{\Xi^{*0}}),
\non\\
\Br(\bar B^0_s\to p\overline{\Delta^+})
   &=&\Br(\bar B^0_s\to n\overline{\Delta^0}),
\en
and
\be
\Delta_{CP}(B^-\to n\overline{\Delta^+})
   &=&-\Delta_{CP}(B^-\to\Xi^{0}\overline{\Sigma^{*+}}),
\non\\      
\Delta_{CP}(B^-\to\Xi^{-}\overline{\Xi^{*0}})
   &=&2 \Delta_{CP}(B^-\to\Sigma^-\overline{\Sigma^{*0}})
      =-2\Delta_{CP}(B^-\to\Xi^{-}\overline{\Sigma^{*0}})
\non\\   
   &=&-\Delta_{CP}(B^-\to\Sigma^-\overline{\Delta^0}),
\non\\
\Delta_{CP}(B^-\to p\overline{\Delta^{++}})
   &=&-
\Delta_{CP}(B^-\to \Sigma^+\overline{\Delta^{++}}),
   \non\\   
\Delta_{CP}(\bar B^0\to n\overline{\Delta^0})
   &=&-\Delta_{CP}(\bar B^0_s\to\Xi^{0}\overline{\Xi^{*0}}),
\non\\   
3\Delta_{CP}(\bar B^0\to\Sigma^{-}\overline{\Sigma^{*-}})
   &=&3 \Delta_{CP}(\bar B^0\to\Xi^{-}\overline{\Xi^{*-}})
   =3 \Delta_{CP}(\bar B^0_s\to\Sigma^{-}\overline{\Xi^{*-}})
\non\\   
   &=&\Delta_{CP}(\bar B^0_s\to\Xi^{-}\overline{\Omega^-})
   = -3 \Delta_{CP}(\bar B^0_s\to\Sigma^{-}\overline{\Sigma^{*-}})
\non\\   
   &=&-3 \Delta_{CP}(\bar B^0_s\to\Xi^{-}\overline{\Xi^{*-}})
     =-3 \Delta_{CP}(\bar B^0\to\Xi^{-}\overline{\Sigma^{*-}})
\non\\
   &=&-\Delta_{CP}(\bar B^0\to\Sigma^{-}\overline{\Delta^-}),
   \non\\      
\Delta_{CP}(\bar B^0\to\Sigma^{+}\overline{\Sigma^{*+}})
    &=&\Delta_{CP}(\bar B^0\to\Xi^{0}\overline{\Xi^{*0}})
       =-\Delta_{CP}(\bar B^0_s\to p\overline{\Delta^+})
\non\\
   &=&-\Delta_{CP}(\bar B^0_s\to n\overline{\Delta^0}),
   \non\\   
\Delta_{CP}(\bar B^0\to p\overline{\Delta^+})
   &=&-\Delta_{CP}(\bar B^0_s\to\Sigma^{+}\overline{\Sigma^{*+}}),
\non\\
\Delta_{CP}(\bar B^0_s\to n\overline{\Sigma^{*0}})
    &=&-\Delta_{CP}(\bar B^0\to\Xi^{0}\overline{\Sigma^{*0}})
\non\\
\Delta_{CP}(\bar B^0_s\to p\overline{\Sigma^{*+}})
    &=&-\Delta_{CP}(\bar B^0\to \Sigma^{+}\overline{\Delta^+}).
\en


For decuplet-antioctet modes, we have
\be
3\tau_{B_d}\Br(\bar B^0\to\Sigma^{*-}\overline{\Sigma^{-}})
  & =&3\tau_{B_d}\Br(\bar B^0\to\Xi^{*-}\overline{\Xi^{-}})
   = \tau_{B_s}\Br(\bar B^0_s\to\Delta^-\overline{\Sigma^{-}})
   \non\\
   &=&3\tau_{B_s}\Br(\bar B^0_s\to\Sigma^{*-}\overline{\Xi^{-}}),
\non\\ 
3\tau_{B_s}\Br(\bar B^0_s\to\Sigma^{*-}\overline{\Sigma^{-}})
   &=&3\tau_{B_s}\Br(\bar B^0_s\to\Xi^{*-}\overline{\Xi^{-}})
   =3 \tau_{B_d}\Br(\bar B^0\to\Xi^{*-}\overline{\Sigma^{-}})
\non\\   
   &=& \tau_{B_d}\Br(\bar B^0\to\Omega^-\overline{\Xi^{-}}),
\non\\
\Br(\bar B^0\to\Sigma^{*+}\overline{\Sigma^{+}})
   &=&\Br(\bar B^0\to\Xi^{*0}\overline{\Xi^{0}}),
   \non\\
\Br(\bar B^0_s\to\Delta^+\overline{p})
   &=&\Br(\bar B^0_s\to\Delta^0\overline{n}),
\en
and
\be
\Delta_{CP}(B^-\to\Delta^0\overline{p})
   &=&
     2\Delta_{CP}(B^-\to\Sigma^{*0}\overline{\Sigma^{+}})
   =-2 \Delta_{CP}(B^-\to\Sigma^{*0}\overline{p})
   \non\\
   &=&-\Delta_{CP}(B^-\to\Xi^{*0}\overline{\Sigma^{+}}),
   \non\\
\Delta_{CP}(B^-\to\Delta^-\overline{n})
   &=& 
   =3 \Delta_{CP}(B^-\to\Xi^{*-}\overline{\Xi^{0}})
  =6 \Delta_{CP}(B^-\to\Sigma^{*-}\overline{\Sigma^{0}})
\non\\
   &=&2 \Delta_{CP}(B^-\to\Sigma^{*-}\overline{\Lambda})
   =  
-3\Delta_{CP}(B^-\to\Sigma^{*-}\overline{n})
\non\\
   &=&
-6 \Delta_{CP}(B^-\to\Xi^{*-}\overline{\Sigma^{0}})
   =   
-\Delta_{CP}(B^-\to\Omega^-\overline{\Xi^{0}})
\non\\
    &=&
-2 \Delta_{CP}(B^-\to\Xi^{*-}\overline{\Lambda}),   
\non\\
\Delta_{CP}(\bar B^0\to\Delta^+\overline{p})
   &=&
-\Delta_{CP}(\bar B^0_s\to\Sigma^{*+}\overline{\Sigma^{+}}),
\non\\
3\Delta_{CP}(\bar B^0\to\Sigma^{*-}\overline{\Sigma^{-}})
  & =&3\Delta_{CP}(\bar B^0\to\Xi^{*-}\overline{\Xi^{-}})
   = \Delta_{CP}(\bar B^0_s\to\Delta^-\overline{\Sigma^{-}})
\non\\  
   &=&3\Delta_{CP}(\bar B^0_s\to\Sigma^{*-}\overline{\Xi^{-}})  
   =-3\Delta_{CP}(\bar B^0_s\to\Sigma^{*-}\overline{\Sigma^{-}})
\non\\   
   &=&-3\Delta_{CP}(\bar B^0_s\to\Xi^{*-}\overline{\Xi^{-}})  
   =-3 \Delta_{CP}(\bar B^0\to\Xi^{*-}\overline{\Sigma^{-}})
\non\\   
   &=& -\Delta_{CP}(\bar B^0\to\Omega^-\overline{\Xi^{-}}),
\non\\   
\Delta_{CP}(\bar B^0\to\Sigma^{*+}\overline{\Sigma^{+}})
   &=&\Delta_{CP}(\bar B^0\to\Xi^{*0}\overline{\Xi^{0}})
   =-\Delta_{CP}(\bar B^0_s\to\Delta^+\overline{p})
\non\\   
   &=&-\Delta_{CP}(\bar B^0_s\to\Delta^0\overline{n}),
\non\\   
\Delta_{CP}(\bar B^0\to\Delta^0\overline{n})
   &=&
-\Delta_{CP}(\bar B^0_s\to\Xi^{*0}\overline{\Xi^{0}}),   
\non\\
\Delta_{CP}(\bar B^0_s\to \Delta^+\overline{\Sigma^{+}})
   &=&
-\Delta_{CP}(\bar B^0\to \Sigma^{*+}\overline{p}),
 \non\\
\Delta_{CP}(\bar B^0_s\to\Sigma^{*0}\overline{\Xi^{0}})
   &=&
-\Delta_{CP}(\bar B^0\to\Sigma^{*0}\overline{n}).
\en

For octet-octet modes, there are no relations for the averaged branching ratios, when the full topological amplitudes are used (see Appendix B).~\footnote{For approximated relations, using only the dominating terms in the amplitudes, one is referred to \cite{Chua:2003it}.}
However for $\Delta_{CP}$, we have 
\be
\Delta_{CP}(B^-\to n\overline{p})
   &=&-\Delta_{CP}(B^-\to\Xi^{0}\overline{\Sigma^{+}}),
\non\\
\Delta_{CP}(B^-\to\Xi^{-}\overline{\Xi^{0}})
   &=&-\Delta_{CP}(B^-\to\Sigma^{-}\overline{n})
\non\\
\Delta_{CP}(\bar B^0\to p\overline{p})
   &=&
-\Delta_{CP}(\bar B^0_s\to\Sigma^{+}\overline{\Sigma^{+}}),
\non\\   
\Delta_{CP}(\bar B^0\to n\overline{n})
   &=&
-\Delta_{CP}(\bar B^0_s\to\Xi^{0}\overline{\Xi^{0}}),
\non\\ 
\Delta_{CP}(\bar B^0\to\Sigma^{+}\overline{\Sigma^{+}})
   &=&-\Delta_{CP}(\bar B^0_s\to p\overline{p}),
\non\\
\Delta_{CP}(\bar B^0\to\Sigma^{-}\overline{\Sigma^{-}})
   &=&-\Delta_{CP}(\bar B^0_s\to\Xi^{-}\overline{\Xi^{-}}),
\non\\      
\Delta_{CP}(\bar B^0\to\Xi^{0}\overline{\Xi^{0}})
   &=&-\Delta_{CP}(\bar B^0_s\to n\overline{n}),
\non\\ 
\Delta_{CP}(\bar B^0\to\Xi^{-}\overline{\Xi^{-}})
   &=&-\Delta_{CP}(\bar B^0_s\to\Sigma^{-}\overline{\Sigma^{-}}),
\non\\
\Delta_{CP}(\bar B^0_s\to p\overline{\Sigma^{+}})
   &=&-\Delta_{CP}(\bar B^0\to \Sigma^{+}\overline{p}),
   \non\\
\Delta_{CP}(\bar B^0_s\to\Sigma^{-}\overline{\Xi^{-}})
   &=&-\Delta_{CP}(\bar B^0\to\Xi^{-}\overline{\Sigma^{-}}).
\en
All of the above relations are obtained without using the large $m_B$ limits and are ready to be checked experimentally.

\subsection{Triangle relations on amplitudes}

In the previous subsection we only make use of some of the relations on amplitudes.
There are, in fact, much more relations on amplitudes. For example, for $\Delta S=0,$ $\overline B_q$ to decuplet-antidecuplet decay, we can have isospin relations,
\be
&\sq2 A(B^-\to\Delta^+\overline{\Delta^{++}})
=
\sq6 A(B^-\to\Delta^0\overline{\Delta^{+}})
-\sq2A(B^-\to\Delta^-\overline{\Delta^{0}}),
&
\non\\
&\sq3 A(\overline{B^0_s}\to\Delta^+\overline{\Sigma^{*+}})
=\sq6 A(\overline{B^0_s}\to\Delta^{0}\overline{\Sigma^{*0}})
-A(\overline{B^0_s}\to\Delta^{-}\overline{\Sigma^{*-}}),
&\non\\
&A(\overline{B^0}\to\Delta^{++}\overline{\Delta^{++}})
-3A(\overline{B^0}\to\Delta^+\overline{\Delta^{+}})
+3A(\overline{B^0}\to\Delta^0\overline{\Delta^{0}})
-A(\overline{B^0}\to\Delta^-\overline{\Delta^{-}})
=0,
&
\en
and, 
\be
2A(\overline{B^0}\to\Xi^{*-}\overline{\Xi^{*-}})
&=&A(\overline{B^0_s}\to\Sigma^{*-}\overline{\Xi^{*-}})
+2A(\overline{B^0}\to\Omega^-\overline{\Omega^-}),
\non\\
2A(\overline{B^0}\to\Sigma^{*0}\overline{\Sigma^{*0}})
&=&
\sq2 A(\overline{B^0_s}\to \Sigma^{*0}\overline{\Xi^{*0}})
+A(\overline{B^0}\to \Xi^{*0}\overline{\Xi^{*0}}),
\non\\
A(\overline{B^0}\to\Delta^{++}\overline{\Delta^{++}})
&=&
3A(\overline{B^0}\to \Xi^{*0}\overline{\Xi^{*0}})
-2A(\overline{B^0}\to\Omega^-\overline{\Omega^-}),
\non\\
A(\overline{B^0}\to\Sigma^{*+}\overline{\Sigma^{*+}})
&=&
2A(\overline{B^0}\to \Xi^{*0}\overline{\Xi^{*0}})
-A(\overline{B^0}\to\Omega^-\overline{\Omega^-}),
\non\\
A(\overline{B^0}\to\Delta^{-}\overline{\Delta^{-}})
&=&3A(\overline{B^0}\to\Xi^{*-}\overline{\Xi^{*-}})
-2A(\overline{B^0}\to\Omega^-\overline{\Omega^-}),
\non\\
A(\overline{B^0}\to\Sigma^{*-}\overline{\Sigma^{*-}})
&=&2A(\overline{B^0}\to\Xi^{*-}\overline{\Xi^{*-}})
-A(\overline{B^0}\to\Omega^-\overline{\Omega^-}),
\non\\
A(\overline{B^0}\to\Delta^+\overline{\Delta^{+}})
&=&A(\overline{B^0_s}\to\Delta^+\overline{\Sigma^{*+}})
+A(\overline{B^0}\to\Sigma^{*+}\overline{\Sigma^{*+}}),
\non\\
A(\overline{B^0}\to\Delta^0\overline{\Delta^{0}})
&=&
\sq2 A(\overline{B^0_s}\to\Delta^{0}\overline{\Sigma^{*0}})
+A(\overline{B^0}\to \Xi^{*0}\overline{\Xi^{*0}}).
\en
These can be easily obtained by using the full decay amplitudes given in the previous section or in Appendix B.

There are many similar relations for amplitudes within decuplet-antidecuplet, octet-antidecuplet, decuplet-antioctet and octet-antioctet modes. The interested reader can work them out using formulas in Appendix B. 
In below we only give two examples of the relations on octet-antioctet amplitues:
\be
A(\bar B^0\to p\overline{p})
   &=&-A(\bar B^0_s\to p\overline{\Sigma^{+}})
   +A(\bar B^0\to\Sigma^{+}\overline{\Sigma^{+}}),
\en
and
\be
A(\bar B^0_s\to p\overline{p})
   &=&
A(\bar B^0_s\to\Sigma^{+}\overline{\Sigma^{+}})
   +A(\bar B^0\to \Sigma^{+}\overline{p}).
\en

For more relations on amplitudes in various limits, one is referred to \cite{Chua:2003it}.

\subsection{Numerical results on rates}


In our numerical analysis, 
masses and lifetimes of hadrons are taken from \cite{PDG}, while values of Wolfenstein parameters for the CKM matrix are from \cite{CKMfitter}. 
Our strategy is to fit the asymptotic amplitude using the experimental $\overline B{}^0\to p\bar p$ rate, and try to predict rates on other baryonic modes with estimations on the corrections to the asymptotic relations and contributions from sub-leading terms. In principle, we can extract the full topological amplitudes directly from data, but at the moment since only one mode is found, we can only start from the asymptotic limit, as the number of parameters is highly reduced, and consider reasonable corrections to it. As we shall see, the prediction on rates are within a factor of 2. The accuracy can be systematically improved when more modes are observed.

Fitting to the experimental result on $\overline B{}^0\to p\bar p$ rate using the topological amplitude, Eq.(\ref{eq:BBB0}), but in the asymptotic forms, Eqs. (\ref{eq:asymptoticrelations}) and (\ref{eq:asymptotic1}),  we obtain
\be
\chi=(3.57^{+0.78}_{-0.71})\times 10^{-3}~{\rm GeV}^2.
\label{eq:correction0}
\en
Rates on other modes in the asymptotic limit can be obtained by using formulas in Sec. II.B and Eqs. (\ref{eq:asymptoticrelations}) and (\ref{eq:asymptotic1}).

In reality the topological amplitudes are, however, not in the asymptotic limit.
Corrections are expected and can be estimated as following.
(i) The correction on $T^{(\prime)}_i$, $P^{(\prime)}_i$ and $P^{(\prime)}_{EWi}$ are estimated to be of order $m_{\cal B}/m_B$ (the baryon and $B$ meson mass ratio), which is roughly, 0.2, hence, we have
\be
T^{(\prime)}_i=(1+t^{(\prime)}_i) T^{(\prime)},
\,
P^{(\prime)}_i=(1+p^{(\prime)}_i) P^{(\prime)}_,
\,
P^{(\prime)}_{EW i}=(1+p^{(\prime)}_{ewi}) P^{(\prime)}_{EW},
\en
with
\be
-0.2\leq t^{(\prime)}_i, p^{(\prime)}_i, p^{(\prime)}_{ewi}\leq 0.2,
\label{eq:correction1}
\en
which parametrize the correction.
(ii) Furthermore,
since the Fierz transformation of $O_{5,6,7,8}$ are different from $O_{1,2,3,4}$, 
the relation of the contributions from these two sets of operators may be distorted when we move away from the asymptotic limit. We assign a coefficient $\kappa$ in front of $c_5-c_6$ and $c_7-c_8$ in Eq.~(\ref{eq:asymptotic1}) with $\kappa$ having a 100\% uncertainty: 
\be
\kappa=1\pm1,
\label{eq:correction2}
\en
to model the correction.
(iii) For subleading terms, such as annihilation, penguin annihilation, exchange amplitude, we have
\be
E^{(\prime)}_i\equiv\eta_i \frac{f_B}{m_B} \frac{m_{\cal B}}{m_B} T^{(\prime)},\,
A^{(\prime)}_j\equiv\eta_j \frac{f_B}{m_B} \frac{m_{\cal B}}{m_B}T^{(\prime)},\,
PA^{(\prime)}_k\equiv\eta_k \frac{f_B}{m_B} \frac{m_{\cal B}}{m_B}P^{(\prime)},
\en
where the ratio $f_B/m_B$ is from the usual estimation~\cite{Gronau:1995hn}, the factor $m_{\cal B}/m_B$ is from the chirality structure, and $|\eta_{i,j,k}|$ are estimated to be of order 1.
Explicitly, we take
\be
0\leq|\eta_{i,j,k}|\leq |\eta|=1,
\label{eq:correction3}
\en
where we set the bound $|\eta|$ to 1 in our numerical results. We will return to this point later when confronting the $\overline B{}^0_s\to p\bar p$ data.
Note that some SU(3) breaking effects in rates are included, as the physical hadron masses \cite{PDG} are used in the numerical analysis.

Before we show our results, we comment on the detectability of baryonic final states. As shown in Appendix C, we note that, (i) $\Delta^{++,0}$, $\Lambda$, $\Xi^-$, $\Sigma^{*\pm}$, $\Xi^{*0}$ and $\Omega^-$ have non-suppressed decay modes of final states with all charged particles, (ii) $\Delta^+$, $\Sigma^{+,0}$, $\Xi^0$, $\Sigma^{*0}$ and $\Xi^{*-}$ can be detected by detecting a $\pi^0$ or $\gamma$, (iii) while one needs to deal with $n$ in detecting $\Delta^-$ and $\Sigma^-$. Modes with final states from the first group or even the second group and with unsuppressed $B$ decay rates should be experimentally accessible.

\begin{table}[t!]
\caption{\label{tab:DDDS=0} Decay rates for $\Delta S=0$,
$\overline B_q\to\DD$ modes. 
The first uncertainty is from the uncertainty of the asymptotic amplitude, $\chi$, and from relaxing the asymptotic relations, by varying $t_i, p_i, p_{ewi}$ (see Eqs. (\ref{eq:correction0}) and (\ref{eq:correction1})), the second uncertainty is from $\delta\kappa$ (see Eq.~(\ref{eq:correction2})), 
and the last uncertainty is from sub-leading contributions, terms with $\eta_{i,j,k}$ (see Eq. (\ref{eq:correction3})). Occasionally the last uncertainty is shown to 
larger decimal place.}
\begin{ruledtabular}
\begin{tabular}{llll}
Mode
          & ${\mathcal B}(10^{-8})$
          & Mode
          & ${\mathcal B}(10^{-8})$
          \\
\hline $B^-\to \Delta^+ \overline{\Delta^{++}}$
          & $17.15^{+19.62}_{-10.15}{}^{+0.81}_{-0.47}\pm0.22$
          & $\overline B{}^0_s\to \Delta^{+} \overline{\Sigma^{*+}}$
          & $5.18^{+5.92}_{-3.07}{}^{+0.24}_{-0.14}\pm0$
           \\
$B^-\to \Delta^0 \overline{\Delta^{+}}$
          & $6.42^{+7.34}_{-3.80}{}^{+1.13}_{-0.68}\pm0.15$
          & $\overline B{}^0_s\to \Delta^{0} \overline{\Sigma^{*0}}$
          & $2.91^{+3.32}_{-1.72}{}^{+0.51}_{-0.31}\pm0$
           \\ 
$B^-\to \Delta^- \overline{\Delta^{0}}$
          & $0.75^{+0.85}_{-0.44}{}^{+0.89}_{-0.55}\pm0.05$
          &$\overline B{}^0_s\to \Delta^{-} \overline{\Sigma^{*-}}$
          & $0.68^{+0.77}_{-0.40}{}^{+0.81}_{-0.49}\pm0$
           \\            
$B^-\to \Sigma^{*0} \overline{\Sigma^{*+}}$
          & $2.99^{+3.42}_{-1.77}{}^{+0.53}_{-0.32}\pm0.07$
          & $\overline B{}^0_s\to \Sigma^{*0} \overline{\Xi^{*0}}$
          & $2.70^{+3.09}_{-1.60}{}^{+0.48}_{-0.29}\pm0$
           \\ 
$B^-\to \Sigma^{*-} \overline{\Sigma^{*0}}$
          & $0.47^{+0.53}_{-0.27}{}^{+0.55}_{-0.34}\pm0.03$
          & $\overline B{}^0_s\to \Sigma^{*-} \overline{\Xi^{*-}}$
          & $0.84^{+0.96}_{-0.50}{}^{+1.00}_{-0.61}\pm0$
           \\ 
$B^-\to \Xi^{*-} \overline{\Xi^{*0}}$
          & $0.21^{+0.24}_{-0.13}{}^{+0.25}_{-0.16}\pm0.01$
          & $\overline B{}^0_s\to \Xi^{*-} \overline{\Omega^{-}}$
          & $0.58^{+0.66}_{-0.34}{}^{+0.69}_{-0.42}\pm0$
           \\ 
$\overline B{}^0\to \Delta^{++} \overline{\Delta^{++}}$
          & $0\pm0\pm0{}^{+0.0053}_{-0}$
          & $\overline B{}^0\to \Sigma^{*+} \overline{\Sigma^{*+}}$
          & $0\pm0\pm0{}^{+0.0031}_{-0}$
           \\
$\overline B{}^0\to \Delta^{+} \overline{\Delta^{+}}$
          & $5.29^{+6.05}_{-3.13}{}^{+0.25}_{-0.15}{}^{+0.27}_{-0.26}$
          & $\overline B{}^0\to \Sigma^{*0} \overline{\Sigma^{*0}}$
          & $1.39^{+1.58}_{-0.82}{}^{+0.24}_{-0.15}\pm0.09$
           \\
$\overline B{}^0\to \Delta^{0} \overline{\Delta^{0}}$
          & $5.94^{+6.79}_{-3.52}{}^{+1.05}_{-0.63}{}^{+0.20}_{-0.19}$
          & $\overline B{}^0\to \Sigma^{*-} \overline{\Sigma^{*-}}$
          & $0.86^{+0.98}_{-0.51}{}^{+1.02}_{-0.63}\pm0.05$
           \\ 
$\overline B{}^0\to \Delta^{-} \overline{\Delta^{-}}$
          & $2.08^{+2.37}_{-1.23}{}^{+2.47}_{-1.52}\pm0.08$
          & $\overline B{}^0\to \Xi^{*0} \overline{\Xi^{*0}}$
          & $0\pm0\pm0{}^{+0.0016}_{-0}$
           \\ 
$\overline B{}^0\to \Omega^{-} \overline{\Omega^{-}}$
          & $0\pm0\pm0{}^{+0.0006}_{-0}$
          & $\overline B{}^0\to \Xi^{*-} \overline{\Xi^{*-}}$
          & $0.20^{+0.22}_{-0.12}{}^{+0.24}_{-0.14}\pm0.02$
           \\                                                                                                                                                         
\end{tabular}
\end{ruledtabular}
\end{table}


We are now ready to discuss our numerical results.
Predictions on $\Delta S=0$, $\overline B_q\to\DD$ decay rates are shown in Table~\ref{tab:DDDS=0}.
The first uncertainty is from the uncertainty of the asymptotic amplitude, $\chi$, and from relaxing the asymptotic relations, by varying $t_i, p_i, p_{ewi}$ (see Eqs. (\ref{eq:correction0}) and (\ref{eq:correction1})), the second uncertainty is from $\delta\kappa$ (see Eq.~(\ref{eq:correction2})), 
and the last uncertainty is from sub-leading contributions, terms with $\eta_{i,j,k}$ (see Eq. (\ref{eq:correction3})). Occasionally the last uncertainty is shown to 
larger decimal place.

There are modes that will cascadely decay to all charged final states, such as $p\overline p$ with one or more charge pions or kaons (see Appendix C). 
These include
$\overline B{}^0\to \Delta^{++}\overline{\Delta^{++}}$, $\Delta^0\overline{\Delta^0}$, $\Omega^-\overline{\Omega^-}$, $\Sigma^{*+}\overline{\Sigma^{*+}}$, $\Sigma^{*-}\overline{\Sigma^{*-}}$ and $\Xi^{*0}\overline{\Xi^{*0}}$ decays.
Among them, we note that
$\overline{B}{}^0\to \Delta^0\overline{\Delta^0}$ 
and 
$\overline{B}{}^0\to \Sigma^{*-}\overline{\Sigma^{*-}}$ 
rates are at $10^{-8}$ level. These two modes are relatively easy to be detected, while
other modes are suppressed.

\begin{table}[t!]
\caption{\label{tab:DDDS=-1} Same as Table~\ref{tab:DDDS=0}, but with $\Delta S=-1$, $\overline B_q\to\DD$ modes.}
\begin{ruledtabular}
\begin{tabular}{llll}
Mode
          & ${\mathcal B}(10^{-8})$
          & Mode
          & ${\mathcal B}(10^{-8})$
          \\
\hline $B^-\to \Sigma^{*+} \overline{\Delta^{++}}$
          & $13.94^{+18.20}_{-8.87}{}^{+17.08}_{-10.09}\pm0.038$
          & $\overline B{}^0\to \Sigma^{*+} \overline{\Delta^{+}}$
          & $4.30^{+5.62}_{-2.74}{}^{+5.27}_{-3.11}\pm0$
           \\
$B^-\to \Sigma^{*0} \overline{\Delta^{+}}$
          & $9.74^{+11.49}_{-5.87}{}^{+11.92}_{-7.18}\pm0.028$
          & $\overline B{}^0\to \Sigma^{*0} \overline{\Delta^{0}}$
          & $9.02^{+10.64}_{-5.43}{}^{+11.03}_{-6.64}\pm0$
           \\ 
$B^-\to \Sigma^{*-} \overline{\Delta^{0}}$
          & $5.24^{+5.96}_{-3.09}{}^{+6.22}_{-3.82}\pm0.015$
          &$\overline B{}^0\to \Sigma^{*-} \overline{\Delta^{-}}$
          & $14.54^{+16.54}_{-8.58}{}^{+17.28}_{-10.60}\pm0$
           \\            
$B^-\to \Xi^{*0} \overline{\Sigma^{*+}}$
          & $18.07^{+21.31}_{-10.88}{}^{+22.10}_{-13.31}\pm0.052$
          & $\overline B{}^0\to \Xi^{*0} \overline{\Sigma^{*0}}$
          & $8.37^{+9.86}_{-5.04}{}^{+10.23}_{-6.16}\pm0$
           \\ 
$B^-\to \Xi^{*-} \overline{\Sigma^{*0}}$
          & $9.71^{+11.05}_{-5.73}{}^{+11.54}_{-7.08}\pm0.028$
          & $\overline B{}^0\to \Xi^{*-} \overline{\Sigma^{*-}}$
          & $17.96^{+20.43}_{-10.60}{}^{+21.34}_{-13.09}\pm0$
           \\ 
$B^-\to \Omega^{-} \overline{\Xi^{*0}}$
          & $13.38^{+15.21}_{-7.90}{}^{+15.89}_{-9.75}\pm0.039$
          & $\overline B{}^0\to \Omega^{-} \overline{\Xi^{*-}}$
          & $12.37^{+14.07}_{-7.30}{}^{+14.70}_{-9.02}\pm0$
           \\ 
$\overline B{}^0_s\to \Delta^{++} \overline{\Delta^{++}}$
          & $0\pm0\pm0{}^{+0.019}_{-0}$
          & $\overline B{}^0_s\to \Sigma^{*+} \overline{\Sigma^{*+}}$
          & $4.21^{+5.50}_{-2.68}{}^{+5.16}_{-3.05}{}^{+0.54}_{-0.50}$
           \\
$\overline B{}^0_s\to \Delta^{+} \overline{\Delta^{+}}$
          & $0\pm0\pm0{}^{+0.018}_{-0}$
          & $\overline B{}^0_s\to \Sigma^{*0} \overline{\Sigma^{*0}}$
          & $4.42^{+5.20}_{-2.66}{}^{+5.40}_{-3.25}{}^{+0.55}_{-0.52}$
           \\
$\overline B{}^0_s\to \Delta^{0} \overline{\Delta^{0}}$
          & $0\pm0\pm0{}^{+0.017}_{-0}$
          & $\overline B{}^0_s\to \Sigma^{*-} \overline{\Sigma^{*-}}$
          & $4.74^{+5.39}_{-2.80}{}^{+5.63}_{-3.45}{}^{+0.56}_{-0.53}$
           \\ 
$\overline B{}^0_s\to \Delta^{-} \overline{\Delta^{-}}$
          & $0\pm0\pm0{}^{+0.017}_{-0}$
          & $\overline B{}^0_s\to \Xi^{*0} \overline{\Xi^{*0}}$
          & $16.33^{+19.25}_{-9.83}{}^{+19.97}_{-12.03}{}^{+1.00}_{-0.97}$
           \\ 
$\overline B{}^0_s\to \Omega^{-} \overline{\Omega^{-}}$
          & $36.24^{+41.22}_{-21.39}{}^{+43.06}_{-26.41}{}^{+1.40}_{-1.37}$
          & $\overline B{}^0_s\to \Xi^{*-} \overline{\Xi^{*-}}$
          & $17.53^{+19.94}_{-10.35}{}^{+20.83}_{-12.78}{}^{+1.02}_{-0.99}$
           \\                                                                                                                                                         
\end{tabular}
\end{ruledtabular}
\end{table}

Modes need one $\pi^0$ or one $\gamma$ for detections are: 
$B^-\to\Delta^+\overline{\Delta^{++}}$, 
$\Delta^0\overline{\Delta^+}$, 
$\Sigma^{*0}\overline{\Sigma^{*+}}$, 
$\Sigma^{*-}\overline{\Sigma^{*0}}$,
$\Xi^{*-}\overline{\Xi^{*0}}$, 
$\overline B{}^0_s\to\Delta^+\overline{\Sigma^{*+}}$, 
$\Delta^0\overline{\Sigma^{*0}}$,
$\Sigma^{*0}\overline{\Xi^{*0}}$, 
$\Sigma^{*-}\overline{\Xi^{*-}}$ and
$\Xi^{*-}\overline{\Omega^-}$ decays.
Among them, we have
${\cal B}(B^-\to \Delta^+\overline{\Delta^{++}})\simeq 2\times 10^{-7}$ and it reduces $\sim30\%$ in producing
$p\,\pi^0\,\overline p\,\pi^-$ final state.

Modes with more than one $\pi^0$ or $\gamma$ are more difficult to detect. They are
$\overline B{}^0\to\Delta^+\overline{\Delta^+}$, $\Sigma^{*0}\overline{\Sigma^{*0}}$, $\Xi^{*-}\overline{\Xi^{*-}}$ decays. The first two have rates of order $10^{-8}$.
Some modes need $n$ for detection and they are very difficult to be observed. They are
$B^-\to\Delta^-\overline{\Delta^0}$, 
$\overline B{}^0\to\Delta^-\overline{\Delta^-}$ and
$\overline B{}^0_s\to\Delta^-\overline{\Sigma^{*-}}$ decays.

Predictions on $\Delta S=-1$, $\overline B_q\to\DD$ decay rates are shown in Table~\ref{tab:DDDS=-1}.
From the table we see that: (i) Modes having all charge final states in cascade decays with rates ranging from $10^{-8}$ to $10^{-7}$ are
$B^-\to\Sigma^{*+}\overline {\Delta^{++}}$, $\Sigma^{*-}\overline{\Delta^0}$, $\Omega^-\overline{\Xi^{*0}}$, $\overline B{}^0_s\to\Omega^-\overline{\Omega^-}$, $\Xi^{*0}\overline{\Xi^{*0}}$, $\Sigma^{*0}\overline{\Sigma^{*-}}$ and $\Sigma^{*+}\overline{\Sigma^{*+}}$ decays.
Note that $\overline B {}^0_s\to\Omega^-\overline{\Omega^-}$ decay has the largest rate. 
(ii) With one $\pi^0$ or $\gamma$ for detection, we have
$B^-\to \Xi^{*-}\overline{\Sigma^{*0}}$, $\overline B{}^0\to \Xi^{*-}\overline{\Sigma^{*-}}$, $\Omega^-\overline{\Xi^{*-}}$ with rate at $10^{-7}$.
(iii) All other modes are either too small in rates or need more than one $\pi^0$ or one $\gamma$ or even $n$ for detection.


\begin{table}[t!]
\caption{\label{tab:BDDS=0} Same as Table~\ref{tab:DDDS=0}, but with $\Delta S=0$, $\overline B_q\to\BD$ modes.
The latest experimental result is given in the parenthesis. 
}
\begin{ruledtabular}
\begin{tabular}{llll}
Mode
          & ${\mathcal B}(10^{-8})$
          & Mode
          & ${\mathcal B}(10^{-8})$
          \\
\hline $B^-\to p\overline{\Delta^{++}}$%
          & $7.50^{+20.63}_{-6.67}{}^{+0.35}_{-0.21}\pm0.10$
          & $\overline B{}^0_s\to p\overline{\Sigma^{*+}}$%
          & $2.28^{+6.28}_{-2.03}{}^{+0.11}_{-0.06}\pm0$
           \\
           & ($<14$)~\cite{Wei:2007fg}
          \\
$B^-\to n\overline{\Delta^+}$%
          & $2.54^{+2.91}_{-1.50}{}^{+0.14}_{-0.09}\pm0.03$
          & $\overline B{}^0_s\to n\overline{\Sigma^{*0}}$%
          & $1.16^{+1.33}_{-0.69}{}^{+0.07}_{-0.04}\pm0$
           \\ 
$B^-\to\Sigma^0\overline{\Sigma^{*+}}$%
          & $4.12^{+4.70}_{-2.44}{}^{+0.04}_{-0.02}\pm0.03$
          &$\overline B{}^0_s\to \Sigma^{0}\overline{\Xi^{*0}}$%
          & $3.74^{+4.27}_{-2.21}{}^{+0.04}_{-0.02}\pm0$
           \\            
$B^-\to\Sigma^-\overline{\Sigma^{*0}}$%
          & $0.05^{+0.05}_{-0.03}{}^{+0.05}_{-0.03}\pm0.003$
          & $\overline B{}^0_s\to \Sigma^{-}\overline{\Xi^{*-}}$%
          & $0.08^{+0.10}_{-0.05}{}^{+0.10}_{-0.06}\pm0$
           \\ 
$B^-\to\Xi^{-}\overline{\Xi^{*0}}$%
          & $0.08^{+0.09}_{-0.05}{}^{+0.10}_{-0.06}\pm0.005$
          & $\overline B{}^0_s\to \Xi^{-}\overline{\Omega^-}$%
          & $0.22^{+0.25}_{-0.13}{}^{+0.25}_{-0.16}\pm0$
           \\ 
$B^-\to\Lambda\overline{\Sigma^{*+}}$%
          & $0.14^{+0.52}_{-0.09}{}^{+0.17}_{-0.10}\pm0.009$
          & $\overline B{}^0_s\to \Lambda\overline{\Xi^{*0}}$%
          & $0.13^{+0.47}_{-0.08}{}^{+0.15}_{-0.10}\pm0$
           \\ 
$\overline B{}^0\to p\overline{\Delta^+}$%
          & $2.31^{+6.37}_{-2.06}{}^{+0.11}_{-0.06}\pm0.03$
          & $\overline B{}^0\to \Sigma^{+}\overline{\Sigma^{*+}}$%
          & $0\pm0\pm0{}^{+0.0001}_{-0}$
           \\
$\overline B{}^0\to n\overline{\Delta^0}$%
          & $2.35^{+2.69}_{-1.39}{}^{+0.13}_{-0.08}\pm0.03$
          & $\overline B{}^0\to \Sigma^{0}\overline{\Sigma^{*0}}$%
          & $1.91^{+2.18}_{-1.13}{}^{+0.02}_{-0.01}\pm0.01$
           \\
$\overline B{}^0\to \Xi^{0}\overline{\Xi^{*0}}$%
          & $0\pm0\pm0{}^{+0.0001}_{-0}$
          & $\overline B{}^0\to \Sigma^{-}\overline{\Sigma^{*-}}$%
          & $0.09^{+0.10}_{-0.05}{}^{+0.10}_{-0.06}\pm0$
           \\      
$\overline B{}^0\to \Xi^{-}\overline{\Xi^{*-}}$%
          & $0.07^{+0.08}_{-0.04}{}^{+0.09}_{-0.05}\pm0$
          & $\overline B{}^0\to \Lambda\overline{\Sigma^{*0}}$%
          & $0.07^{+0.24}_{-0.04}{}^{+0.08}_{-0.05}\pm0.004$
           \\                                                                                                                                                                
\end{tabular}
\end{ruledtabular}
\end{table}

Predictions on rates of  $\Delta S=0$, $\overline B_q\to\BD$ decays are shown in Table~\ref{tab:BDDS=0}.
We note that for modes having all charge final states in cascade decays, the central value of the $B^-\to p\overline{\Delta^{++}}$ predicted rate is only half of the experimental upper bound. It should be searchable in the near future. Furthermore, the measurement of this mode will be useful to reduce the theoretical uncertainty. Another all charge cascade decay final state mode $\overline B{}^0_s\to p\overline{\Sigma^{*+}}$ has rate at $10^{-8}$ order, while all other states with similar cascade decay final states are suppressed. 
With one $\pi^0$ or $\gamma$, one may search for $B^-\to \Sigma^0\overline{\Sigma^{*+}}$, $\overline B{}^0\to p\overline{\Delta^+}$ and $\overline B{}^0_s\to\Sigma^0\overline{\Xi^{*0}}$. All other modes are either suppressed or are more difficult to be detected.  

\begin{table}[t!]
\caption{\label{tab:BDDS=-1} Same as Table~\ref{tab:DDDS=0}, but with $\Delta S=-1$, $\overline B_q\to\BD$ modes.
The latest experimental results are given in parentheses. 
}
\begin{ruledtabular}
\begin{tabular}{llll}
Mode
          & ${\mathcal B}(10^{-8})$
          & Mode
          & ${\mathcal B}(10^{-8})$
          \\
\hline $B^-\to \Sigma^+\overline{\Delta^{++}}$%
          & $5.75^{+8.76}_{-3.94}{}^{+7.04}_{-4.16}\pm0.02$
          & $\overline B{}^0\to \Sigma^{+}\overline{\Delta^+}$%
          & $1.77^{+2.70}_{-1.22}{}^{+2.17}_{-1.28}\pm0$
           \\
$B^-\to \Sigma^0\overline{\Delta^+}$%
          & $4.01^{+5.07}_{-2.50}{}^{+4.91}_{-2.96}\pm0.01$
          & $\overline B{}^0\to \Sigma^{0}\overline{\Delta^0}$%
          & $3.71^{+4.69}_{-2.32}{}^{+4.54}_{-2.74}\pm0$
           \\ 
$B^-\to \Sigma^-\overline{\Delta^0}$%
          & $2.15^{+2.45}_{-1.27}{}^{+2.56}_{-1.57}\pm0.006$
          &$\overline B{}^0\to \Sigma^{-}\overline{\Delta^-}$%
          & $5.98^{+6.80}_{-3.53}{}^{+7.11}_{-4.36}\pm0$
           \\            
$B^-\to \Xi^{0}\overline{\Sigma^{*+}}$%
          & $2.32^{+3.13}_{-1.48}{}^{+2.46}_{-1.55}\pm0.006$
          & $\overline B{}^0\to \Xi^{0}\overline{\Sigma^{*0}}$%
          & $1.07^{+1.45}_{-0.69}{}^{+1.14}_{-0.72}\pm0$
           \\ 
$B^-\to \Xi^{-}\overline{\Sigma^{*0}}$%
          & $0.95^{+1.08}_{-0.56}{}^{+1.12}_{-0.69}\pm0.003$
          & $\overline B{}^0\to \Xi^{-}\overline{\Sigma^{*-}}$%
          & $1.75^{+1.99}_{-1.03}{}^{+2.08}_{-1.27}\pm0$
           \\ 
$B^-\to \Lambda\overline{\Delta^{+}}$%
          & $0.18^{+0.23}_{-0.11}\pm0.005\pm0$
          & $\overline B{}^0\to \Lambda\overline{\Delta^0}$%
          & $0.17^{+0.21}_{-0.10}\pm0.004\pm0$
           \\
           & ($<82$)~\cite{Wang:2007as}
           &
           & ($<93$)~\cite{Wang:2007as} 
          \\ 
$\overline B{}^0_s\to p\overline{\Delta^+}$%
          & $0\pm0\pm0{}^{+0.00001}_{-0}$
          & $\overline B{}^0_s\to \Sigma^{+}\overline{\Sigma^{*+}}$%
          & $1.75^{+2.67}_{-1.20}{}^{+2.14}_{-1.27}\pm0.005$
           \\
$\overline B{}^0_s\to n\overline{\Delta^0}$%
          & $0\pm0\pm0{}^{+0.00001}_{-0}$
          & $\overline B{}^0_s\to \Sigma^{0}\overline{\Sigma^{*0}}$%
          & $1.83^{+2.31}_{-1.14}{}^{+2.24}_{-1.35}\pm0.003$
           \\
$\overline B{}^0_s\to \Xi^{0}\overline{\Xi^{*0}}$%
          & $2.11^{+2.84}_{-1.35}{}^{+2.23}_{-1.41}\pm0.006$
          & $\overline B{}^0_s\to \Sigma^{-}\overline{\Sigma^{*-}}$%
          & $1.96^{+2.23}_{-1.16}{}^{+2.33}_{-1.43}\pm0$
           \\ 
$\overline B{}^0_s\to \Xi^{-}\overline{\Xi^{*-}}$%
          & $1.72^{+1.95}_{-1.01}{}^{+2.04}_{-1.25}\pm0$
          & $\overline B{}^0_s\to \Lambda\overline{\Sigma^{*0}}$%
          & $0.08^{+0.10}_{-0.05}\pm0.002\pm0.0008$
           \\                                                                                                                                                        
\end{tabular}
\end{ruledtabular}
\end{table}

Predictions on $\Delta S=-1$, $\overline B_q\to\BD$ decay rates are shown in Table \ref{tab:BDDS=-1}. 
There are only two modes having all charge final states in cascade decays, namely $\overline B{}^0\to \Xi^-\overline{\Sigma^{*-}}$ and $\Lambda\overline{\Delta^0}$. The former has rate of $10^{-8}$ order, while the latter is of order $10^{-9}$ and is two order of magnitude smaller than the experimental upper limit. 
With one $\pi^0$ or $\gamma$ one may search for $B^-\to \Sigma^+\overline{\Delta^{++}}$ and $\overline B{}^0\to \Sigma^0\overline{\Delta^0}$. Note that $B^-\to \Lambda\overline{\Delta^+}$ is lower than the experimental upper limit by one to two orders of magnitudes.


\begin{table}[t!]
\caption{\label{tab:DBDS=0} Same as Table~\ref{tab:DDDS=0}, but with $\Delta S=0$, $\overline B_q\to\DB$ modes.
The latest experimental result is given in the parenthesis. 
}
\begin{ruledtabular}
\begin{tabular}{llll}
Mode
          & ${\mathcal B}(10^{-8})$
          & Mode
          & ${\mathcal B}(10^{-8})$
          \\
\hline $B^-\to \Delta^0\overline{p}$%
          & $2.54^{+2.90}_{-1.50}{}^{+0.14}_{-0.09}\pm0.03$
          & $\overline B{}^0_s\to \Delta^+\overline{\Sigma^{+}}$%
          & $2.13^{+2.44}_{-1.26}{}^{+0.10}_{-0.06}\pm0$
           \\
           & ($<138$)~ \cite{Wei:2007fg}
          \\
$B^-\to \Delta^-\overline{n}$%
          & $0.33^{+0.37}_{-0.19}{}^{+0.39}_{-0.24}\pm0.02$
          & $\overline B{}^0_s\to \Delta^0\overline{\Sigma^{0}}$%
          & $1.19^{+1.37}_{-0.71}{}^{+0.21}_{-0.13}\pm0$
           \\ 
$B^-\to \Sigma^{*0}\overline{\Sigma^{+}}$%
          & $1.08^{+1.22}_{-0.64}{}^{+0.06}_{-0.04}\pm0.013$
          &$\overline B{}^0_s\to \Delta^-\overline{\Sigma^{-}}$%
          & $0.28^{+0.32}_{-0.16}{}^{+0.33}_{-0.20}\pm0$
           \\            
$B^-\to \Sigma^{*-}\overline{\Sigma^{0}}$%
          & $0.05^{+0.05}_{-0.03}{}^{+0.05}_{-0.03}\pm0.003$%
          & $\overline B{}^0_s\to \Sigma^{*0}\overline{\Xi^{0}}$%
          & $3.65^{+4.16}_{-2.16}{}^{+0.04}_{-0.02}\pm0$
           \\ 
$B^-\to \Xi^{*-}\overline{\Xi^{0}}$%
          & $0.08^{+0.09}_{-0.05}{}^{+0.10}_{-0.06}\pm0.005$
          & $\overline B{}^0_s\to \Sigma^{*-}\overline{\Xi^{-}}$%
          & $0.08^{+0.09}_{-0.05}{}^{+0.10}_{-0.06}\pm0$
           \\ 
$B^-\to \Sigma^{*-}\overline{\Lambda}$%
          & $0.14^{+0.16}_{-0.08}{}^{+0.17}_{-0.10}\pm0.01$
          & $\overline B{}^0_s\to \Delta^0\overline{\Lambda}$%
          & $3.17^{+3.61}_{-1.87}{}^{+0.00}_{-0.00}\pm0$
           \\ 
$\overline B{}^0\to \Delta^+\overline{p}$%
          & $2.31^{+2.65}_{-1.37}{}^{+0.11}_{-0.06}\pm0.03$
          & $\overline B{}^0\to \Sigma^{*+}\overline{\Sigma^{+}}$%
          & $0\pm0\pm0{}^{+0.0001}_{-0}$
           \\
$\overline B{}^0\to \Delta^0\overline{n}$%
          & $8.99^{+10.25}_{-5.31}{}^{+0.10}_{-0.06}\pm0.06$
          & $\overline B{}^0\to \Sigma^{*0}\overline{\Sigma^{0}}$%
          & $0.50^{+0.57}_{-0.29}{}^{+0.03}_{-0.02}\pm0.006$
           \\
$\overline B{}^0\to \Xi^{*0}\overline{\Xi^{0}}$%
          & $0\pm0\pm0{}^{+0.0001}_{-0}$
          & $\overline B{}^0\to \Sigma^{*-}\overline{\Sigma^{-}}$%
          & $0.09^{+0.10}_{-0.05}{}^{+0.10}_{-0.06}\pm0$
           \\      
$\overline B{}^0\to \Xi^{*-}\overline{\Xi^{-}}$%
          & $0.07^{+0.08}_{-0.04}{}^{+0.09}_{-0.05}\pm0$
          & $\overline B{}^0\to \Sigma^{*0}\overline{\Lambda}$%
           & $1.52^{+1.74}_{-0.90}{}^{+0.07}_{-0.04}\pm0.02$
           \\                                                                                                                                                                
\end{tabular}
\end{ruledtabular}
\end{table}

Predictions on $\Delta S=0$, $\overline B_q\to \DB$ decay rates are shown in Table \ref{tab:DBDS=0}. 
Note  that $B^-\to \Delta^0\overline p$ and $\overline B{}^0_s\to\Delta^0\overline\Lambda$ decays are modes that having all charge final states in cascade decays and with rates of order $10^{-8}$. The former is one to two orders of magnitudes below the present experimental limit.

\begin{table}[t!]
\caption{\label{tab:DBDS=-1} Same as Table~\ref{tab:DDDS=0}, but with $\Delta S=-1$, $\overline B_q\to\DB$ modes.}
\begin{ruledtabular}
\begin{tabular}{llll}
Mode
          & ${\mathcal B}(10^{-8})$
          & Mode
          & ${\mathcal B}(10^{-8})$
          \\
\hline $B^-\to \Sigma^{*0}\overline{p}$%
          & $1.36^{+1.63}_{-0.81}{}^{+1.43}_{-0.91}\pm0.004$
          & $\overline B{}^0\to \Sigma^{*+}\overline{p}$%
          & $1.82^{+2.38}_{-1.16}{}^{+2.23}_{-1.32}\pm0$
           \\
$B^-\to \Sigma^{*-}\overline{n}$%
          & $2.21^{+2.52}_{-1.31}{}^{+2.63}_{-1.61}\pm0.006$
          & $\overline B{}^0\to \Sigma^{*0}\overline{n}$%
          & $0.90^{+1.39}_{-0.62}{}^{+1.01}_{-0.57}\pm0$
           \\ 
$B^-\to \Xi^{*0}\overline{\Sigma^{+}}$%
          & $2.27^{+2.73}_{-1.37}{}^{+2.41}_{-1.52}\pm0.006$
          &$\overline B{}^0\to \Xi^{*0}\overline{\Sigma^{0}}$%
          & $1.05^{+1.26}_{-0.63}{}^{+1.11}_{-0.70}\pm0$
           \\            
$B^-\to \Xi^{*-}\overline{\Sigma^{0}}$%
          & $0.93^{+1.06}_{-0.55}{}^{+1.10}_{-0.68}\pm0.003$
          & $\overline B{}^0\to \Xi^{*-}\overline{\Sigma^{-}}$%
          & $1.71^{+1.95}_{-1.01}{}^{+2.04}_{-1.25}\pm0$
           \\ 
$B^-\to \Omega^-\overline{\Xi^{0}}$%
          & $4.81^{+5.47}_{-2.84}{}^{+5.72}_{-3.51}\pm0.014$
          & $\overline B{}^0\to \Omega^-\overline{\Xi^{-}}$%
          & $4.44^{+5.05}_{-2.62}{}^{+5.27}_{-3.24}\pm0$
           \\ 
$B^-\to \Xi^{*-}\overline{\Lambda}$%
          & $2.88^{+3.28}_{-1.70}{}^{+3.42}_{-2.10}\pm0.008$
          & $\overline B{}^0\to \Xi^{*0}\overline{\Lambda}$%
          & $2.37^{+3.09}_{-1.51}{}^{+2.90}_{-1.71}\pm0$
           \\ 
$\overline B{}^0_s\to \Delta^+\overline{p}$%
          & $0\pm0\pm0{}^{+0.00001}_{-0}$
          & $\overline B{}^0_s\to \Sigma^{*+}\overline{\Sigma^{+}}$%
          & $1.67^{+2.18}_{-1.06}{}^{+2.05}_{-1.21}\pm0.005$
           \\
$\overline B{}^0_s\to \Delta^0\overline{n}$%
          & $0\pm0\pm0{}^{+0.00001}_{-0}$
          & $\overline B{}^0_s\to \Sigma^{*0}\overline{\Sigma^{0}}$%
          & $1.75^{+2.07}_{-1.05}{}^{+2.14}_{-1.29}\pm0.003$
           \\
$\overline B{}^0_s\to \Xi^{*0}\overline{\Xi^{0}}$%
          & $1.45^{+2.22}_{-1.00}{}^{+1.62}_{-0.92}\pm0.004$
          & $\overline B{}^0_s\to \Sigma^{*-}\overline{\Sigma^{-}}$%
          & $1.88^{+2.13}_{-1.11}{}^{+2.23}_{-1.37}\pm0$
           \\ 
$\overline B{}^0_s\to \Xi^{*-}\overline{\Xi^{-}}$%
          & $1.64^{+1.86}_{-0.97}{}^{+1.94}_{-1.19}\pm0$
          & $\overline B{}^0_s\to \Sigma^{*0}\overline{\Lambda}$%
          & $0.08^{+0.09}_{-0.05}{}^{+0.00}_{-0.00}\pm0.001$
           \\                                                                                                                                                  
\end{tabular}
\end{ruledtabular}
\end{table}

Predictions on $\Delta S=-1$, $\overline B_q\to \DB$ decay rates are shown in Table \ref{tab:DBDS=-1}. 
Note  that $\overline B{}^0\to \Sigma^{*-}\overline p$, $\Omega^-\overline{\Xi^-}$ and $\Xi^{*0}\overline\Lambda$ are modes that having all charge final states in cascade decays and have rates of order $10^{-8}$.

\begin{table}[t!]
\caption{\label{tab:BBDS=0} Same as Table~\ref{tab:DDDS=0}, but with $\Delta S=0$, $\overline B_q\to\BB$ modes.
The latest experimental results are given in parentheses under the theoretical results.}
\begin{ruledtabular}
\begin{tabular}{llll}
Mode
          & ${\mathcal B}(10^{-8})$
          & Mode
          & ${\mathcal B}(10^{-8})$
          \\
\hline $B^-\to n\overline{p}$%
          & $3.20^{+3.69}_{-1.90}{}^{+2.02}_{-1.23}{}^{+0.11}_{-0.10}$
          & $\overline B{}^0_s\to p\overline{\Sigma^{+}}$%
          & $1.42^{+3.91}_{-1.26}{}^{+0.07}_{-0.04}\pm0$
           \\
$B^-\to \Sigma^{0}\overline{\Sigma^{+}}$%
          & $3.26^{+3.92}_{-1.98}{}^{+0.57}_{-0.35}{}^{+0.12}_{-0.11}$
          & $\overline B{}^0_s\to n\overline{\Sigma^{0}}$%
          & $0.72^{+0.83}_{-0.43}{}^{+0.04}_{-0.03}\pm0$
           \\ 
$B^-\to \Sigma^{-}\overline{\Sigma^{0}}$%
          & $0.51^{+0.78}_{-0.35}{}^{+0.60}_{-0.37}{}^{+0.05}_{-0.05}$
          &$\overline B{}^0_s\to n\overline{\Lambda}$%
          & $2.88^{+3.44}_{-1.74}{}^{+0.98}_{-0.59}\pm0$
           \\            
$B^-\to \Sigma^{-}\overline{\Lambda}$%
          & $0.39^{+0.44}_{-0.23}{}^{+0.46}_{-0.28}{}^{+0.02}_{-0.02}$
          & $\overline B{}^0_s\to \Sigma^{0}\overline{\Xi^{0}}$%
          & $10.84^{+12.40}_{-6.42}{}^{+0.79}_{-0.47}\pm0$
           \\ 
$B^-\to \Xi^{-}\overline{\Xi^{0}}$%
          & $0.06^{+0.07}_{-0.04}{}^{+0.07}_{-0.04}{}^{+0.004}_{-0.004}$
          & $\overline B{}^0_s\to \Sigma^{-}\overline{\Xi^{-}}$%
          & $1.44^{+1.67}_{-0.86}{}^{+1.72}_{-1.05}\pm0$%
           \\ 
$B^-\to \Lambda\overline{\Sigma^+}$%
          & $0.39^{+0.69}_{-0.23}{}^{+0.46}_{-0.28}{}^{+0.02}_{-0.02}$
          & $\overline B{}^0_s\to \Lambda\overline{\Xi^0}$%
          & $0.09^{+1.04}_{-0.07}{}^{+0.10}_{-0.06}\pm0$
           \\ 
$\overline B{}^0\to p\overline{p}$%
          & $1.47^{+4.04}_{-1.31}{}^{+0.07}_{-0.04}{}^{+0.15}_{-0.15}$
          & $\overline B{}^0\to \Sigma^{+}\overline{\Sigma^{+}}$%
          & $0\pm0\pm0{}^{+0.004(+1.66)}_{-0}$
           \\
           &($1.47{}^{+0.62+0.35}_{-0.51-0.14}$)\footnotemark[1]~\cite{Aaij:2013fta}
           \\           
$\overline B{}^0\to n\overline{n}$%
          & $6.60^{+7.98}_{-4.01}{}^{+1.16}_{-0.70}{}^{+0.11(+2.55)}_{-0.11}$
          & $\overline B{}^0\to \Sigma^{0}\overline{\Sigma^{0}}$%
          & $1.51^{+1.82}_{-0.92}{}^{+0.27}_{-0.16}{}^{+0.09(+2.45)}_{-0.09}$
           \\
$\overline B{}^0\to \Xi^{0}\overline{\Xi^{0}}$%
          & $0\pm0\pm0{}^{+0.0004}_{-0}$%
          & $\overline B{}^0\to \Sigma^{-}\overline{\Sigma^{-}}$%
          & $0.94^{+1.44}_{-0.65}{}^{+1.11}_{-0.68}{}^{+0.03}_{-0.03}$
           \\      
$\overline B{}^0\to \Xi^{-}\overline{\Xi^{-}}$%
          & $0.06^{+0.06}_{-0.03}{}^{+0.07}_{-0.04}{}^{+0.01}_{-0.01}$
          & $\overline B{}^0\to \Sigma^{0}\overline{\Lambda}$%
           & $4.10^{+4.68}_{-2.42}{}^{+0.19}_{-0.11}{}^{+0.05}_{-0.05}$
           \\
$\overline B{}^0\to \Lambda\overline{\Lambda}$%
          & $0^{+0.33}_{-0}\pm0{}^{+0.0007(0.34)}_{-0}$%
          & $\overline B{}^0\to \Lambda\overline{\Sigma^{0}}$%
           & $0.18^{+0.32}_{-0.11}{}^{+0.21}_{-0.13}{}^{+0.01}_{-0.01}$
           \\ 
           & ($<32$)~\cite{Tsai:2007pp}
          \\                                                                                                                                                                                                     
\end{tabular}
\end{ruledtabular}
\footnotetext[1]{Taken as the input of our numerical analysis.}
\end{table}

Predictions on $\Delta S=0$, $\overline B_q\to \BB$ decay rates are shown in Table \ref{tab:BBDS=0}. 
We see from the table that the $\overline B{}^0\to p\bar p$ decay has the highest rate among modes that have all charge final states in cascade decays. Although there are rates higher than it, they require detection of $\pi^0$ and/or $\gamma$ for observations. For example, the $\overline B{}^0_s\to\Sigma^0\overline{\Xi^0}$ decay rate is of the order of $10^{-7}$, but one needs $\gamma$ and $\pi^0$ for detection. For $\Delta S=0$, $\overline B_q\to \BB$ decays, the $\overline B{}^0\to p\bar p$ decay is the most accessible mode among them. Therefore, it is not surprise that it is the first $\overline B_q\to \BB$ mode being found. Furthermore, we note that $\overline B{}^0\to\Xi^-\overline{\Xi^-}$ and $\Lambda\overline{\Lambda}$ decays having all charge final states in cascade decays are predicted to be highly suppressed. In fact, the latter is several orders of magnitudes below the present experimental limit.
These predictions can be checked experimentally.

\begin{table}[t!]
\caption{\label{tab:BBDS=-1} Same as Table~\ref{tab:DDDS=0}, but with $\Delta S=-1$, $\overline B_q\to\BB$ modes.
The latest experimental result is given in the parenthesis under the theoretical results.}
\begin{ruledtabular}
\begin{tabular}{llll}
Mode
          & ${\mathcal B}(10^{-8})$
          & Mode
          & ${\mathcal B}(10^{-8})$
          \\
\hline $B^-\to \Sigma^{0}\overline{p}$%
          & $0.88^{+1.13}_{-0.53}{}^{+0.93}_{-0.59}{}^{+0.002}_{-0.002}$
          & $\overline B{}^0\to \Sigma^{+}\overline{p}$%
          & $1.18^{+1.81}_{-0.81}{}^{+1.45}_{-0.86}\pm0$
           \\
$B^-\to \Sigma^{-}\overline{n}$%
          & $1.44^{+1.64}_{-0.85}{}^{+1.71}_{-1.05}{}^{+0.004}_{-0.004}$
          & $\overline B{}^0\to \Sigma^{0}\overline{n}$%
          & $0.59^{+1.26}_{-0.47}{}^{+0.66}_{-0.37}\pm0$
           \\ 
$B^-\to \Xi^{0}\overline{\Sigma^{+}}$%
          & $32.54^{+38.07}_{-19.51}{}^{+39.22}_{-23.89}{}^{+0.09}_{-0.09}$
          &$\overline B{}^0\to \Xi^{0}\overline{\Sigma^{0}}$%
          & $15.05^{+17.61}_{-9.02}{}^{+18.14}_{-11.05}\pm0$
           \\            
$B^-\to\Xi^{-}\overline{\Sigma^{0}}$%
          & $16.76^{+19.33}_{-9.97}{}^{+19.91}_{-12.21}{}^{+0.05}_{-0.05}$
          & $\overline B{}^0\to \Xi^{0}\overline{\Lambda}$%
          & $1.68^{+4.35}_{-1.46}{}^{+2.06}_{-1.22}\pm0$
           \\ 
$B^-\to \Xi^{-}\overline{\Lambda}$%
          & $2.04^{+4.55}_{-1.68}{}^{+2.43}_{-1.49}{}^{+0.01}_{-0.01}$
          & $\overline B{}^0\to  \Xi^{-}\overline{\Sigma^{-}}$%
          & $31.00^{+35.75}_{-18.44}{}^{+36.83}_{-22.59}\pm0$
           \\ 
$B^-\to \Lambda\overline{p}$%
          & $18.78^{+25.16}_{-12.11}{}^{+22.80}_{-13.82}{}^{+0.07}_{-0.07}$
          & $\overline B{}^0\to \Lambda\overline{n}$%
          & $16.68^{+23.54}_{-11.06}{}^{+20.48}_{-12.26}\pm0$
           \\ 
           & ($<32$)~\cite{Tsai:2007pp}
          \\                    
$\overline B{}^0_s\to p\overline{p}$%
          & $0\pm0\pm0{}^{+0.006}_{-0}$
          & $\overline B{}^0_s\to \Sigma^{+}\overline{\Sigma^{+}}$%
          & $1.14^{+1.75}_{-0.78}{}^{+1.40}_{-0.83}{}^{+0.16}_{-0.15}$
           \\
           & ($2.84{}^{+2.03+0.85}_{-1.68-0.18}$)~\cite{Aaij:2013fta}
          \\
$\overline B{}^0_s\to n\overline{n}$%
          & $0\pm0\pm0{}^{+0.005}_{-0}$
          & $\overline B{}^0_s\to \Sigma^{0}\overline{\Sigma^{0}}$%
          & $1.20^{+1.51}_{-0.75}{}^{+1.47}_{-0.88}{}^{+0.16}_{-0.15}$
           \\
$\overline B{}^0_s\to \Xi^{0}\overline{\Xi^{0}}$%
          & $18.23^{+29.78}_{-13.02}{}^{+22.29}_{-13.43}{}^{+0.55}_{-0.55}$
          & $\overline B{}^0_s\to \Sigma^{-}\overline{\Sigma^{-}}$%
          & $1.29^{+1.46}_{-0.76}{}^{+1.53}_{-0.94}{}^{+0.15}_{-0.14}$
           \\ 
$\overline B{}^0_s\to \Xi^{-}\overline{\Xi^{-}}$%
          & $19.55^{+29.91}_{-13.51}{}^{+23.22}_{-14.24}{}^{+0.56}_{-0.55}$
          & $\overline B{}^0_s\to \Sigma^{0}\overline{\Lambda}$%
          & $0.05^{+0.13}_{-0.04}{}^{+0.00}_{-0.00}{}^{+0.001}_{-0.001}$
           \\    
$\overline B{}^0_s\to \Lambda\overline{\Lambda}$%
          & $11.10^{+15.01}_{-7.20}{}^{+13.58}_{-8.18}{}^{+0.46}_{-0.45}$
          & $\overline B{}^0_s\to \Lambda\overline{\Sigma^{0}}$%
          & $0.05^{+0.07}_{-0.03}{}^{+0.00}_{-0.00}{}^{+0.001}_{-0.001}$
           \\                                                                                                                                                             
\end{tabular}
\end{ruledtabular}
\end{table}

Predictions on $\Delta S=-1$, $\overline B_q\to \BB$ decay rates are shown in Table \ref{tab:BBDS=-1}. 
The results can be summarized as following. (i) $B^-\to \Lambda\bar p$, $\Xi^-\overline{\Lambda}$ and $\overline B{}^0_s\to \Lambda\overline{\Lambda}$, $\Xi^-\overline{\Xi^-}$ decays are unsuppressed modes having all charge final states in cascade decays. (ii) In fact, since $B^-\to \Lambda\bar p$ and $\overline B{}^0_s\to \Lambda\overline{\Lambda}$ decays having rates at $10^{-7}$ level and do not lost much in producing $p\bar p\pi^-$ (reduced 26\%) and $p\bar p\pi^+\pi^-$ (reduced 60\%) final states, respectively, they are interesting modes to search for. Indeed the predicted $B^-\to\Lambda\overline p$ rate is close to the present experimental upper limit. It could be the second $\overline B\to\BB$ mode to be observed. (iii) Although $B^-\to \Xi^-\overline{\Sigma^0}$ has rate of the order of $10^{-7}$, it needs $\gamma$ for detection.

We now comment on the $\overline B{}^0_s\to p\bar p$ mode.
The predicted rate is several order smaller than the present experimental result, which, however, has large uncertainty.
To accommodate the central value of the experimental result on $B_s\to p\bar p$ rate, 
one need to scale $|\eta|$ from 1 [see Eq.~(\ref{eq:correction3})] up to 20.54. 
Although it is unlikely that for $|\eta|$ to be enhanced by factor 20, some enhancement is possible if final state rescattering is present~\cite{FSI}.
Note that the last entries of rates for modes with vanishing central values in Tables~\ref{tab:BBDS=0} and \ref{tab:BBDS=-1}, scale with $|\eta|^2$, while those with non-vanishing central values, roughly scale with $|\eta|$. By naively scaling up $|\eta|$ by a factor of 20.54, we find that the contribution of the ``subleading terms" (term with $\eta$) will give rate five time of the tree contribution in $\overline B{}^0\to p\bar p$ rate.~\footnote{The last uncertainty in the $\overline B{}^0\to p\bar p$ rate changes from $\pm 0.15$ to ${}^{+5.03}_{-1.47}$.} This is highly un-nature and unlikely.
We certainly need more data to clarify the situation.

We give a summary of our suggestions before ending this section.
We shall concentrate on modes that will cascadely decay to all charged final states and have large decay rates.
(i) For $\overline B_q\to\DD$, $\Delta S=0$ decays, we have
$\overline{B}{}^0\to \Delta^0\overline{\Delta^0}$ 
and 
$\overline{B}{}^0\to \Sigma^{*-}\overline{\Sigma^{*-}}$ having 
rates at $10^{-8}$ level. 
(ii) For $\Delta S=-1$, $\overline B_q\to\DD$ decays,
$B^-\to\Sigma^{*+}\overline {\Delta^{++}}$, $\Sigma^{*-}\overline{\Delta^0}$, $\Omega^-\overline{\Xi^{*0}}$ and $\overline B{}^0_s\to\Omega^-\overline{\Omega^-}$, $\Xi^{*0}\overline{\Xi^{*0}}$, $\Sigma^{*0}\overline{\Sigma^{*-}}$, $\Sigma^{*+}\overline{\Sigma^{*+}}$ decays have rates ranging from $10^{-8}$ to $10^{-7}$, where the
$\overline B {}^0_s\to\Omega^-\overline{\Omega^-}$ decay has the largest rate. 
(iii) For $\Delta S=0$, $\overline B_q\to\BD$ decays, the central value of the $B^-\to p\overline{\Delta^{++}}$ rate is only half of the experimental upper bound and should be searchable in the near future,  
while another all charge final state $\overline B{}^0_s\to p\overline{\Sigma^{*+}}$ has rate at $10^{-8}$ order.
(iv) For $\Delta S=-1$, $\overline B_q\to\BD$ decays,
$\overline B{}^0\to \Xi^-\overline{\Sigma^{*-}}$ decay rate is at $10^{-8}$ order.
(v) For $\Delta S=0$, $\overline B_q\to \DB$ decays,
$B^-\to \Delta^0\overline p$ and $\overline B{}^0_s\to\Delta^0\overline\Lambda$ decays have rates of order $10^{-8}$. 
(vi) For $\Delta S=-1$, $\overline B_q\to \DB$ decays,
$\overline B{}^0\to \Sigma^{*-}\overline p$, $\Omega^-\overline{\Xi^-}$ and $\Xi^{*0}\overline\Lambda$ rates are at the order of $10^{-8}$.
(vii) For $\Delta S=0$, $\overline B_q\to \BB$ decays, $\overline B{}^0\to p\bar p$ is the most accessible mode. It is not surprise that it is the first $\overline B_q\to \BB$ mode being observed. 
(viii) For $\Delta S=-1$, $\overline B_q\to \BB$ decays, $B^-\to \Lambda\bar p$ and $\overline B{}^0_s\to \Lambda\overline{\Lambda}$ have rates at $10^{-7}$ level and do not lost much in producing $p\bar p\pi^-$ and $p\bar p\pi^+\pi^-$ final states, respectively. They are interesting modes to search for. The $B^-\to\Lambda\overline p$ decay could be the second $\overline B\to\BB$ mode to be observed as its rate is close to the present experimental upper limit.
(ix)The predicted $\overline B{}^0_s\to p\bar p$ rate is several order smaller than the present experimental result.
The central value of the experimental result can be reproduced only with a unnaturally scale up $|\eta|$.
By naively scaling up $|\eta|$, we find that the contribution of the ``subleading terms" (term with $\eta$) will give rate five time of the tree contribution in $\overline B{}^0\to p\bar p$ rate. We need more data to clarify the situation.

\section{Conclusion}

In this work, we study charmless two-body baryonic $B_{u,d,s}$ decays using the topological amplitude approach.
We extend previous work~\cite{Chua:2003it} to include all ground state octet and decuplet final states with full topological amplitudes. Relations on rates and CP asymmetries are obtained using these amplitudes.

There are in general more than one tree and one penguin amplitudes in
the baryonic decays. However, by considering the chirality nature of weak
interaction and asymptotic relations~\cite{Brodsky:1980sx}, the 
number of independent amplitudes is significantly reduced~\cite{Chua:2003it}.

With the long awaited $\overline B{}^0\to p\bar p$ data~\cite{Aaij:2013fta}, we can finally extract information on the topological amplitudes. Using ratio of the Wilson coefficients, we estimate the penguin to tree amplitude ratio and be able to predict rates of all other modes in the asymptotic limit.
Corrections to the amplitudes by relaxing the asymptotic relations and including sub-leading contributions are estimated. The predicted rates on decay rates are in general with uncertainties of a factor of two.
 

We point out some modes that will cascadely decay to all charged final states and have large decay rates.
(i) For $\overline B_q\to\DD$, $\Delta S=0$ decays, we have
$\overline{B}{}^0\to \Delta^0\overline{\Delta^0}$ 
and 
$\overline{B}{}^0\to \Sigma^{*-}\overline{\Sigma^{*-}}$ having 
rates at $10^{-8}$ level. 
(ii) For $\Delta S=-1$, $\overline B_q\to\DD$ decays,
$B^-\to\Sigma^{*+}\overline {\Delta^{++}}$, $\Sigma^{*-}\overline{\Delta^0}$, $\Omega^-\overline{\Xi^{*0}}$ and $\overline B{}^0_s\to\Omega^-\overline{\Omega^-}$, $\Xi^{*0}\overline{\Xi^{*0}}$, $\Sigma^{*0}\overline{\Sigma^{*-}}$, $\Sigma^{*+}\overline{\Sigma^{*+}}$ decays have rates ranging from $10^{-8}$ to $10^{-7}$, where
the $\overline B {}^0_s\to\Omega^-\overline{\Omega^-}$ decay has the largest rate. 
(iii) For $\Delta S=0$, $\overline B_q\to\BD$ decays, $\overline B{}^0_s\to p\overline{\Sigma^{*+}}$ has rate at $10^{-8}$ order, while the predicted $B^-\to p\overline{\Delta^{++}}$ rate is close to the experimental upper bound and should be searchable in the near future.
(iv) For $\Delta S=-1$, $\overline B_q\to\BD$ decays,
$\overline B{}^0\to \Xi^-\overline{\Sigma^{*-}}$ decay rate is at $10^{-8}$ order.
(v) For $\Delta S=0$, $\overline B_q\to \DB$ decays,
$B^-\to \Delta^0\overline p$ and $\overline B{}^0_s\to\Delta^0\overline\Lambda$ decays have rates of order $10^{-8}$. 
(vi) For $\Delta S=-1$, $\overline B_q\to \DB$ decays,
$\overline B{}^0\to \Sigma^{*-}\overline p$, $\Omega^-\overline{\Xi^-}$ and $\Xi^{*0}\overline\Lambda$ rates are at the order of $10^{-8}$.
(vii) For $\Delta S=0$, $\overline B_q\to \BB$ decays, $\overline B{}^0\to p\bar p$ is the most accessible mode. It is not surprise that it is the first $\overline B_q\to \BB$ mode being found. 
(viii) For $\Delta S=-1$, $\overline B_q\to \BB$ decays, $B^-\to \Lambda\bar p$ and $\overline B{}^0_s\to \Lambda\overline{\Lambda}$ have rates at $10^{-7}$ level and do not lost much in cascade decays.
They are interesting modes to search for. In fact, the $B^-\to\Lambda\overline p$ decay could be the second $\overline B\to\BB$ mode to be observed as its rate is close to the present experimental upper limit.

With the detection of $\pi^0$ and/or $\gamma$ many other unsuppressed modes can be searched for.

The predicted $\overline B{}^0_s\to p\bar p$ rate is several order smaller than the present experimental result.
The central value of the experimental result can be reproduced only with a unnaturally scaled up $|\eta|$.
By naively scaling up $|\eta|$, we find that the contribution of the ``subleading terms" (term with $\eta$) will give rate five time of the tree contribution in $\overline B{}^0\to p\bar p$ rate. We need more data to clarify the situation.

The analysis presented in this work can be systematically improved when more measurements on decay rates become available.

%
%

\begin{acknowledgments}
The author thanks Paoti Chang and Eduardo Rodrigues for discussions. This work is supported in part by
the National Science Council of R.O.C. under Grant
NSC100-2112-M-033-001-MY3 and the National Center for Theoretical Sciences.
\end{acknowledgments}

\appendix

\section{Asymptotic relations in the large $m_B$ limit}\label{append:A}

In this appendix we summarize the main procedures to obtain asymptotic relations as in Ref.~\cite{Chua:2002yd} and further extend it to include discussion on electroweak penguins.
In general the decay amplitudes of $\overline B$ to final states with octet
baryon (${\cal B}$) and decuplet baryons (${\cal D}$) can be
expressed as~\cite{Jarfi:1990ej}
\begin{eqnarray}
A(\overline B\to {\cal B}_1 \overline {\cal B}_2)&=&\bar
u_1(A_{{\cal B}\overline {\cal B}}+\gamma_5 B_{{\cal B}\overline
{\cal B}}) v_2,
\nonumber\\
A(\overline B\to {\cal D}_1 \overline {\cal B}_2)&=&i
\frac{q^\mu}{m_B} \bar u^\mu_1(A_{{\cal D}\overline {\cal
B}}+\gamma_5 B_{{\cal D}\overline {\cal B}}) v_2,
\nonumber\\
A(\overline B\to {\cal B}_1 \overline {\cal D}_2)&=&i
\frac{q^\mu}{m_B}\bar u_1(A_{{\cal B}\overline {\cal D}}+\gamma_5
B_{{\cal B}\overline {\cal D}}) v^\mu_2,
\nonumber\\
A(\overline B\to {\cal D}_1 \overline {\cal D}_2)&=&\bar
u^\mu_1(A_{{\cal D}\overline {\cal D}}+\gamma_5 B_{{\cal
D}\overline {\cal D}}) v_{2\mu}+\frac{q^\mu q^\nu}{m^2_B}\bar
u^\mu_1(C_{{\cal D}\overline {\cal D}}+\gamma_5 D_{{\cal
D}\overline {\cal D}}) v_{2\nu},
\end{eqnarray}
where $q=p_1-p_2$ and $u^\mu,\,v^\mu$ are the Rarita-Schwinger
vector spinors for a spin-$\frac{3}{2}$ particle. The vector
spinor can be expressed as~\cite{Moroi:1995fs}
$u_\mu(\pm\frac{3}{2})=\epsilon_\mu(\pm1) u(\pm\frac{1}{2})$ and
$u_\mu(\pm\frac{1}{2})=(\epsilon_\mu(\pm1)
u(\mp\frac{1}{2})+\sqrt{2}\,\epsilon_\mu(0)
u(\pm\frac{1}{2}))/\sqrt3$, where $\epsilon_\mu(\lambda)$ and
$u(s)$ are the usual polarization vector and spinor, respectively.
By using
$q\cdot\epsilon(\lambda)_{1,2}=\mp\,\delta_{\lambda,0}\,m_B
p_c/m_{1,2}$, where $p_c$ is the baryon momentum in the $B$ rest
frame and the fact that
$\epsilon^*_1(0)\cdot\epsilon_2(0)=(m_B^2-m^2_1-m^2_2)/2m_1 m_2$
is the largest product among $\epsilon^*_1(\lambda_1)\cdot
\epsilon_2(\lambda_2)$, we have
\begin{eqnarray}
A(\overline B\to {\cal D}_1 \overline {\cal B}_2)&=&-i
\sqrt{\frac{2}{3}}\frac{p_c}{m_1} \bar u_1(A_{{\cal D}\overline
{\cal B}}+\gamma_5 B_{{\cal D}\overline {\cal B}}) v_2,
\nonumber\\
A(\overline B\to {\cal B}_1 \overline {\cal D}_2)&=&i
\sqrt{\frac{2}{3}}\frac{p_c}{m_2}\bar u_1(A_{{\cal B}\overline
{\cal D}}+\gamma_5 B_{{\cal B}\overline {\cal D}}) v_2,
\nonumber\\
A(\overline B\to {\cal D}_1 \overline {\cal
D}_2)&\simeq&\frac{m_B^2}{3m_1m_2}\bar u_1(A^\prime_{{\cal
D}\overline {\cal D}}+\gamma_5 B^\prime_{{\cal D}\overline {\cal
D}})v_2, \label{eq:largemB}
\end{eqnarray}
where $A^\prime_{{\cal D}\overline {\cal D}}=A_{{\cal D}\overline
{\cal D}}-2(p_c/m_B)^2 C_{{\cal D}\overline {\cal D}}$ and
$B^\prime_{{\cal D}\overline {\cal D}}=B_{{\cal D}\overline {\cal
D}}-2(p_c/m_B)^2 D_{{\cal D}\overline {\cal D}}$ and decuplets can
only in $\pm\frac{1}{2}$-helicity states. 
%
%
All four $\overline B\to {\mathbf B}_1\overline {\mathbf B}_2$
(${\mathbf B}\overline {\mathbf B}={\cal B} \overline {\cal B}$,
${\cal D} \overline {\cal B}$, ${\cal B} \overline {\cal D}$,
${\cal D} \overline {\cal D}$) decays can be effectively expressed as
\begin{equation}
A(\overline B\to {\mathbf B}_1\overline {\mathbf B}_2)
 =\bar u_1(A+\gamma_5 B)v_2.
 \label{eq:asymptoticform}
\end{equation}
%
%
%
The chiral structure of weak interaction provide further
information on $A$ and $B$. For example, in the $\Delta S=0$
processes, we have either $b\to u_L\bar u_R d_L$ or $b\to
q_{L(R)}\bar q_{R(L)} d_L$ decays, therefore the produced $d_L$
quark is left-handed. Furthermore, as strong interaction is
chirality conserving, the pop up quark pair $q^\prime\bar
q^\prime$ should have $q^\prime_{L(R)}\bar q^\prime_{R(L)}$. From
the conservation of helicity, the produced baryon must be in a
left-helicity state and the produced anti-baryon must be in a
right-helicity state. In large $m_B$ limit, as the spinor helicity
identify to chirality, we should have $B\to -A$ in above
equations. 

We follow Ref.~\cite{Brodsky:1980sx,Chua:2002yd} to obtain the
asymptotic relations these coefficients ($A$ and $B$). As noted we
only need to consider helicity $\pm\frac{1}{2}$ states. The wave
function of a right-handed (helicity$=\frac{1}{2}$) baryon can be
expressed as
\begin{equation}
|{\mathbf B}\,;\uparrow\rangle\sim \frac{1}{\sqrt3}(|{\mathbf
B}\,;\uparrow\downarrow\uparrow\rangle
                +|{\mathbf B}\,;\uparrow\uparrow\downarrow\rangle
                +|{\mathbf B}\,;\downarrow\uparrow\uparrow\rangle),
\end{equation}
i.e. composed of 13-, 12- and 23-symmetric terms, respectively.
For ${\mathbf B}=p,\,n,\,\Sigma^0,\,\Lambda$, we have
\begin{eqnarray}
|\Delta^{++};\uparrow\downarrow\uparrow\rangle&=&u(1)u(2)u(3)|\uparrow\downarrow\uparrow\rangle,\qquad\qquad
|\Delta^{-};\uparrow\downarrow\uparrow\rangle=d(1)d(2)d(3)|\uparrow\downarrow\uparrow\rangle,
\nonumber\\
|\Delta^{+};\uparrow\downarrow\uparrow\rangle&=&
\frac{1}{\sqrt3}[u(1)u(2)d(3)+u(1)d(2)u(3)+d(1)u(2)u(3)]|\uparrow\downarrow\uparrow\rangle,
\nonumber\\
|\Delta^{0};\uparrow\downarrow\uparrow\rangle&=&(|\Delta^{+};\uparrow\downarrow\uparrow\rangle\,\,{\rm
with}\,\,u \leftrightarrow d),\qquad
|\Sigma^{*+};\uparrow\downarrow\uparrow\rangle=(|\Delta^{+};\uparrow\downarrow\uparrow\rangle\,\,{\rm
with}\,\,d \leftrightarrow s),
\nonumber\\
|\Sigma^{*0};\uparrow\downarrow\uparrow\rangle&=&\frac{1}{\sqrt6}[u(1)d(2)s(3)+{\rm
permutation}]|\uparrow\downarrow\uparrow\rangle,
\nonumber\\
 |p\,;\uparrow\downarrow\uparrow\rangle&=&
\left[\frac{d(1)u(3)+u(1)d(3)}{\sqrt6} u(2)
 -\sqrt{\frac{2}{3}} u(1)d(2)u(3)\right]
|\uparrow\downarrow\uparrow\rangle,
\nonumber\\
|n\,;\uparrow\downarrow\uparrow\rangle&=&
(-|p\,;\uparrow\downarrow\uparrow\rangle \,\,{\rm with}\,\,u
\leftrightarrow d),
\nonumber\\
|\Sigma^0\,;\uparrow\downarrow\uparrow\rangle&=&
\bigg[-\frac{u(1)d(3)+d(1)u(3)}{\sqrt3}\,s(2)
      +\frac{u(2)d(3)+d(2)u(3)}{2\sqrt3}\,s(1)
\nonumber\\
      &&\,\,+\frac{u(1)d(2)+d(1)u(2)}{2\sqrt3}\,s(3)\bigg]
|\uparrow\downarrow\uparrow\rangle,
\nonumber\\
|\Lambda\,;\uparrow\downarrow\uparrow\rangle&=&
\bigg[\frac{d(2)u(3)-u(2)d(3)}{2}\,s(1)
      +\frac{u(1)d(2)-d(1)u(2)}{2}\,s(3)\bigg]
|\uparrow\downarrow\uparrow\rangle, \label{eq:wavefunction}
\end{eqnarray}
for the corresponding $|{\mathbf
B}\,;\uparrow\downarrow\uparrow\rangle$ parts, while the 12- and
23-symmetric parts can be obtained by permutation. 

Following Ref.~\cite{Brodsky:1980sx} and using the above helicity
argument, asymptotically we have
\begin{eqnarray}
\langle {\mathbf B}(p_1)| {\cal O}|\overline B\,{\mathbf
B}^\prime(p_2)\rangle
&=&\bar u(p_1)\left[
                   \frac{1-\gamma_5}{2}\,F(t)\right] u(p_2),
\nonumber\\
F(t)&=&\sum_{i=T,P_L,P_R} e_{i}
               ({\mathbf B}^\prime -B-{\mathbf B})
               \,F_{i}(t),
\end{eqnarray}
where ${\cal O}$ are the operators in $H_{\rm eff}$. For
simplicity, we illustrate with the space-like case. Note that the
above equation is obtained in the large $t(=(p_1-p_2)^2)$ limit,
where we may take a large $m_B$ limit. 
Quark mass dependent terms are sub-leading and are neglected.
%
\begin{figure}[t]
\centering
 \subfigure[]{
  \includegraphics[width=0.5\textwidth]{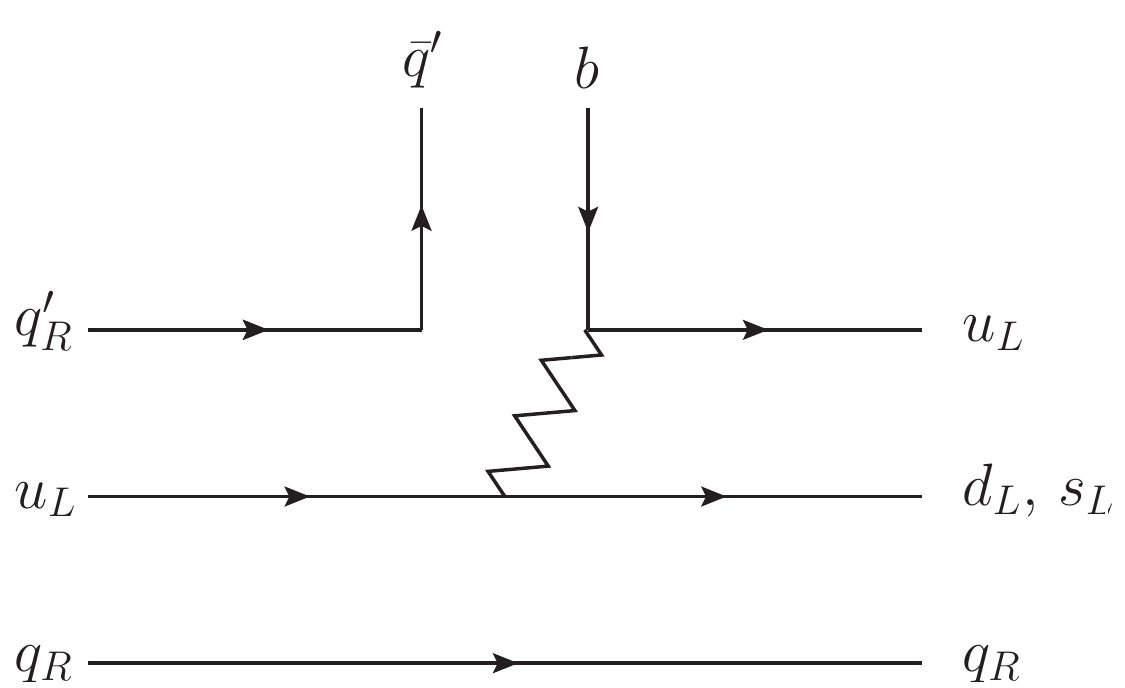}
}\subfigure[]{
  \includegraphics[width=0.5\textwidth]{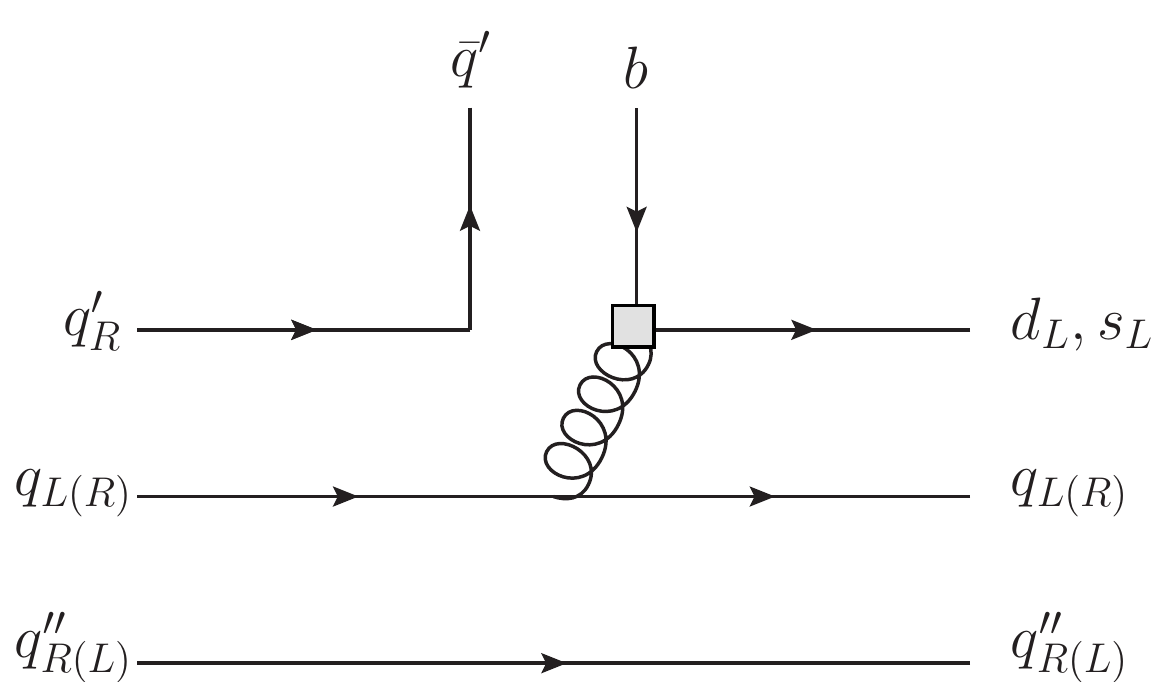}
}\\\subfigure[]{
  \includegraphics[width=0.5\textwidth]{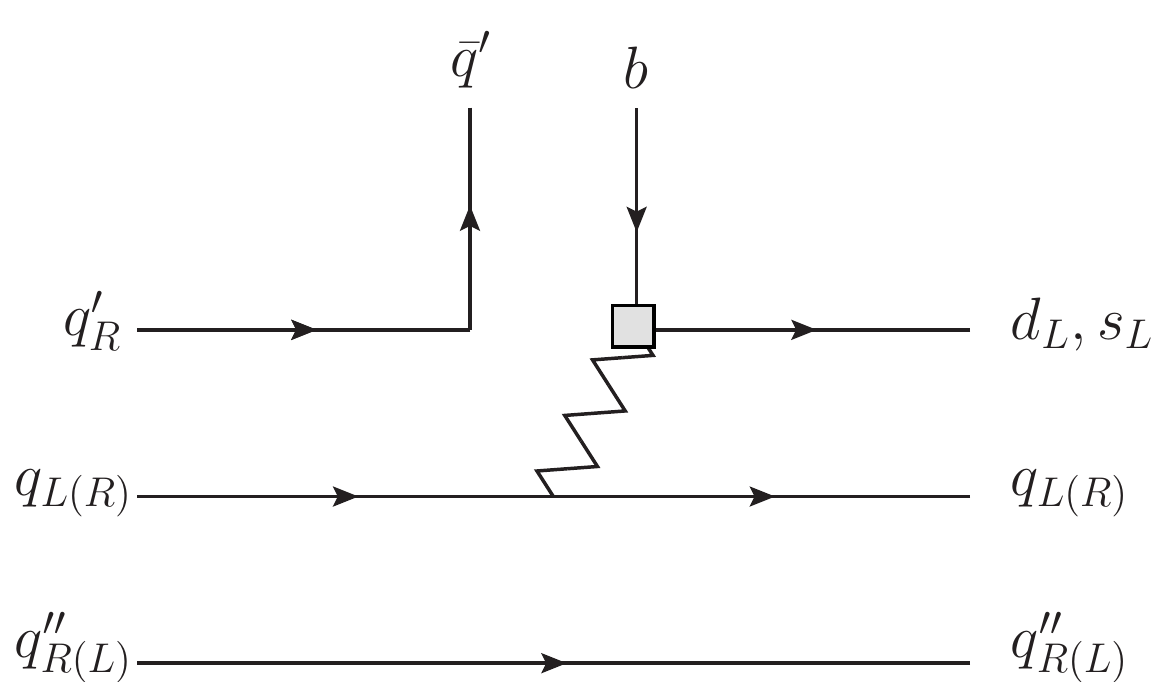}
}
\caption{(a) Tree, (b) penguin and (c) electroweak penguin
  ${\mathbf B}^\prime -\overline B-{\mathbf
B}$ diagrams in the asymptotic limit. } 
\label{fig:eTePePEW}
\end{figure}

As shown in Fig.~\ref{fig:eTePePEW}(a) the ${\mathbf
B}^\prime(q^\prime_R\, u_L\, q_R)$--$\overline B(\overline
q^\prime_L\,b )$--${\mathbf B}(u_L \,d_L\, q_R)$ coupling is
governed by the the tree operator $(\bar u b)_{V-A} (\bar d
u)_{V-A}$. The corresponding coefficient $e_{T}({\mathbf B}^\prime
-\overline B-{\mathbf B})$ is given by
\begin{eqnarray}
e_{T}({\mathbf B}^\prime -\overline B-{\mathbf B})
 &=&\langle{\mathbf B};\,\downarrow\downarrow\uparrow|
     {\cal Q}[q_R^\prime(1)\to u_L(1);u_L(2)\to d_L(2)]
    |{\mathbf B}^\prime\,;\uparrow\downarrow\uparrow\rangle
\nonumber\\
&&+\langle {\mathbf B};\,\uparrow\downarrow\downarrow|
     {\cal Q}[q_R^\prime(3)\to u_L(3);u_L(2)\to d_L(2)]
   |{\mathbf B}^\prime\,;\uparrow\downarrow\uparrow\rangle,
\label{eq:eT}
\end{eqnarray}
where ${\cal Q}[q^\prime_R(1(3))\to u_L(1,3);u_L(2)\to d_L(2)]$ changes
the parallel spin $q^\prime(1(3))|\uparrow\rangle \otimes
u(2)|\downarrow\rangle$ part of $|{\mathbf
B}^\prime;\uparrow\downarrow\uparrow\rangle$ to the
$u(1(3))|\downarrow\rangle\otimes d(2)|\downarrow\rangle$ part.

\begin{table}[t!]
\caption{\label{tab:eTePePEW} The coefficients $e_{T,P}({\mathbf
B}^\prime -\overline B-{\mathbf B})$ for various modes obtained
from Eqs.~(\ref{eq:eT}), (\ref{eq:eP}) and (\ref{eq:ePEW}).}
\begin{ruledtabular}
\begin{tabular}{cccccccc}
${\mathbf B}^\prime$-$\overline B$-${\mathbf B}$
          & $e_T$
          & $e_P$
          & $e_{EWP}$
          & ${\mathbf B}^\prime$-$\overline B$-${\mathbf B}$
          & $e_T$
          & $e_P$
          & $e_{EWP}$
          \\
\hline $\Sigma^{*0}$-$\overline B {}^0$-$\Sigma^{*0}$
          & $1/3$
          & $2/3$
          & $1/9$
          & $\Delta^0$-$B^-$-$\Delta^-$
          & $0$
          & $2/\sqrt3$
          & $-2/3\sqrt3$
          \\
$\Delta^+$-$B^-$-$\Delta^0$
          & $2/3$
          & $4/3$
          & $2/9$
          & $\Delta^{+}$-$B^-$-$\Sigma^{*0}$
          & $\sqrt2/3$
          & $2\sqrt2/3$
          & $\sq2/9$
          \\
\hline $\Delta^{++}$-$B^-$-$p$
          & $\sqrt{2/3}$
          & $\sqrt{2/3}$
          & $2\sq2/3\sq3$
          & $\Sigma^{*0}$-$\overline B {}^0$-$\Lambda$
          & $0$
          & $-1/\sqrt6$
          & $1/3\sq6$
          \\
$\Delta^+$-$\overline B {}^0$-$p$
          & $\sqrt2/3$
          & $\sqrt2/3$
          & $2\sq2/9$
          & $\Delta^{+}$-$B^-$-$n$
          & $-\sqrt2/3$
          & $\sqrt2/3$
          & $-4\sq2/9$
          \\
$\Sigma^{*+}$-$\overline B {}^0_s$-$p$
          & $\sqrt2/3$
          & $\sqrt2/3$
          & $2\sq2/9$
          & $\Delta^{+}$-$B^-$-$\Sigma^{0}$
          & $1/3$
          & $2/3$
          & $1/9$
          \\
\hline $p$-$B^-$-$\Delta^0$
          & $\sqrt2/3$
          & $-\sqrt2/3$
          & $4\sq2/9$
          & $p$-$\overline B {}^0$-$\Delta^+$
          & $\sqrt2/3$
          & $\sqrt2/3$
          & $2\sq2/9$
          \\
$n$-$B^-$-$\Delta^-$
          & $0$
          & $-\sqrt{2/3}$
          & $\sq2/3\sq3$
          & $\Lambda$-$\overline B {}^0$-$\Sigma^{*0}$
          & $-1/\sqrt6$
          & $-1/\sqrt6$
          & $-\sq2/3\sq3$
          \\
$p$-$B^-$-$\Sigma^{*0}$
          & $1/3$
          & $-1/3$
          & $4/9$
          & $n$-$\overline B {}^0$-$\Sigma^{*0}$
          & $2/3$
          & $1/3$
          & $5/9$
          \\
\hline $p$-$\overline B {}^0$-$p$
          & $1/3$
          & $1/3$
          & $2/9$
          & $p$-$B^-$- $n$
          & $-1/3$
          & $-5/3$
          & $2/9$
          \\
$n$-$\overline B {}^0$-$n$
          & $-2/3$
          & $-4/3$
          & $-2/9$
          & $\Lambda$-$\overline B {}^0$-$\Sigma^0$
          & $1/\sqrt3$
          & $1/\sqrt3$
          & $2/3\sq3$
          \\
$\Sigma^0$-$\overline B {}^0$-$\Sigma^0$
          & $-1/3$
          & $-2/3$
          & $-1/9$
          & $\Sigma^0$-$\overline B {}^0$-$\Lambda$
          & $0$
          & $1/\sqrt3$
          & $-1/3\sq3$
          \\
$\Lambda$-$\overline B {}^0$-$\Lambda$
          & $0$
          & $0$
          & $0$
          & $n$-$\overline B {}^0$-$\Lambda$
          & $\sqrt{2/3}$
          & $\sqrt{3/2}$
          & $1/\sq6$
          \\
$p$-$B^-$-$\Sigma^0$
          & $1/3\sqrt2$
          & $-1/3\sqrt2$
          & $2\sq2/9$
          & $p$-$B^-$-$\Lambda$
          & $1/\sqrt6$
          & $\sqrt{3/2}$
          & $0$
          \\
\end{tabular}
\end{ruledtabular}
\end{table}

Similarly coefficients $e_{P_L,P_R}({\mathbf B}^\prime -\overline
B-{\mathbf B})$ for the ${\mathbf B}^\prime(q^\prime_R\, q_L\,
q^{\prime\prime}_R)$--$\overline B(\overline
q^\prime_L\,b)$--${\mathbf B}(d_L\, q_L\, q^{\prime\prime}_R)$ and
${\mathbf B}^\prime(q^\prime_R\, q_R
\,q^{\prime\prime}_L)$--$\overline B(\overline
q^\prime_L\,b)$--${\mathbf B}(d_L \,q_R\, q^{\prime\prime}_L)$
couplings governed respectively by the penguin operators $(\bar d
b)_{V-A} (\bar q q)_{V\mp A}$ are given by
\begin{eqnarray}
e_{P_L}({\mathbf B}^\prime -\overline B-{\mathbf B})
 &=&\langle{\mathbf B};\,\downarrow\downarrow\uparrow|
     {\cal Q}[q_R^\prime(1)\to d_L(1);q_L(2)\to q_L(2)]
    |{\mathbf B}^\prime\,;\uparrow\downarrow\uparrow\rangle
\nonumber\\
&&+\langle {\mathbf B};\,\uparrow\downarrow\downarrow|
     {\cal Q}[q_R^\prime(3)\to d_L(3);q_L(2)\to q_L(2)]
   |{\mathbf B}^\prime\,;\uparrow\downarrow\uparrow\rangle,
\nonumber\\
e_{P}({\mathbf B}^\prime -\overline B-{\mathbf B})
 &\equiv& e_{P_L}({\mathbf B}^\prime -\overline B-{\mathbf B})
 =e_{P_R}({\mathbf B}^\prime -\overline B-{\mathbf B}).
 \label{eq:eP}
\end{eqnarray}
The corresponding diagram is shown in Fig.~\ref{fig:eTePePEW}(b). Note
that that $e_{P_L}$ is similar to $e_T$ with the
$q^\prime_R(1,3)\to u_L(1,3)$ and $u_L(2)\to d_L(2)$ operations
replaced by the $q^\prime_R(1,3)\to d_L(1,3)$ and $q_L(2)\to
q_L(2)$ operations, respectively. The equality of $e_{P_R}$ and
$e_{P_L}$ can be understood by interchanging $q\leftrightarrow
q^{\prime\prime}$ in ${\mathbf B}^\prime(q^\prime_R\,q_L\,
q^{\prime\prime}_R)$--$\overline B(\overline
q^\prime_L\,b)$--${\mathbf B}(d_L\, q_L\,q^{\prime\prime}_R)$ and
${\mathbf
B}^\prime(q^\prime_R\,q_R\,q^{\prime\prime}_L)$--$\overline
B(\overline q^\prime_L\,b)$--${\mathbf
B}(d_L\,q_R\,q^{\prime\prime}_L)$. The coefficients for the
$|\Delta S|=1$ case can be obtained by the suitable replacement of
$d_L\to s_L$ in the $\mathbf B$ content in Eqs.~(\ref{eq:eT},
\ref{eq:eP}).
Similarly for electroweak penguin, we have
\begin{eqnarray}
e_{EWP_L}({\mathbf B}^\prime -\overline B-{\mathbf B})
 &=&Q(q)[\langle{\mathbf B};\,\downarrow\downarrow\uparrow|
     {\cal Q}[q_R^\prime(1)\to d_L(1);q_L(2)\to q_L(2)]
    |{\mathbf B}^\prime\,;\uparrow\downarrow\uparrow\rangle
\nonumber\\
&&+\langle {\mathbf B};\,\uparrow\downarrow\downarrow|
     {\cal Q}[q_R^\prime(3)\to d_L(3);q_L(2)\to q_L(2)]
   |{\mathbf B}^\prime\,;\uparrow\downarrow\uparrow\rangle],
   \nonumber\\
e_{EWP}({\mathbf B}^\prime -\overline B-{\mathbf B})
 &\equiv& e_{EWP_L}({\mathbf B}^\prime -\overline B-{\mathbf B})
 =e_{EWP_R}({\mathbf B}^\prime -\overline B-{\mathbf B}).
 \label{eq:ePEW}
\end{eqnarray}
where $Q(q)$ is the electric charge of quark $q$. 
Note that we do not include factor $3/2$ 
in the above formulas.
The corresponding diagram is shown in Fig.~\ref{fig:eTePePEW}(c).

By using the above equations, it is straightforward to obtain the
coefficients of various modes as shown in Table~\ref{tab:eTePePEW}. 
Comparing these results to the decay amplitudes in terms of topological amplitudes, we obtain the asymptotic amplitudes shown in Eqs. (\ref{eq:asymptoticrelations}) and (\ref{eq:asymptotic1}).



\section{Independent amplitudes}\label{append:B}

The number of independent amplitudes are in general less then the one of topological amplitudes. 
In this appendix we express decay amplitudes in terms of independent amplitudes. 
Although the physical interpretations and size estimations of these independent amplitudes are not as clear as the topological amplitudes, they are useful in finding relations of decay amplitudes, where some examples are given in Sec. III.A. Readers can use the following expressions to work out additional relations.

For $\overline B_q\to\DD$ decays, we have
\be
&&\sq2A(B^-\to\Delta^-\overline{\Delta^{0}})
=\sq3 A(B^-\to\Sigma^{*-}\overline{\Sigma^{*0}})
=\sq6 A(B^-\to\Xi^{*-}\overline{\Xi^{*0}})
\equiv\sq6 A_A,
\non\\
&&A(B^-\to\Delta^0\overline{\Delta^{+}})
=\sq2 A(B^-\to\Sigma^{*0}\overline{\Sigma^{*+}})
=A_T-A_P+2A_A,
\non\\
&&
A(B^-\to\Delta^+\overline{\Delta^{++}})
=\sq3 (A_T-A_P+A_A),
\en
\be
A(\overline{B^0}\to \Xi^{*0}\overline{\Xi^{*0}})
&\equiv&A_E,
\non\\
A(\overline{B^0}\to\Xi^{*-}\overline{\Xi^{*-}})
&=&A_P+A_{PA},
\non\\
2A(\overline{B^0}\to\Sigma^{*0}\overline{\Sigma^{*0}})
&=&A_T+A_P+2A_E,
\non\\
A(\overline{B^0}\to\Omega^-\overline{\Omega^-})
&\equiv&A_{PA},
\non\\
A(\overline{B^0}\to\Delta^{++}\overline{\Delta^{++}})
&=&3A_E
-2A_{PA},
\non\\
A(\overline{B^0}\to\Sigma^{*+}\overline{\Sigma^{*+}})
&=&2 A_E
-A_{PA},
\non\\
A(\overline{B^0}\to\Delta^{-}\overline{\Delta^{-}})
&=&3A_P
+A_{PA},
\non\\
A(\overline{B^0}\to\Sigma^{*-}\overline{\Sigma^{*-}})
&=&2A_P
+A_{PA},
\non\\
A(\overline{B^0}\to\Delta^+\overline{\Delta^{+}})
&=& A_T+2A_E-A_{PA},
\non\\
A(\overline{B^0}\to\Delta^0\overline{\Delta^{0}})
&=& A_T+A_P+A_E,
\en
and
\be
\sq2A(\overline{B^0_s}\to\Delta^{0}\overline{\Sigma^{*0}})
&=&\sq2A(\overline{B^0_s}\to \Sigma^{*0}\overline{\Xi^{*0}})
=A_T+A_P,
\non\\
2A(\overline{B^0_s}\to\Delta^{-}\overline{\Sigma^{*-}})
&=&2A(\overline{B^0_s}\to\Xi^{*-}\overline{\Omega^-})
=\sq3A(\overline{B^0_s}\to\Sigma^{*-}\overline{\Xi^{*-}})
\equiv2 \sq3 A_P,
\non\\
A(\overline{B^0_s}\to\Delta^+\overline{\Sigma^{*+}})
&\equiv&A_T,
\en
where these $A_{T,P,A,E,PA}$ can be easily read out by comparing the decay amplitudes to those shown in Sec.~\ref{sec:formalism}.B. 
It is important to strees that the labels $T,P,A,E,PA$ of these $A$s are for the purpose of book keeping, they not necessarily correspond to tree, penguin, annihilation, exchange and penguin annihilation amplitudes. 
These remarks are also true for the following discussion.
Note that there are only five independent amplitudes for these modes.


Similarly for $\Delta S=-1$ transition, we have
\be
\sq2 A(B^-\to\Sigma^{*0}\overline{\Delta^+})
&=&
A(B^-\to\Xi^{*0}\overline{\Sigma^{*+}})
=A'_T-A'_P+2A'_A,
   \non\\
\sq6 A(B^-\to\Sigma^{*-}\overline{\Delta^0})
   &=&   
\sq2 A(B^-\to\Omega^-\overline{\Xi^{*0}})
   =\sq3 A(B^-\to\Xi^{*-}\overline{\Sigma^{*0}})
   =\sq6 A'_A,
\non\\
A(B^-\to \Sigma^{*+}\overline{\Delta^{++}})
&=&\sq3(A'_T-A'_P+A'_A),
\en
\be
   \sq2 A(\bar B^0\to\Sigma^{*0}\overline{\Delta^0})
   &=&
   \sq2 A(\bar B^0\to\Xi^{*0}\overline{\Sigma^{*0}})
   =(A'_T+A'_P),
   \non\\
  2 A(\bar B^0\to\Sigma^{*-}\overline{\Delta^-})
   &=&
   2A(\bar B^0\to\Xi^{*-}\overline{\Sigma^{*-}})
   = \sq3 A(\bar B^0\to\Omega^-\overline{\Xi^{*-}})
   =2\sq3 A'_P,
   \non\\
A(\bar B^0\to \Sigma^{*+}\overline{\Delta^+})
   &=&A'_T.
   \en
and   
\be
   A(\bar B^0_s\to\Delta^0\overline{\Delta^0})
   &=&A'_E,
\non\\
   A(\bar B^0_s\to\Delta^-\overline{\Delta^-})
   &=&A'_{PA},
   \non\\
   2A(\bar B^0_s\to\Sigma^{*0}\overline{\Sigma^{*0}})
   &=&A'_T+A'_P+2A'_E,
 \non\\
    A(\bar B^0_s\to\Sigma^{*-}\overline{\Sigma^{*-}})
    &=&A'_P+A'_{PA} ,
   \non\\      
A(\bar B^0_s\to \Delta^{++}\overline{\Delta^{++}})
   &=&3 A'_E-2A'_{PA},
   \non\\
A(\bar B^0_s\to\Delta^+\overline{\Delta^+})
   &=&2 A'_E-A'_{PA},
   \non\\
A(\bar B^0_s\to\Sigma^{*+}\overline{\Sigma^{*+}})
   &=&
   A'_T+2A'_E-A'_{PA},
    \non\\
A(\bar B^0_s\to\Xi^{*0}\overline{\Xi^{*0}})
   &=&
    A'_T+A'_P+A'_E,
   \non\\
A(\bar B^0_s\to\Xi^{*-}\overline{\Xi^{*-}})
   &=&
   2A'_P+A'_{PA},
   \non\\
A(\bar B^0_s\to\Omega^{-}\overline{\Omega^{-}})
   &=&
   3A'_P+A'_{PA}.
\en
There are five independent amplitudes.


For $\overline B_q\to\BD$ decays, we have 
\be
A(B^-\to n\overline{\Delta^+})
   &=&\sq2 (B_{1T}+B_P-B_A),
\non\\
   A(B^-\to\Sigma^0\overline{\Sigma^{*+}})
    &=&B_{2T}-B_P,
   \non\\      
\sq2 B_A&\equiv&
   A(B^-\to\Xi^{-}\overline{\Xi^{*0}})
   =\sq2 A(B^-\to\Sigma^-\overline{\Sigma^{*0}}),
   \non\\
\sq3 A(B^-\to\Lambda\overline{\Sigma^{*+}})
   &=&-2 B_{1T}+B_{2T}-3B_P+3B_A,
   \non\\
A(B^-\to p\overline{\Delta^{++}})
   &=&\sq6 (B_{1T}-B_{2T}+2 B_P-B_A),  
\en
\be
A(\bar B^0\to n\overline{\Delta^0})
   &=&\sq2 (B_{1T}-B_E),
   \non\\
\sq2 A(\bar B^0\to\Sigma^{0}\overline{\Sigma^{*0}})
   &=&B_{2T}-B_E,
   \non\\ 
\sq2 B_P
  &\equiv& 
   A(\bar B^0\to\Sigma^{-}\overline{\Sigma^{*-}})
   =A(\bar B^0\to\Xi^{-}\overline{\Xi^{*-}}),
   \non\\     
\sq2 B_E
   &\equiv& A(\bar B^0\to\Sigma^{+}\overline{\Sigma^{*+}})
    =A(\bar B^0\to\Xi^{0}\overline{\Xi^{*0}}),
   \non\\
A(\bar B^0\to p\overline{\Delta^+})
   &=&\sq2(B_{1T}-B_{2T}+B_P-B_E),
   \non\\
\sq6 A(\bar B^0\to\Lambda\overline{\Sigma^{*0}})
   &=&-2B_{1T}+B_{2T}+3B_E,  
\en
and
\be
B_{1T}
    &\equiv& A(\bar B^0_s\to n\overline{\Sigma^{*0}}),
    \non\\
B_{2T}
     &\equiv& A(\bar B^0_s\to\Sigma^{0}\overline{\Xi^{*0}}),
    \non\\
\sq3 A(\bar B^0_s\to\Sigma^{-}\overline{\Xi^{*-}})
   &=&A(\bar B^0_s\to\Xi^{-}\overline{\Omega^-})
   =\sq6 B_P,
   \non\\
A(\bar B^0_s\to p\overline{\Sigma^{*+}})
   &=&\sq2 (B_{1T}-B_{2T}+B_P), 
   \non\\
\sq3 A(\bar B^0_s\to\Lambda\overline{\Xi^{*0}})
   &=&-2B_{1T}+B_{2T},
\en
where we have five independent amplitudes for these modes.
Those for $|\Delta S|=1$ transitions are given by
\be
A(B^-\to \Sigma^+\overline{\Delta^{++}})
   &=&-\sq6 (B'_{1T}-B'_{2T}+2B'_P-B'_A),
   \non\\
\sq2A(B^-\to\Sigma^0\overline{\Delta^+})
   &=&B'_{1T}-B'_{2T}+2B'_P-2B'_A,
   \non\\
A(B^-\to\Xi^{0}\overline{\Sigma^{*+}})
   &=&-\sq2(B'_{1T}+B'_P-B'_A),
   \non\\   
\sq2A(B^-\to\Xi^{-}\overline{\Sigma^{*0}})
&=&A(B^-\to\Sigma^-\overline{\Delta^0})=-\sq2 B'_A,
\non\\
\sq3 A(B^-\to\Lambda\overline{\Delta^{+}})
   &=&-B'_{1T}-B'_{2T},
\en
\be
A(\bar B^0\to \Sigma^{+}\overline{\Delta^+})
   &=&-\sq2 (B'_{1T}-B'_{2T}+B'_P),
   \non\\
A(\bar B^0\to\Sigma^{0}\overline{\Delta^0})
   &=&B'_{1T}-B'_{2T},
   \non\\
A(\bar B^0\to\Xi^{0}\overline{\Sigma^{*0}})
   &=&-B'_{1T},
   \non\\
\sq3 A(\bar B^0\to\Xi^{-}\overline{\Sigma^{*-}})
   &=&A(\bar B^0\to\Sigma^{-}\overline{\Delta^-})
   =-\sq6 B'_P,
   \non\\   
\sq3 A(\bar B^0\to\Lambda\overline{\Delta^0})
   &=&-B'_{1T}-B'_{2T},
\en
and
\be
A(\bar B^0_s\to p\overline{\Delta^+})
   &=&A(\bar B^0_s\to n\overline{\Delta^0})
   =-\sq2 B'_E,
   \non\\
A(\bar B^0_s\to\Sigma^{+}\overline{\Sigma^{*+}})
   &=&-\sq2 (B'_{1T}-B'_{2T}+B'_P-B'_E),
   \non\\
\sq2 A(\bar B^0_s\to\Sigma^{0}\overline{\Sigma^{*0}})
   &=&B'_{1T}-2 B'_{2T}-B'_E,
   \non\\
A(\bar B^0_s\to\Sigma^{-}\overline{\Sigma^{*-}})
   &=&A(\bar B^0_s\to\Xi^{-}\overline{\Xi^{*-}})
   =-\sq2 B'_P,
   \non\\   
A(\bar B^0_s\to\Xi^{0}\overline{\Xi^{*0}})
   &=&-\sq2 (B'_{1T}-B'_E),
   \non\\
\sq6 A(\bar B^0_s\to\Lambda\overline{\Sigma^{*0}})
   &=&-B'_{1T}-B'_{2T}+3B'_E.
\en

For $\overline B_q\to \DB$ decays, we have
\be
A(B^-\to\Delta^0\overline{p})
   &=&
     -\sq2A(B^-\to\Sigma^{*0}\overline{\Sigma^{+}})
   =\sq2(C_{1T}-C_P+C_A),
\non\\   
A(B^-\to\Delta^-\overline{n})
   &=& 
   =-\sq3 A(B^-\to\Xi^{*-}\overline{\Xi^{0}})
  =\sq6 A(B^-\to\Sigma^{*-}\overline{\Sigma^{0}})
\non\\
   &=&-\sq2 A(B^-\to\Sigma^{*-}\overline{\Lambda})
   =\sq6 C_A,
\en
\be
A(\bar B^0\to\Delta^+\overline{p})
   &=&\sq2(C_{2T}+C_P-C_E),
\non\\
\sq2 A(\bar B^0\to\Sigma^{*0}\overline{\Sigma^{0}})
   &=&
   C_{1T}-C_E,
   \non\\
\sq3A(\bar B^0\to\Sigma^{*-}\overline{\Sigma^{-}})
  & =&\sq3A(\bar B^0\to\Xi^{*-}\overline{\Xi^{-}})
   =-A(\bar B^0_s\to\Delta^-\overline{\Sigma^{-}})
\non\\   
   &=&
   -\sq3A(\bar B^0_s\to\Sigma^{*-}\overline{\Xi^{-}})
   =\sq6 C_P,
\non\\   
A(\bar B^0\to\Sigma^{*+}\overline{\Sigma^{+}})
   &=&A(\bar B^0\to\Xi^{*0}\overline{\Xi^{0}})
   =\sq2 C_E,
   \non\\
A(\bar B^0\to\Delta^0\overline{n})
   &=&\sq2(C_{1T}+C_{2T}-C_E),
   \non\\
\sq6  A(\bar B^0\to\Sigma^{*0}\overline{\Lambda})
   &=&C_{1T}+2C_{2T}-3C_E,   
\en
and
\be
A(\bar B^0_s\to \Delta^+\overline{\Sigma^{+}})
   &=&-\sq2 (C_{2T}+C_P),
\non\\
A(\bar B^0_s\to\Delta^0\overline{\Sigma^{0}})
   &=&C_{2T},
\non\\
A(\bar B^0_s\to\Sigma^{*0}\overline{\Xi^{0}})
   &=&-(C_{1T}+C_{2T}),
\non\\   
\sq3 A(\bar B^0_s\to\Delta^0\overline{\Lambda})
   &=&-(2C_{1T}+C_{2T}),
\en
where we have five independent amplitudes.
Similarly, the amplitudes for $|\Delta S|=1$ transitions are given by
\be
\sq2 A(B^-\to\Sigma^{*0}\overline{p})
   &=&-A(B^-\to\Xi^{*0}\overline{\Sigma^{+}})
   =\sq2 (C'_{1T}-C'_P+C'_A),
   \non\\
\sq3A(B^-\to\Sigma^{*-}\overline{n})
   &=&
\sq6 A(B^-\to\Xi^{*-}\overline{\Sigma^{0}})
   =   
-A(B^-\to\Omega^-\overline{\Xi^{0}})
\non\\
    &=&
-\sq2 A(B^-\to\Xi^{*-}\overline{\Lambda})
   =\sq6 C'_A,    
\en
\be
A(\bar B^0\to \Sigma^{*+}\overline{p})
   &=&\sq2 (C'_{2T}+C'_P),
 \non\\
A(\bar B^0\to\Sigma^{*0}\overline{n})
   &=&C'_{1T}+C'_{2T},
   \non\\
A(\bar B^0\to\Xi^{*0}\overline{\Sigma^{0}})
   &=&C'_{1T},
   \non\\
\sq3 A(\bar B^0\to\Xi^{*-}\overline{\Sigma^{-}})
   &=&  
   A(\bar B^0\to\Omega^-\overline{\Xi^{-}})
   =-\sq3A(\bar B^0_s\to\Sigma^{*-}\overline{\Sigma^{-}})
 \non\\  
   &=&-\sq3A(\bar B^0_s\to\Xi^{*-}\overline{\Xi^{-}})
   =\sq6 C'_P,
\non\\      
\sq3 A(\bar B^0\to\Xi^{*0}\overline{\Lambda})
   &=&-(C'_{1T}+2C'_{2T}),
\en
and
\be
A(\bar B^0_s\to\Delta^+\overline{p})
   &=&
   A(\bar B^0_s\to\Delta^0\overline{n})
   =-\sq2 C'_E,
\non\\
A(\bar B^0_s\to\Sigma^{*+}\overline{\Sigma^{+}})
   &=&-\sq2(C'_{2T}+C'_P-C'_E),
\non\\
\sq2 A(\bar B^0_s\to\Sigma^{*0}\overline{\Sigma^{0}})
   &=& C'_{2T}-C'_E,
\non\\   
A(\bar B^0_s\to\Xi^{*0}\overline{\Xi^{0}})
   &=&-\sq2(C'_{1T}+C'_{2T}-C'_E),
\non\\
\sq6 A(\bar B^0_s\to\Sigma^{*0}\overline{\Lambda})
   &=&-(2 C'_{1T}+C'_{2T}-3C'_E).
\en


For $\overline B_q\to\BB$ decays, we have
\be
A(B^-\to n\overline{p})
   &=&-D_{1T}+ D_P+D_{1A},
\non\\
\sq2 A(B^-\to\Sigma^{0}\overline{\Sigma^{+}})
   &=&2 D_{3T}-D_P-D_{1A}+D_{2A},
\non\\   
\sq2 A(B^-\to\Sigma^{-}\overline{\Sigma^{0}})
   &=&D_{1A}-D_{2A},
\non\\
\sq6 A(B^-\to\Sigma^{-}\overline{\Lambda})
   &=&D_{1A}+D_{2A},
\non\\  
A(B^-\to\Xi^{-}\overline{\Xi^{0}})
   &=&D_{2A},
\non\\
\sq6 A(B^-\to\Lambda\overline{\Sigma^+})
   &=&-2D_{1T}+2D_{3T}+D_P+D_{1A}+D_{2A},
\en
\be
A(\bar B^0\to p\overline{p})
   &=&-D_{2T}+2D_{4T}
          +D_{1E}+D_{2E},
   \non\\
A(\bar B^0\to n\overline{n})
   &=&-D_{1T}-D_{2T}+D_{2E},
\non\\
A(\bar B^0\to\Sigma^{+}\overline{\Sigma^{+}})
   &=&D_{1E}+D_{2E},
\non\\
2A(\bar B^0\to\Sigma^{0}\overline{\Sigma^{0}})
   &=&-2D_{3T}+D_{1E}+D_{2E}+D_{PA},
\non\\
2\sq3 A(\bar B^0\to\Sigma^{0}\overline{\Lambda})
   &=&2D_{3T}+4D_{4T}+D_{1E}-D_{2E}+D_{PA},
\non\\
A(\bar B^0\to\Sigma^{-}\overline{\Sigma^{-}})
   &=&-D_P+D_{PA},
\non\\   
A(\bar B^0\to\Xi^{0}\overline{\Xi^{0}})
   &=&D_{2E},
 \non\\
A(\bar B^0\to\Xi^{-}\overline{\Xi^{-}})
   &=&D_{PA},
   \non\\
2\sq3 A(\bar B^0\to\Lambda\overline{\Sigma^{0}})
   &=&2D_{1T}-2D_{3T}+D_{1E}-D_{2E}+D_{PA},
\non\\
6A(\bar B^0\to\Lambda\overline{\Lambda})
   &=&-2D_{1T}-4D_{2T}+2D_{3T}+4D_{4T}+D_{1E}+5D_{2E}+D_{PA},  
\en
and
\be
A(\bar B^0_s\to p\overline{\Sigma^{+}})
   &=&D_{2T}-2D_{4T},
\non\\
\sq2 A(\bar B^0_s\to n\overline{\Sigma^{0}})
   &=&-D_{2T},
\non\\
\sq6 A(\bar B^0_s\to n\overline{\Lambda})
   &=&2D_{1T}+D_{2T}   
\non\\   
A(\bar B^0_s\to\Sigma^{0}\overline{\Xi^{0}})
   &=&\sq2(D_{3T}+D_{4T}),
\non\\
A(\bar B^0_s\to\Sigma^{-}\overline{\Xi^{-}})
   &=&-D_P,
   \non\\   
\sq3A(\bar B^0_s\to\Lambda\overline{\Xi^0})
   &=&\sq2(-D_{1T}-D_{2T}+D_{3T}+D_{4T}),
\en
where we need ten independent amplitudes for these modes.
Similarly the amplitudes for $|\Delta S|=1$ transitions are given by
\be
\sq2 A(B^-\to\Sigma^{0}\overline{p})
   &=&-D'_{1T}+2D'_{3T}+D'_{2A},
\non\\
A(B^-\to\Sigma^{-}\overline{n})
   &=&D'_{2A},
\non\\
A(B^-\to\Xi^{0}\overline{\Sigma^{+}})
   &=&-D'_{1T}+D'_P+D'_{1A},
\non\\
\sq2A(B^-\to\Xi^{-}\overline{\Sigma^{0}})
   &=&D'_{1A},
\non\\  
\sq6 A(B^-\to\Xi^{-}\overline{\Lambda})
   &=&D'_{1A}-2D'_{2A},
\non\\      
\sq6 A(B^-\to\Lambda\overline{p})
   &=&D'_{1T}+2D'_{3T}-2D'_P-2D'_{1A}+D'_{2A},       
\en
\be
A(\bar B^0\to \Sigma^{+}\overline{p})
   &=& D'_{2T}-2D'_{4T},
   \non\\
\sq2 A(\bar B^0\to\Sigma^{0}\overline{n})
   &=&-D'_{1T}-D'_{2T}+2D'_{3T}+2D'_{4T},
\non\\
\sq2 A(\bar B^0\to\Xi^{0}\overline{\Sigma^{0}})
   &=&D'_{1T},
\non\\
\sq6 A(\bar B^0\to\Xi^{0}\overline{\Lambda})
   &=&-D'_{1T}-2D'_{2T},
   \non\\ 
A(\bar B^0\to\Xi^{-}\overline{\Sigma^{-}})
   &=&-D'_P,
\non\\   
\sq6 A(\bar B^0\to\Lambda\overline{n})
   &=&D'_{1T}+D'_{2T}+2D'_{3T}+2D'_{4T},
\en
and
\be
A(\bar B^0_s\to p\overline{p})
   &=&D'_{1E}+D'_{2E},
\non\\
A(\bar B^0_s\to n\overline{n})
   &=&D'_{2E},
\non\\
A(\bar B^0_s\to\Sigma^{+}\overline{\Sigma^{+}})
   &=&-D'_{2T}+2D'_{4T}
           +D'_{1E}+D'_{2E},
\non\\
2A(\bar B^0_s\to\Sigma^{0}\overline{\Sigma^{0}})
   &=&-D'_{2T}+2D'_{4T}+D'_{1E}+D'_{2E}+D'_{PA},
\non\\
2\sq3 A(\bar B^0_s\to\Sigma^{0}\overline{\Lambda})
   &=&2D'_{1T}+D'_{2T}-4D'_{3T}-2D'_{4T}
           +D'_{1E}-D'_{2E}+D'_{PA},
\non\\     
A(\bar B^0_s\to\Sigma^{-}\overline{\Sigma^{-}})
   &=&D'_{PA},
\non\\   
A(\bar B^0_s\to\Xi^{0}\overline{\Xi^{0}})
   &=&-D'_{1T}-D'_{2T}+D'_{2E},
\non\\
A(\bar B^0_s\to\Xi^{-}\overline{\Xi^{-}})
   &=&-D'_P+D'_{PA},
\non\\     
2\sq3A(\bar B^0_s\to\Lambda\overline{\Sigma^{0}})
   &=&D'_{2T}+2D'_{4T}+D'_{1E}-D'_{2E}+D'_{PA},
   \non\\
6A(\bar B^0_s\to\Lambda\overline{\Lambda})
   &=&-2D'_{1T}-D'_{2T}-4D'_{3T}-2D'_{4T}
           +D'_{1E}+5D'_{2E}+D'_{PA}.
\en
\begin{table}[t!]
\footnotesize{
\caption{\label{tab:DBD} Branching ratios of baryon dominant decay modes \cite{PDG}. The $n$ in $\pi^{n,n\pm1}$ follows the charge of the decaying baryon.
}
\begin{ruledtabular}
\begin{tabular}{lllllll}
          & $p \pi^+$
          & $p \pi^0$
          & $p \pi^-$
          & $n\pi^+$
          & $n\pi^0$
          & $n \pi^-$
          \\
\hline 
$\Delta^{++} $
          & 100\%
          & 
           \\
$\Delta^{+} $
          &
          & 2/3
          &
          & 1/3
          \\
$\Delta^{0} $
          &
          &
          & 1/3
          &
          & 2/3
          \\
$\Delta^- $
          &
          &
          &
          &
          &
          & 100\%
           \\                   
$\Lambda$
          &
          &
          & $(63.9\pm0.5)$\%
          &
          &  $(48.31\pm0.30)$\%
           \\
$\Sigma^{+} $
          &
          & $(51.57\pm0.30)$\%
          &
          & $(48.31\pm0.30)$\% 
          \\ 
$\Sigma^{-} $
          &
          &
          &
          &
          & 
          & $(99.848\pm0.005)$\% 
          \\ 
\hline          
          & $\Lambda\pi^n$
          & $\Lambda\gamma$
          & $\Sigma^+\pi^{n-1}$
          & $\Sigma^0\pi^n$
          & $\Sigma^-\pi^{n+1}$
          \\
\hline
$\Sigma^0$
          &
          & 100\%
          \\
$\Xi^0$
          & $(99.525\pm0.012)$\%
          &
          \\
$\Xi^-$
         & $(99.887\pm0.035)$\%
         &
         \\
$\Sigma^{*+}$
          & $(87.0\pm1.5)$\%
          &
          &$(5.8\pm0.8)$\%
          &$(5.8\pm0.8)$\%
          &
          \\
$\Sigma^{*0}$
          & $(87.0\pm1.5)$\%
          & $(1.25^{+0.13}_{-0.12})$\%
          &$(5.8\pm0.8)$\%
          &
          &$(5.8\pm0.8)$\%
          \\ 
$\Sigma^{*-}$
          & $(87.0\pm1.5)$\%
          &
          &
          &$(5.8\pm0.8)$\%
          &$(5.8\pm0.8)$\%
          \\
\hline          
          & $\Lambda K^-$
          & $\Xi^0\pi^0$
          & $\Xi^0\pi^-$
          & $\Xi^-\pi^{+}$
          & $\Xi^-\pi^{0}$
          \\            
\hline
$\Xi^{*0}$
          &
          & 1/3
          &
          & 2/3
          \\
$\Xi^{*-}$
         &
         & 
         & 2/3
         &
         & 1/3
         \\
$\Omega^-$
          & $(67.8\pm0.7)$\%
          &
          & $(23.6\pm0.6)$\%
          &
          &$(8.6\pm0.4)$\%
          \\                                                                                                                                                                    
\end{tabular}
\end{ruledtabular}
}
\end{table}

\section{Branching ratios of baryon dominant decay modes}

In this appendix we collect dominant decay branching ratios of ground state octet and ducuplet baryons. The informations are shown in Table~XI. They will be useful in the discussions of the accessibilities of searching of the charmless two-body baryonic modes. 
We note that (i) $\Delta^{++,0}$, $\Lambda$, $\Xi^-$, $\Sigma^{*\pm}$, $\Xi^{*0}$ and $\Omega^-$ have non-suppressed decay modes of final states with all charged particles, (ii) $\Delta^+$, $\Sigma^{+,0}$, $\Xi^0$, $\Sigma^{*0}$ and $\Xi^{*-}$ can be detected by detecting a $\pi^0$ or $\gamma$, (iii) while one needs to deal with $n$ in detecting $\Delta^-$ and $\Sigma^-$.




\begin{thebibliography}{99}

\bibitem{Aaij:2013fla} 
  R. Aaij {\it et al.}  [LHCb Collaboration],
  Phys.\ Rev.\ D {\bf 88}, 052015 (2013)
  [arXiv:1307.6165 [hep-ex]].

\bibitem{Aaij:2013fta} 
  R. Aaij {\it et al.}  [LHCb Collaboration],
  JHEP {\bf 1310}, 005 (2013)
  [arXiv:1308.0961 [hep-ex]].

\bibitem{Tsai:2007pp} 
  Y.~-T.~Tsai {\it et al.}  [BELLE Collaboration],
  Phys.\ Rev.\ D {\bf 75}, 111101 (2007)
  [hep-ex/0703048].

\bibitem{Wang:2007as} 
  M.~-Z.~Wang {\it et al.}  [Belle Collaboration],
  Phys.\ Rev.\ D {\bf 76}, 052004 (2007)
  [arXiv:0704.2672 [hep-ex]].

\bibitem{Wei:2007fg} 
  J.~T.~Wei {\it et al.}  [BELLE Collaboration],
  Phys.\ Lett.\ B {\bf 659}, 80 (2008)
  [arXiv:0706.4167 [hep-ex]].


\bibitem{PDG}
J.~Beringer {\it et al.}  [Particle Data Group Collaboration],
  Phys.\ Rev.\ D {\bf 86}, 010001 (2012).

\bibitem{Hou:2000bz}
W.S.~Hou and A.~Soni,
Phys.\ Rev.\ Lett.\  {\bf 86}, 4247 (2001) [hep-ph/0008079].

\bibitem{Chua:2001vh}
C.K.~Chua, W.S.~Hou and S.Y.~Tsai,
Phys.\ Rev.\ D {\bf 65}, 034003 (2002) [hep-ph/0107110].

\bibitem{Chua:2003it} 
  C.~-K.~Chua,
  Phys.\ Rev.\ D {\bf 68}, 074001 (2003)
  [hep-ph/0306092].

\bibitem{Chua:2001xn}
C.K.~Chua, W.S.~Hou and S.Y.~Tsai,
Phys.\ Lett.\ B {\bf 528}, 233 (2002) [hep-ph/0108068].

\bibitem{Cheng:2001tr}
H.Y.~Cheng and K.C.~Yang,
Phys.\ Rev.\ D {\bf 66}, 014020 (2002) [hep-ph/0112245].

\bibitem{Cheng:2002fp}
H.Y.~Cheng and K.C.~Yang,
Phys.\ Rev.\ D {\bf 66}, 094009 (2002) [hep-ph/0208185].

\bibitem{Chua:2002wn}
C.K.~Chua, W.S.~Hou and S.Y.~Tsai,
Phys.\ Rev.\ D {\bf 66}, 054004 (2002) [hep-ph/0204185].

\bibitem{Chua:2002yd}
C.~-K.~Chua and W.~-S.~Hou,
  Eur.\ Phys.\ J.\ C {\bf 29}, 27 (2003)
  [hep-ph/0211240].
 
\bibitem{Geng}
C.~Q.~Geng and Y.~K.~Hsiao,
  Phys.\ Lett.\ B {\bf 610}, 67 (2005)
  [hep-ph/0405283];
  Phys.\ Rev.\ D {\bf 75}, 094013 (2007)
  [hep-ph/0702249];
  Phys.\ Rev.\ D {\bf 74}, 094023 (2006)
  [hep-ph/0606141].



\bibitem{review}
H.~-Y.~Cheng and J.~G.~Smith,
  Ann.\ Rev.\ Nucl.\ Part.\ Sci.\  {\bf 59}, 215 (2009)
  [arXiv:0901.4396 [hep-ph]].

\bibitem{review1}
 H.~-Y.~Cheng,
  Int.\ J.\ Mod.\ Phys.\ A {\bf 21}, 4209 (2006)
  [hep-ph/0603003].

\bibitem{Deshpande:1987nc}
N.~G.~Deshpande, J.~Trampetic and A.~Soni,
Mod.\ Phys.\ Lett.\  {\bf 3A}, 749 (1988).

\bibitem{Jarfi:1990ej}
M.~Jarfi, O.~Lazrak, A.~Le Yaouanc, L.~Oliver, O.~Pene and
J.~C.~Raynal,
Phys.\ Rev.\ D {\bf 43}, 1599 (1991).


\bibitem{Cheng:2001ub}
H.~Y.~Cheng and K.~C.~Yang,
Phys.\ Rev.\ D {\bf 65}, 054028 (2002) [Erratum-ibid.\ D {\bf 65},
099901 (2002)] [arXiv:hep-ph/0110263].

\bibitem{Chernyak:ag}
V.~L.~Chernyak and I.~R.~Zhitnitsky,
Nucl.\ Phys.\ B {\bf 345}, 137 (1990).

\bibitem{Ball:1990fw}
P.~Ball and H.~G.~Dosch,
Z.\ Phys.\ C {\bf 51}, 445 (1991).

\bibitem{Chang:2001jt}
C.~H.~Chang and W.~S.~Hou,
Eur.\ Phys.\ J.\ C {\bf 23}, 691 (2002) [arXiv:hep-ph/0112219].

\bibitem{Gronau:1987xq}
M.~Gronau and J.~L.~Rosner,
Phys.\ Rev.\ D {\bf 37}, 688 (1988).


\bibitem{He:re}
X.~G.~He, B.~H.~McKellar and D.~d.~Wu,
Phys.\ Rev.\ D {\bf 41}, 2141 (1990).

\bibitem{Sheikholeslami:fa}
S.~M.~Sheikholeslami and M.~P.~Khanna,
Phys.\ Rev.\ D {\bf 44}, 770 (1991).

\bibitem{Luo:2003pv}
Z.~Luo and J.~L.~Rosner,
  Phys.\ Rev.\ D {\bf 67}, 094017 (2003)
  [hep-ph/0302110].

\bibitem{Zeppenfeld:1980ex}
D.~Zeppenfeld,
Z.\ Phys.\ C {\bf 8}, 77 (1981).

\bibitem{Chau:tk}
L.~L.~Chau and H.~Y.~Cheng,
Phys.\ Rev.\ D {\bf 36}, 137 (1987).

\bibitem{Chau:1990ay}
L.~L.~Chau, H.~Y.~Cheng, W.~K.~Sze, H.~Yao and B.~Tseng,
Phys.\ Rev.\ D {\bf 43}, 2176 (1991) [Erratum-ibid.\ D {\bf 58},
019902 (1998)].

\bibitem{Gronau:1994rj}
M.~Gronau, O.~F.~Hernandez, D.~London and J.~L.~Rosner,
Phys.\ Rev.\ D {\bf 50}, 4529 (1994) [arXiv:hep-ph/9404283].


\bibitem{Gronau:1995hn}
M.~Gronau, O.~F.~Hernandez, D.~London and J.~L.~Rosner,
Phys.\ Rev.\ D {\bf 52}, 6374 (1995) [arXiv:hep-ph/9504327].

\bibitem{Savage:ub}
M.~J.~Savage and M.~B.~Wise,
Phys.\ Rev.\ D {\bf 39}, 3346 (1989) [Erratum-ibid.\ D {\bf 40},
3127 (1989)].

\bibitem{Brodsky:1980sx}
S.J.~Brodsky, G.P.~Lepage and S.A.~Zaidi,
Phys.\ Rev.\ D {\bf 23}, 1152 (1981).



\bibitem{Wang:2003yi}
M.~Z.~Wang {\it et al.}  [Belle Collaboration],
  Phys.\ Rev.\ Lett.\  {\bf 90}, 201802 (2003)
  [hep-ex/0302024].

\bibitem{Lepage:1979za}
G.P.~Lepage and S.J.~Brodsky,
Phys.\ Rev.\ Lett.\  {\bf 43}, 545 (1979) [Erratum-ibid.\  {\bf
43}, 1625 (1979)].





\bibitem{Buras} 
  A.~J.~Buras,
  hep-ph/9806471.


\bibitem{Beneke:2001ev} 
  M.~Beneke, G.~Buchalla, M.~Neubert and C.~T.~Sachrajda,
  Nucl.\ Phys.\ B {\bf 606}, 245 (2001)
  [hep-ph/0104110].

\bibitem{text}
T.~D.~Lee,
Contemp.\ Concepts Phys.\  {\bf 1}, 1 (1981);
H.~Georgi, {\it Weak Interactions And Modern Particle Theory},
Benjamin/Cummings, 1984.

\bibitem{Uspin}
N.~G.~Deshpande and X.~-G.~He,
  Phys.\ Rev.\ Lett.\  {\bf 75}, 1703 (1995)
  [hep-ph/9412393];
  M.~Gronau,
  Phys.\ Lett.\ B {\bf 492}, 297 (2000)
  [hep-ph/0008292];

\bibitem{CKMfitter}
J.~Charles {\it et al.}  [CKMfitter Group Collaboration],
  Eur.\ Phys.\ J.\ C {\bf 41}, 1 (2005)
  [hep-ph/0406184];
  updated results at http://ckmfitter.in2p3.fr.

\bibitem{FSI}
  H.~-Y.~Cheng, C.~-K.~Chua and A.~Soni,
  Phys.\ Rev.\ D {\bf 71}, 014030 (2005)
  [hep-ph/0409317];
  C.~-K.~Chua,
  Phys.\ Rev.\ D {\bf 78}, 076002 (2008)
  [arXiv:0712.4187 [hep-ph]].


\bibitem{Moroi:1995fs}
T.~Moroi,
arXiv:hep-ph/9503210;
S.~Deser, V.~Pascalutsa and A.~Waldron,
Phys.\ Rev.\ D {\bf 62}, 105031 (2000) [arXiv:hep-th/0003011].





\end{thebibliography}
\end{document}